\documentclass[10pt, aps, prd, amsmath, floats, floatfix, twocolumn, notitlepage,
superscriptaddress, nofootinbib, showpacs, longbibliography]{revtex4-2}
\usepackage{booktabs}
\usepackage[utf8]{inputenc}
\usepackage[T1]{fontenc}
\usepackage{subfigure}
\usepackage{comment}
\usepackage{bm}
\usepackage{multirow}
\usepackage[normalem]
{ulem}\usepackage{mathtools,amsmath,amssymb,amsfonts,mathrsfs,eucal,graphicx,tensor,csquotes,accents,commath,chngcntr,siunitx}
\usepackage[unicode]{hyperref}
\hypersetup{colorlinks=true, citecolor=MidnightBlue,
            linkcolor=MidnightBlue, urlcolor=MidnightBlue, linktocpage=true}
\usepackage[normalem]{ulem}
\usepackage{orcidlink}
\usepackage{natbib}

\usepackage{braket}

\DeclareMathAlphabet{\mathpzc}{OT1}{pzc}{m}{it}
\definecolor{darkgreen}{rgb}{0.0, 0.6, 0.0}

\newcommand{\Ord}{\mathcal{O}}

\newcommand{\mc}{\mathcal}

\newcommand{\mr}{\mathrm}
\newcommand{\lb}{\left(}
\newcommand{\rb}{\right)}

\newcommand{\ls}{\left[}
\newcommand{\rs}{\right]}
\newcommand{\nnm}{\nonumber}

\newcommand{\thorn}{\mathord{\text{þ}}}

\usepackage[table,xcdraw,svgnames,dvipsnames]{xcolor}

\begin{document}

\date{\today}

\title{Strong-field Gravitational-Wave Lensing in the Kerr Background}

%\author{}
%\email{}

% \author{Rajes, Saketh, Anuj}
\author{M. V. S. Saketh\orcidlink{0000-0003-2777-7891}}%
\email{venkata.saketh@icts.res.in}
\affiliation{International Centre for Theoretical Sciences, Tata Institute of Fundamental Research, Bangalore 560089, India.}%
\author{Rajes Ghosh\orcidlink{0000-0002-1264-938X}}%
\email{rajes.ghosh@icts.res.in}
\affiliation{International Centre for Theoretical Sciences, Tata Institute of Fundamental Research, Bangalore 560089, India.}%
\author{Anuj Mishra\orcidlink{0000-0002-2580-2339}}%
\email{anuj.mishra@icts.res.in}
\affiliation{International Centre for Theoretical Sciences, Tata Institute of Fundamental Research, Bangalore 560089, India.}%

\begin{abstract}
Gravitational lensing of gravitational waves (GWs) can encode valuable information about the properties of the intervening lens, but most existing studies remain restricted to the small-deflection, weak-field regime. To bridge this crucial gap, this work presents the first systematic analysis of strong-field, wave-optical GW lensing by a Kerr black hole (BH), extending recent results obtained in Ref.~\citep{Chan:2025wgz} for non-rotating lens to the astrophysically more relevant case of spinning-lens. Using the Mano-Suzuki-Takasugi formalism, we define and compute the strong-field scattering factor (SFSF) and show that the the spin produces characteristic modifications to the lensed waveform, and high-frequency incident radiation is not strongly absorbed by the BH lens, contrary to earlier claims in Ref.~\citep{Chan:2025wgz}. We further derive an explicit expression for the observed waveform for the general source-lens-observer configuration, showcasing the distortions produced by the scattering and quantifying their departure from the Schwarzschild case. Specializing to on-axis scattering, a mismatch analysis for a \texttt{GW150914}-like source lensed by a Kerr BH of mass $M=10^2~\mathrm{M}_\odot$ situated $100GM/c^2$ away from the source reveals percentage-level deviations from the direct (unscattered) wave at scattering angles near $30^\circ$, across a range of lens spin values. The mismatch generally decreases as the scattering angle increases, but this behavior can change substantially when polarization mixing induced by scattering becomes significant. In such cases, components that are absent or suppressed in the direct signal may become appreciable once scattering effects are taken into account. For a fixed scattering angle, however, the mismatch shows only a weak dependence on the BH spin in the case of on-axis scattering, which may improve for more general off-axis considerations. The framework developed here offers a unified treatment of strong-field GW scattering in Kerr spacetime and provides tools for interpreting future high-precision observations of compact-object lenses in the wave-optics regime.
\end{abstract}

\maketitle

\section{Introduction} \label{secI}
Since the first direct detection of gravitational waves (GWs) with \texttt{GW150914}~\citep{LIGOScientific:2016aoc}, identifying GW signals from compact binary coalescences (CBCs) with the LIGO~\citep{LIGOScientific:2014pky} and Virgo~\citep{VIRGO:2014yos} has become a routine. 
With more than $200$ confident detections to date~\citep{LIGOScientific:2018mvr,  LIGOScientific:2020ibl, LIGOScientific:2021usb,  KAGRA:2021vkt, LIGOScientific:2025slb}, we are now able to probe the most extreme and dynamic regions of spacetime like never before~\citep{LIGOScientific:2016lio, LIGOScientific:2017zic, LIGOScientific:2019fpa, LIGOScientific:2021sio, LIGOScientific:2025obp, Siegel:2025xgb}. The ever-growing catalog of CBCs has also enabled population studies of stellar-mass black holes (BHs)~\citep{LIGOScientific:2020kqk,KAGRA:2021duu,LIGOScientific:2025pvj}, revealing several surprising features, such as lower/upper mass-gap events~\citep{Mehta:2021fgz,Zevin:2020gma,Edelman_2021}, high-mass and misaligned-spin BHs~\citep{Pierra:2024fbl, LIGOScientific:2025pvj}, the presence of hybrid binaries (e.g., BH-neutron star systems)~\citep{Chattopadhyay:2022cnp, Qin:2024ojw}, and indications of hierarchical mergers~\citep{Li:2022gly, Tagawa:2021ofj}. The GW detection rate, currently several events per month, is expected to increase significantly as the sensitivities of second-generation ground-based detectors, including KAGRA~\cite{Somiya:2011np, Aso:2013eba, KAGRA:2018plz, KAGRA:2020tym}, improve and the addition of LIGO-India~\cite{LigoIndia, Unnikrishnan:2013qwa, Ajith:2024inj} strengthens the global network. The growth will be further accelerated with the advent of next-generation observatories such as the Einstein Telescope (ET)~\cite{Punturo:2010zz, Hild:2010id}, Cosmic Explorer (CE)~\cite{Reitze:2019iox, LIGOScientific:2016wof, Regimbau:2016ike}, and the space-based Laser Interferometer Space Antenna (LISA)~\citep{LISA:2017}. With these advancements, a host of subtle phenomena are expected to become observable. Among these, GW lensing~\citep{1936Sci....84..506E, PhysRev.51.290, 1992grle.book.....S} stands out as a particularly promising probe, allowing us to explore structures across a wide range of astrophysical and cosmological scales.
   
Gravitational lensing of GWs arise when spacetime curvature produced by an intervening mass alters the propagation of the wavefront~\citep{1971NCimB...6..225L, Ohanian:1974ys}. In the limit where the GW wavelength $\lambda$ is much shorter than the characteristic lens scale (i.e., $\lambda \ll 2GM/c^2$, where $M$ is the lens mass), gravitational lensing can be described using ray optics analogous to the standard electromagnetic case, resulting in multiple images of the same source that arrive at the detector with characteristic time delays and relative magnifications~\cite{Bernardeau:1999mh, Takahashi:2016jom, Dai:2017huk, Smith:2017jdz, Liao:2017ioi, Haris:2018vmn, Li:2018prc, Broadhurst:2018saj, Broadhurst:2019ijv, Broadhurst:2020cvm, Dai:2020tpj, Ezquiaga:2020gdt, Caliskan:2022wbh, Vijaykumar:2022dlp, Barsode:2024zwv}. 
However, when the GW wavelength becomes comparable to or larger than the Schwarzschild radius of the lens (i.e., $\lambda \gtrsim 2GM/c^2$), the geometric-optics approximation breaks down, necessitating a full wave-optics treatment~\citep{1986ApJ...307...30D, Nakamura:1997sw, Nakamura:1999uwi, 2003ApJ...595.1039T}. 
% \textcolor{blue}{For current ground-based detectors, an intervening compact object with mass $\sim 10-10^5~$M$_\odot$ can induce observable wave-optics signatures. For space-based detectors such as LISA, which are sensitive to much lower GW frequencies (and hence probe much larger lens masses), wave-optics effects are expected to be particularly prominent~\citep{Caliskan:2022hbu, Pijnenburg:2024btj}. In this regime, interference and diffraction effects become significant and produce characteristic frequency-dependent modulations in the observed GW amplitude and phase, which has been studied extensively for various lens models in the context of parameter estimation~\citep{}. In fact, these effects can bias the estimated parameters or other analyses such as tests of general relativity~\citep{Jung:2017flg, Shan:2020esq, Mishra:2021xzz, Meena:2022unp, Mishra:2023ddt, Mishra:2023vzo, Cremonese:2021puh, Cremonese:2021ahz,  Shan:2023ngi, Rao:2025poe}.} 
In this regime, interference and diffraction effects become significant, producing characteristic frequency-dependent modulations in the observed GW amplitude and phase, which can bias the estimated parameters or other analyses such as tests of general relativity~\citep{Jung:2017flg, Shan:2020esq, Mishra:2021xzz, Cremonese:2021puh, Cremonese:2021ahz, Meena:2022unp, Shan:2023ngi, Mishra:2023ddt, Mishra:2023vzo, Liu:2023emk, Liu:2024xxn, Lo:2024wqm,  Ezquiaga:2025gkd, Rao:2025poe}. For current ground-based detectors, an intervening compact object with mass $\sim 10-10^5~$M$_\odot$ can induce observable wave-optics signatures. For space-based detectors such as LISA, which are sensitive to much lower GW frequencies (and hence probe much larger lens masses), wave-optics effects are expected to be particularly prominent~\citep{Caliskan:2022hbu, Pijnenburg:2024btj}. GW lensing therefore naturally bridges two complementary regimes: the geometric-optics limit relevant for galaxy- and cluster-scale lenses, and the wave-optics limit essential for compact-object lenses such as stellar-mass and supermassive black holes (SMBHs).

When a wave traverses through the strong-field regime of a lens, additional relativistic effects become important as the lensing occurs ``deep'' within the gravitational potential of a compact object~\citep{Virbhadra:1999nm, Bozza:2002zj, Kumar_2022}. 
In this regime, the curvature is strong enough for the GW wavefront to undergo multiple deflections, and even partial orbiting around the lens, giving rise to intricate interference between different propagation paths. The resulting amplification factor encodes detailed information about the lens geometry~\citep{1986ApJ...307...30D, Nakamura:1997sw, Nakamura:1999uwi}. In the electromagnetic case, such strong-field lensing effects give rise to the so-called ``photon rings'' and higher-order images, which can be probed by the Event Horizon Telescope~\citep{EventHorizonTelescope:2019dse, EventHorizonTelescope:2022wkp}. For GWs, however, similar effects manifest as interference fringes and waveform modulations, potentially observable in high signal-to-noise ratio (SNR) events~\citep{2003ApJ...595.1039T, Bondarescu:2022srx, Mishra:2023ddt}.

Driven by these exciting prospects, the theoretical study of GW lensing has advanced considerably in recent years. Early studies primarily focused on point-mass and galactic-scale lenses, employing diffraction integrals and Kirchhoff-type formalisms to compute magnifications and time delays within the wave-optics regime~\citep{Ohanian:1974ys, 1986ApJ...307...30D, 1992grle.book.....S, Nakamura:1997sw, Baraldo:1999ny, 2003ApJ...595.1039T, Biesiada:2021pzo, Dalang:2021qhu, Tambalo:2022plm, Harikumar:2023gzh, Leung:2023lmq,Deka:2024ecp}. These approaches successfully captured the interference patterns characteristic of microlensed GWs and were later generalized to include realistic lens mass profiles, dark matter substructures, and cosmological corrections~\citep{Mishra:2021xzz, Meena:2022unp, Shan:2022xfx, Shan:2023ngi, Liu:2023ikc, Villarrubia-Rojo:2024xcj, Deka:2025vzx, Vujeva:2025nwg, Vujeva:2025kko, Bulashenko:2025vdx}. More recently, investigations of strong-field lensing by Schwarzschild BHs~\citep{Chan:2025wgz} in the long-wavelength regime have revealed a wealth of wave-optical phenomena, such as frequency-dependent amplification, intricate phase structures, and polarization effects absent in the geometric optic limit. However, this study is restricted to non-rotating lenses, leaving open the important question of how BH spin, one of the defining features of astrophysical BHs, affects GW propagation in the strong-field regime.

Including rotation is far more than a mathematical refinement, as lens spin can fundamentally alter its scattering behaviour. A spinning BH described by the Kerr geometry~\citep{Kerr:1963ud, Bardeen1972RotatingBH, Misner:1973prb} introduces frame dragging, breaks spherical symmetry and couples the wave’s polarization to the background rotation~\citep{Farooqui:2013rga, Gelles:2021kti, Frolov:2025bva}. In the geometric-optics limit, this leads to asymmetric deflection angles and time delays between prograde and retrograde trajectories~\citep{Bozza:2005tg, Sereno:2006ss, Sereno:2007gd}. In the wave-optical regime, however, the spin dependence becomes even more intricate as rotation-induced polarization-mixing and helicity-dependent scattering become important~\citep{Teukolsky:1972my, Teukolsky:1974yv, Futterman:1988ni, Glampedakis:2001cx, Dolan:2007ut, Dolan:2008kf}. Although classical analyses of wave scattering in Kerr spacetime~\citep{Kubota:2024zkv, Leite:2017hkm} have long highlighted these features, their detailed imprint within the framework of wave-optical gravitational lensing remains largely unexplored.

Motivated by these considerations, this work presents a systematic study of strong-field GW lensing by Kerr BHs, extending the framework of Ref.~\citep{Chan:2025wgz} beyond spherical symmetry. We aim to quantify the influence of spin on the deflection, amplification, and interference patterns of lensed GWs, thereby identifying potential observational signatures relevant to current and upcoming detectors. The analysis incorporates both the strong-field and long-wavelength limits (i.e., $\lambda \gtrsim 2G M/c^2$) to isolate key physical effects governing the phenomenon. Particular attention is paid to polarization-dependent features, which encode unique signatures of the lens spin parameter. A key outcome of our analysis is a rigorous derivation of the strong-field scattering factor (SFSF), which quantifies the effects of Kerr lensing with respect to the direct (unscattered) wave from the source to observer. Contrary to the earlier claims in Ref.~\cite{Chan:2025wgz}, our analysis demonstrates that the SFSF does not decay at high frequencies irrespective of the choice of lens spin. We also find that the resulting waveform modulations can leave observable signatures even in the era of current-generation detectors. 

In particular, for the special case of on-axis scattering, our mismatch analysis for a \texttt{GW150914}-like event (i.e., a binary with intrinsic parameters similar to those of \texttt{GW150914})
% \textcolor{blue}{(in terms of binary parameters, e.g., component masses and distance)}
shows a percentage-level deviations from the direct wave at scattering angle near $30^\circ$, which falls sharply at larger scattering angles. Here, we must emphasize that a percentage-level mismatch lies close to the detectability threshold $\sim (\text{SNR})^{-2}$, which is around $0.1-1\%$ for typical (and conservative) SNRs in the range of $10-30$ for the current ground-based detectors~\cite{Lindblom:2008cm, Damour:2010zb, Cornish:2011ys, Thompson:2025hhc}. By contrast, next-generation detectors will drive this threshold down to $10^{-4}-10^{-6}$ level, making a percentage-level or even smaller mismatches unambiguously detectable~\cite{Thompson:2025hhc}. Interestingly, at a fixed scattering angle, the mismatch for on-axis scattering shows only a mild sensitivity to the lens spin parameter, which may improve for more general off-axis scattering. Although our formulation is valid for off-axis scattering as well, it posses significant numerical challenges, which we aim to address in a dedicated future work. These results demonstrate that strong-field wave-optical lensing by Kerr BHs can generate potentially detectable distortions in GW signals and should therefore be accounted for in future searches targeting compact-object lensing in dense astrophysical environments.

Astrophysically, such strong-field lensing effects may arise in several realistic scenarios. For example, CBCs formed via dynamical formation channels in dense globular clusters~\citep{Rodriguez:2018rmd, Antonini:2020xnd, Ubach:2025dob} or active galactic nuclei (AGN) disks~\citep{2012MNRAS.425..460M, Tagawa:2019osr, Mckernan:2017ssq, Grobner:2020drr, Ubach:2025dob} may experience strong-field Kerr lensing of their emitted GWs by a central SMBH, where the spin of the lens plays a decisive role. A recent analysis of gravitational lensing of GWs emitted by a stellar-mass compact binary orbiting a spinning SMBH is presented in Ref.~\citep{Santos:2025ass}. Another promising setting involves hierarchical triple systems~\citep{Antonini:2013tea, Silsbee:2016djf, Fragione:2019zhm, Liu:2020gif, Ubach:2025dob}, in which an inner compact binary inspirals within the gravitational potential of an outer spinning BH. In such configurations, the emitted GWs can undergo strong-field Kerr lensing as they propagate through the curved spacetime near the outer companion. Although the overall occurrence rates of these scenarios remain uncertain, even a small fraction of detectable mergers (on the order of a few percent) occurring in such environments could provide valuable probes of gravity in the strong and dynamical regimes predicted by GR. Indeed, events such as \texttt{GW190521} (the most massive binary reported in GWTC-3), which has been suggested to originate within an AGN disk~\citep{Graham:2020gwr, Morton:2023wxg, Leong:2024nnx}, demonstrate that such environments may already be within observational reach of current ground-based detectors. With the advent of next-generation observatories offering enhanced sensitivity at lower frequencies, the prospects for detecting strong-field lensing signatures in GWs will only grow more promising.

The rest of the paper is organized as follows. In Sec.~\ref{secII}, we introduce Kerr scattering computations to derive the lensed waveform, highlighting the role of spin-induced effects. Sec.~\ref{secIII} discusses the observational implications and mismatch studies. We finally conclude in Sec.~\ref{secIV} by briefly summarizing the main results of this work and highlighting the key future directions. The Appendices~\ref{appA}, \ref{appB}, \ref{appC} and \ref{appD} contain various mathematical derivations and details for some of the results used in the main text. Going forward, we shall adopt geometric units with $G=c=1$ and the metric signature $(-,+,+,+)$.

\section{GWs Scattering by Kerr Lens} \label{secII}
\begin{figure}[h!]
\centering
\includegraphics[width=\linewidth]{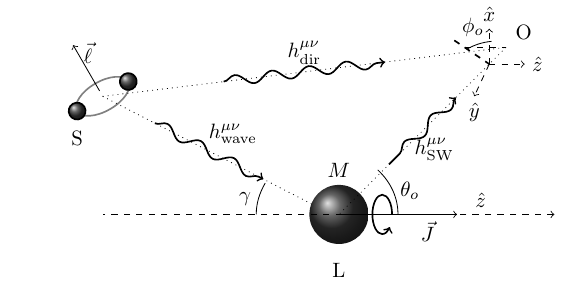}
\caption[Low frequency gravitational Raman/Compton scattering.]{The source (S) is a coalescing binary, which is situated close to a lensing BH (L) of mass $M$ and spin-angular momentum $J$. A portion of the radiation emitted by the source is scattered ($h^{\mu\nu}_{\rm SW}$) by the lens in the direction of the observer (O). The coordinate system is aligned so that the spin of the lensing BH is along the $z$-axis. The $x$-axis is chosen such that the source and lens lie on the $xz$-plane. The angular coordinates of the observer/detector (O) are denoted as $(\theta_o,\phi_o)$.  The scattered wave $h^{\mu\nu}_{\rm SW}$, at the observer can be computed in the framework of BH perturbation theory (BHPT). The radiation from the source can also reach the observer ``directly'' as $h^{\mu\nu}_{\rm dir}$. The direct component is only weakly affected by the lens for large $\theta_o\sim 30^\circ$.}
\label{sec:scatter}
\end{figure}

In order to investigate the gravitational lensing of GWs by the strong field of a Kerr BH, let us first setup a few basic notations. The source of GWs is considered to be a coalescing compact binary, while the lens is taken to be a rotating Kerr BH characterized by its mass $M$ and spin angular momentum $J=a\, M=\chi\, M^2$. We assume the source is sufficiently far away from the lens such that, at the lens location, the incident GWs can be well described as plane waves. We also have $r_{\rm SL} \gg R_L$, where $r_{\rm SL}$ is the physical lens-source distance and $R_L=2M$ is the Schwarzschild radius of the lens. For reasons that will be clear later, we shall primarily focus on configurations in which the source and the lens are relatively nearby, i.e., $r_{\rm SL} \ll r_{\rm SO} \sim r_{\rm LO}$, with $r_{\rm LO}$ ($r_{\rm SO}$) being the physical observer-lens (observer-source) distance. Under this assumption, the apparent angular position of the observer with respect to the lens, $\Omega_{\rm LO}=\{\theta_{\rm LO},\phi_{\rm LO}\}$, can be taken to be approximately the same as that with respect to the source, $\Omega_{\rm SO}=\{\theta_{\rm SO},\phi_{\rm SO}\}$. While treating the scattering problem, we shall use the lens-centred coordinates $x^\mu=\{t,x,y,z\}$ (unit vectors along the spatial directions will be denoted with hats) and their spherical polar analogues, with the $z$-axis chosen as the lens' spin axis. The observer's polar coordinates are then denoted as $\vec{r}_{\rm LO} \equiv (r_{\rm LO},\theta_{\rm LO}=\theta_o,\phi_{\rm LO}=\phi_o)$. The setup is schematically illustrated in Fig.~\ref{sec:scatter}.

\subsection{Incident Wave}
Since the lens is sufficiently away from the source, the emitted GWs will become almost plane-fronted before reaching the lens. In fact, in an intermediate region (asymptotic rest frame) between the source and the lens, one may consider that the incident wave
\begin{alignat}{3} \label{h}
h_{\mr{wave},\, h=\pm 2}^{\mu\nu}=\epsilon~\Re\lb\varepsilon_\pm^\mu \varepsilon_\pm^\nu e^{ i k\cdot x}\rb
\end{alignat}
is propagating over the flat background $\eta_{\mu \nu}$\footnote{This is actually not quite right, owing to the long range nature of the Coulombic $1/r$ potential, but the effect of that can be incorporated later as we shall see.}. Here, $\epsilon=|\mathcal{A}_{\pm}(\hat{n}_{\rm SL})|/r_{\rm SL}$ is a small real quantity associated with the amplitude $\mathcal{A}_{\pm}(\hat{n}_{\rm SL})=|\mathcal{A}_{\pm}(\hat{n}_{\rm SL})|\exp(i\psi_{\rm SL})$ of the wave emitted in the direction of lens ($\hat{n}_{\rm SL}$) and $r_{\rm SL}$ is the physical source-lens distance. Note also that the factor $\exp(i k.x)$ should be supplemented by $\exp(i \omega r_{\rm SL}+i\psi_{\rm SL})$ as well to incorporate the obvious source-to-lens propagation phase shift. However, in the subsequent discussions, we shall omit these kinematical factors, treating $\epsilon$ as a book-keeping parameter, for brevity and only reintroduce them during waveform computation. In addition, we have expressed the polarization tensor $e^{\pm}_{\mu \nu}(\hat{k})$ of the spin-$2$ (helicity $h=\pm 2$) graviton as a product of two spin-$1$ (helicity $\pm 1$) polarization vectors $\varepsilon_\pm^\mu(\hat{k})$. Note that, although a general incident wave will be a linear combination of both helicities, we shall restrict to positive-helicity ($h=+2$) incident wave in what follows, for brevity. The results for negative-helicity ($h=-2$) incident wave (or for a linear combination, in general) can be easily obtained from this result by exploiting the parity and time-reversal symmetries of the Kerr metric~\cite{Saketh:2022wap}.

Moreover, owing to the axisymmetry of the system, we are free to rotate about the the $z$-axis and hence, the spatial wave vector $\vec{k}$ can always be made to lie in the $xz$-plane in the initial asymptotic rest frame, at an angle $\gamma$ to the $z$-axis. In other words, $\vec{k} = \omega(\sin \gamma,0,\cos \gamma) \equiv \omega\, \hat{k}$, with $\omega>0$ being the frequency. Now, to ensure the transverse-traceless (TT) gauge, the spin-$1$ polarization vectors $\varepsilon_\pm^\mu$ must be spatial ($\varepsilon_\pm.\partial_t=0$), null ($\varepsilon_\pm.\varepsilon_\pm=0$), and transverse ($k.\varepsilon_\pm=0$) to the incident wave four-vector $k^\mu = (\omega,\vec{k})$. These conditions uniquely fix (apart from an overall normalization) the polarization vectors as $\varepsilon^\mu_{\pm} = ( 0,\cos\gamma,\pm i, -\sin\gamma)$.

\textit{(a) Parity decomposition:} As the incident plane wave approaches the Kerr BH lens, it will scatter and a portion of the scattered wave will reach the observer. The scattering problem is conveniently handled in the parity basis, rather than the helicity basis, as the parity modes do not mix with each other due to the parity-invariant [under ${\rm P}: (t,\vec{x})\to(t,-\vec{x})$] nature of the background Kerr metric. Going forward, we shall work in the parity basis. These parity modes can be calculated from Eq.~\eqref{h} by applying the operators $(1 \pm {\rm P})/2$ as
\begin{equation} \label{h11}
    \begin{split}
        &h^{\mu\nu,\, {\rm P}=1}_{{\rm wave},\, h=2} = \epsilon\, \Re[\varepsilon_+^{\mu} \varepsilon_+^{\nu}e^{-i\omega t}\cos(\vec{k}\cdot\vec{x})], \\ 
        &h^{\mu\nu,\, {\rm P}=-1}_{{\rm wave},\, h=2} = - \epsilon\, \Im[\varepsilon_+^{\mu} \varepsilon_+^{\nu}e^{-i\omega t}\sin(\vec{k}\cdot\vec{x})],
    \end{split}
\end{equation}
where we have used ${\rm P}\, e^{\mu \nu}_\pm(\hat{k})=e^{\mu \nu}_\mp(-\hat{k})=e^{\mu \nu}_\pm(\hat{k})$, i.e., the parity operator $\rm P$ leaves the polarization tensor invariant.

\textit{(b) Curvature perturbations:} In a static spherically symmetric spacetime, like Schwarzschild, metric perturbations $h_{\rm wave}^{\mu\nu}$ can be decomposed into tensor spherical harmonics and reduce Einstein's equations into two decoupled master equations, namely the Regge-Wheeler~\cite{ReggePhysRev.108.1063} and Zerilli equations~\cite{ZerilliPhysRevLett.24.737}. However, the Kerr metric is only stationary and axisymmetric. As a result, linearized Einstein's equations mix different modes of the metric perturbations and in turn, one fails to obtain tractable decoupled equations for $h_{\rm wave}^{\mu\nu}$. Instead, we can first project the Weyl curvature tensor $C_{\mu \nu \rho \sigma}$ associated with the perturbed Kerr metric onto a null Kinnersley tetrad $\{\ell^\mu,n^\mu,m^\mu,\bar{m}^\mu\}$. On a type-$D$ spacetime, like Kerr, the Weyl scalars $\psi_0=-C_{lmlm}$, $\psi_4=-C_{n\bar{m}n\bar{m}}$ obey two decoupled and separable Teukolsky equations~\cite{Teuk1PhysRevLett.29.1114}. Now, asymptotically (in particular, in the initial asymptotic rest frame), the aforementioned tetrad vectors can be written in polar coordinates $(t,r,\theta,\phi)$ as 
\begin{alignat}{3} \label{tetrad}
&\ell^\mu \approx ( 1,1,0,0), \, \, n^\mu \approx \frac{1}{2}(1,-1,0,0), \\
&m^\mu \approx \frac{1}{\sqrt{2} r} \left( 0,0,1,\frac{i}{\sin\theta}\right), \nnm
\end{alignat}
satisfying $\ell.n=-1$, $m.\bar{m}=1$, and all other products vanish. Now, in the TT gauge, linearized Ricci tensor vanishes trivially and hence, the linearized  Weyl tensor is same as the linearized Riemann tensor. Then, we can project it onto the tetrad vectors to obtain the parity even/odd pieces of $\psi_4$ from the metric perturbations given by Eq.~\eqref{h11} as
\begin{equation} \label{psi411}
    \begin{split}
        &\psi_{4,\, h=2}^{\rm P}(t,r,\theta,\phi) \approx \\
        &-\epsilon\,e^{-i\chi^+}\, \frac{\omega^2\, [\cos(\frac{\phi}{2})\cos(\frac{\gamma-\theta}{2})+ i \sin(\frac{\phi}{2})\cos(\frac{\theta+\gamma}{2})]^4}{4}  \\
        &- {\rm P}\, \epsilon\,e^{i \chi^-}\, \frac{\omega^2\, [\cos(\frac{\phi}{2})\sin(\frac{\gamma-\theta}{2})+ i \sin(\frac{\phi}{2})\sin(\frac{\theta+\gamma}{2})]^4}{4},
    \end{split}
\end{equation} 
where $\chi^{+}= \omega t - \omega r \cos\gamma\cos\theta -\omega r \sin\gamma\sin\theta\cos\phi$, while $\chi^{-}=\omega t + \omega r \cos\gamma\cos\theta + \omega r \sin\gamma\sin\theta\cos\phi$. We emphasize again that these expressions are only valid in the initial asymptotic rest frame, which is sufficient for our purpose.

\textit{(c) Spheroidal harmonic decomposition:} The next step is to decompose $\psi_4^{\rm P}$ into spheroidal harmonics, which are the natural bases for Kerr perturbation. To obtain the final result in a suitable form, let us first rewrite 
\begin{alignat}{3}
\psi^{\rm P}_{4,\, h=2} = -\frac{\epsilon\, \omega^2}{4}\int_{-\infty}^\infty d\omega' e^{-i\omega' t}\, \widetilde{\psi}_{4}^{\rm P}(\omega',r,\theta,\phi),
\label{eq:psi4decomp}
\end{alignat}
where to match with Eq.~\eqref{psi411}, we need
$\widetilde{\psi}_{4}^{\rm P}(\omega',r,\theta,\phi)= \widetilde{\psi}_+\, \delta(\omega-\omega') + {\rm P}\, \widetilde{\psi}_-\,\delta(\omega+\omega').$
Here, we have used the shorthands that
\begin{equation} \label{psipm}
    \begin{split}
    &\widetilde{\psi}_{+}(\omega', r, \theta, \phi) \approx e^{i \omega' r (\cos\gamma\cos\theta+ \sin\gamma\sin\theta\cos\phi)} \times \\
    &\Big[\cos\Big(\frac{\phi}{2}\Big)\cos\Big(\frac{\gamma-\theta}{2}\Big)+ i \sin\Big(\frac{\phi}{2}\Big)\cos\Big(\frac{\theta+\gamma}{2}\Big)\Big]^4,\\[1em]
    &\widetilde{\psi}_{-}(\omega', r, \theta, \phi) \approx e^{-i \omega' r (\cos\gamma\cos\theta+ \sin\gamma\sin\theta\cos\phi)} \times \\
    &\Big[\cos\Big(\frac{\phi}{2}\Big)\sin\Big(\frac{\gamma-\theta}{2}\Big)+ i \sin\Big(\frac{\phi}{2}\Big)\sin\Big(\frac{\theta+\gamma}{2}\Big)\Big]^4,
    \end{split}
\end{equation}
which can be further decomposed in spheroidal harmonics with spin-weight $s=-2$ as
\begin{equation} \label{psip}
\widetilde{\psi}_{\rm P}(\omega', r, \theta, \phi)=\sum_{lm} \widetilde{\psi}^{lm}_{\rm P}(\omega', r)\,{}_{-2}S_{lm}(\theta,a\omega')\, e^{ i m \phi}.
\end{equation}
The sum runs over $l \in [2,\infty) \cap \mathbb{Z}$ and $m \in [-l,l] \cap \mathbb{Z}$ for any fixed $l$. Now, the the asymptotic forms ($r\rightarrow \infty$) the coefficients $\widetilde{\psi}^{lm}_{\rm P}$ can be computed using the stationary phase approximation (SPA) in the limit $\omega' r \gg 1$ as outlined in detail in Appendix-\ref{appA} (see also Ref.~\cite{Bautista:2022wjf}). Here, we shall only quote the final results:
\begin{equation}
    \begin{split}
        &\widetilde{\psi}_{+}^{lm}(\omega', r) \approx -\frac{2\pi i}{\omega' r}\, e^{i\omega' r_*}\, {}_{-2}S_{lm}(\gamma,a\omega'),\\
        &\widetilde{\psi}_{-}^{lm}(\omega', r) \approx (-1)^{m+1}\, \frac{2\pi i}{\omega' r}\, e^{i\omega' r_*}\, {}_{-2}S_{lm}(\pi-\gamma,a\omega').
    \end{split}
\end{equation}
Note, we have replaced $r$ by the tortoise coordinate $r_*$ in the exponent, which is allowed as $r \to r_*$ in spatial infinity. This replacement is useful to identify the outgoing wave at $r \to \infty$ in its most natural form. Note also that $\psi_4$ also contains an ingoing piece at spatial infinity, which decays very rapidly (as $r^{-5}$) which can be obtained from the outgoing piece using the Teukolsky-Starobinsky identities as outlined in Appendix-\ref{appB}. Hence, to make it self-evident, we shall represent the above-calculated $\psi_4$ by $\psi_4^{\rm out}$.

\subsection{Scattering amplitudes}
The partial waves indexed by spheroidal ($l,m$) and parity ($\rm P$) indices do not mix after scattering. The only effects of scattering are to phase-lag the outgoing pieces of $\psi_4$ consistently with the retarded boundary condition and cause helicity mixing, i.e., even though the incident wave was of positive-helicity, the scattered wave will contain both helicities. Now, using Eq.~\eqref{eq:psi4decomp}, let us first rewrite  the outgoing part of the incident $\psi_4$ by adding the specific-parity modes in the following suggestive form
\begin{equation} \label{psi4out}
    \begin{split}
        &\psi_{4}^{\rm out} \approx \frac{-\epsilon\, \omega^2}{4}\int_{-\infty}^{\infty}   d\omega'\sum_{{\rm P} lm} K^{\rm out}_{l m \rm P}(\omega')\, \frac{e^{-i\omega'(t-r^*)}}{\omega' r} \\
        &\kern12em \times {}_{-2}S_{lm}(\theta,a\omega')\, e^{im\phi},
    \end{split}
\end{equation}
where $K^{\rm out}_{lm \rm P} =  \widehat{\psi}_+^{lm}(\omega')\, \delta(\omega'-\omega) + {\rm P}\, \widehat{\psi}_{-}^{lm}(\omega')\, \delta(\omega'+\omega)$ with $\widehat{\psi}_{\rm \rm P}^{lm}(\omega') = \widetilde{\psi}_{\rm P}^{lm}\, \omega' r\, e^{-i \omega' r_*}$. An interesting point to note is that in several of the above equations, the parity operator ($\rm P$) only appears linearly. This arises from the fact that $\rm P$ is a projection operator satisfying $\rm P^2=1$. Consequently, any function of $\rm P$ can always be reduced to a linear combination of $\rm P$ and the identity operator. Then, as discussed earlier, the scattered wave can easily be obtained by introducing a phase shift $K^{\rm out}_{lm \rm P} \to K^{\rm out}_{lm \rm P}\, \eta^{\rm P}_{l m}\, \exp(2i\delta_{\rm P}^{l m})$:
\begin{equation} \label{eq:finwave4}
    \begin{split}
        &\psi_{4}^{\rm SW}(t,r,\theta,\phi) \approx \frac{-\epsilon\, \omega^2}{4} \int_{-\infty}^{\infty}   d\omega'\sum_{{\rm P}lm} \Big[\eta^{\rm P}_{l m}\, e^{2i\delta^{\rm P}_{l m}}-1\Big]\\
        &\kern4em \times K^{\rm out}_{l m \rm P}(\omega')\, \frac{e^{-i\omega'(t-r^*)}}{\omega' r}\, {}_{-2}S_{lm}(\theta,a\omega')e^{im\phi}.
    \end{split}
\end{equation}
Since we only care for the scattered wave $\psi_4^{\rm SW}$, we have subtracted out the contribution from the incident wave. The scattering phase can be easily obtained via the Mano-Suzuki-Takasugi (MST) approach and is given by 
\begin{alignat}{3}
\eta^{\rm P}_{l m } e^{2i\delta^{\rm P}_{l m}} = (-1)^{l+1} \frac{\Re(C)+12iM\omega'{\rm P} (-1)^{l}}{16\omega'^{4}}
\times \frac{B^{(\text{refl})}_{l m\omega'}}{B^{(\text{inc})}_{l m \omega'}}.
\label{sec:eq:ph}
\end{alignat}
For a detailed derivation of this formula using the Teukolsky-Starobinsky identities is outlined in Appendix-\ref{appB}. Here, $B^{(\rm{refl})}_{l m \omega'}$ and $B^{(\rm{inc})}_{l m \omega'}$ are the coefficients that enter the asymptotic expansion of $\psi_4$ for a given $\{l m\omega'\}$-mode as
$B^{\rm (refl)}_{l m\omega'}e^{-i\omega'(t-r^*)}r^{-1} + B^{\rm (inc)}_{l m\omega'}e^{-i\omega'(t+r^*)}\omega'^{-4} r^{-5}$. Eq~\eqref{sec:eq:ph} is consistent with the corresponding expression in Ref.~\cite{Dolan:2008kf}, except for the difference in parity convention. We use $\mathrm{P}=\pm 1$ to denote symmetry/antisymmetry under reflection, while Ref.~\cite{Dolan:2008kf} adopts the convention where axial modes are assigned $\mathrm{P}=-1$ and polar modes are assigned $\mathrm{P}=1$.

Now, substituting Eq.~\eqref{sec:eq:ph} into Eq.~\eqref{eq:finwave4}, summing over parities and integrating over frequency, we finally get
\begin{widetext}
\begin{alignat}{3}
-2\, \omega^{-2}\, &\psi_{4}^{\rm SW}(t,r,\theta,\phi) \approx \nnm \frac{\epsilon\, e^{-i\omega(t-r_*)}}{r}\, \Bigg[\underbrace{\frac{2\pi}{i\omega}\sum_{lm} e^{im\phi}\,{}_{-2}S_{lm}(\gamma,a\omega)\,{}_{-2}S_{lm}(\theta,a\omega)\, \Bigg\{(-1)^{l+1}\frac{\Re(C)}{16 \omega^4} \frac{B^{(\text{refl})}_{l m\omega}}{B^{(\text{inc})}_{l m \omega}} - 1\Bigg\}}_{\equiv\, f(\gamma, \theta, \phi)}\Bigg]\\
&\kern3em +\frac{\epsilon\, e^{i\omega(t-r_*)}}{r} \Bigg[\underbrace{\frac{2\pi}{i\omega}\sum_{lm} (-1)^{l+m}\, e^{-im\phi}\, {}_{-2}S_{lm}(\gamma,a\omega)\, {}_{-2}S_{lm}(\pi-\theta,a\omega)\, \Bigg\{(-1)^{l+1}\frac{3i M \omega}{4\omega^4} \frac{B^{(\text{refl})}_{l m \omega}}{B^{(\text{inc})}_{l m \omega}}\Bigg\}}_{\equiv\, g(\gamma,\theta,\phi)}\Bigg]^*,
\label{eq:psi4sw2}
\end{alignat}
\end{widetext}
where we have used the identity ${}_{-2}S_{lm}(\theta,a\omega)=(-1)^{l+s}{}_{-2}S_{l-m}(\pi-\theta,-a\omega)$, and a property of the radial Teukolsky equation that transforming $\{m,\omega\}\to\{-m,-\omega\}$ is equivalent to performing complex conjugation. The ratio $B_{(\rm refl)}/B_{(\rm inc)}$ can be computed analytically in the MST formalism~\citep{MST,MT,Bautista:2022wjf} for arbitrary frequency.

Few more comments regarding the above result are in order. First, we should remind our readers again that the complex conjugation from the spheroidal harmonics are removed as $\omega$ is real. The first term inside the square braces represents the helicity-preserving piece, whereas the second term represents the helicity-reversing piece. Secondly, for the special case of on-axis scattering ($\gamma=0$), the above formula matches exactly with that in Ref.~\cite{Dolan:2008kf}. In the generic case ($\gamma\neq 0$), it is consistent with the expressions in Ref.~\cite{Bautista:2022wjf}.

\begin{figure*}[h!]
\includegraphics[width=\textwidth]{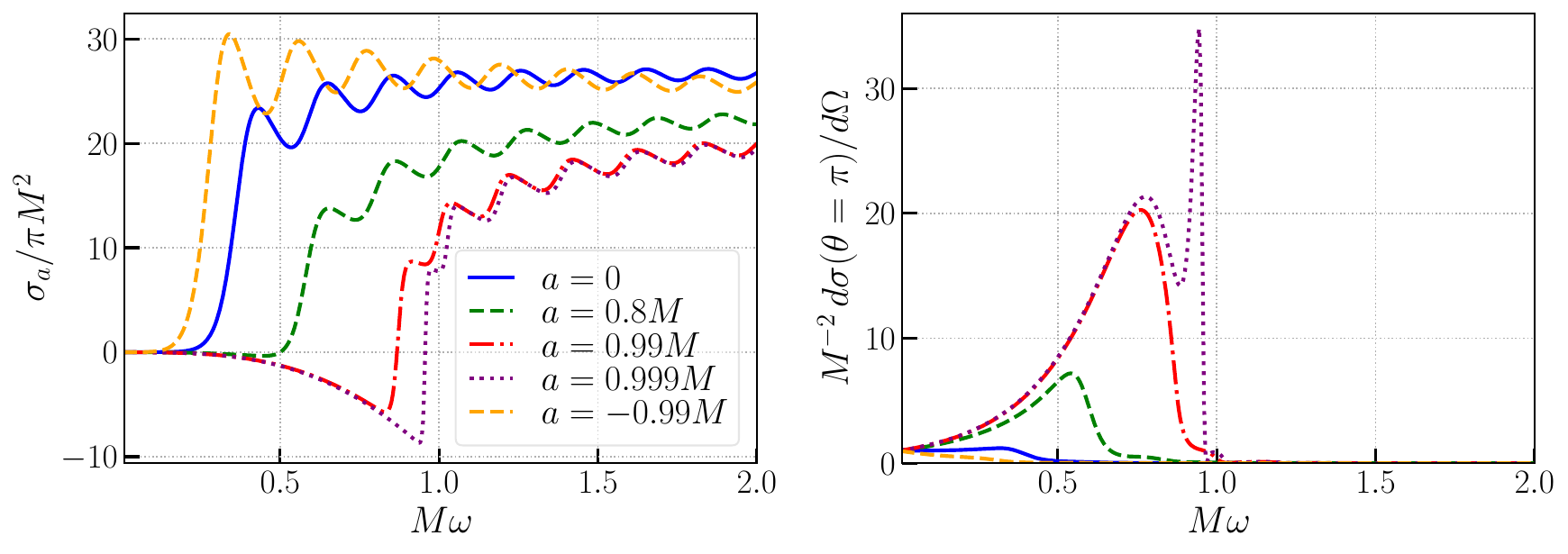}
\caption[Cross sections vs frequencies]{Plots showing the variations of the on-axis ($\gamma=0$) absorption cross section ($\sigma_a$) and differential scattering cross section ($d\sigma/d\Omega$) as a function of frequency ($\omega$) for different spin values and $\theta=\pi$. These plots match very well with that of Ref.~\cite{Dolan:2008kf}. Note that the absorption cross section becomes negative for corotating ($a>0$) configurations at low frequencies. This is due to superradiant amplification.} 
\centering
\label{sec:crossspin}
\end{figure*}

\begin{figure*}[h!]
\includegraphics[width=\textwidth]{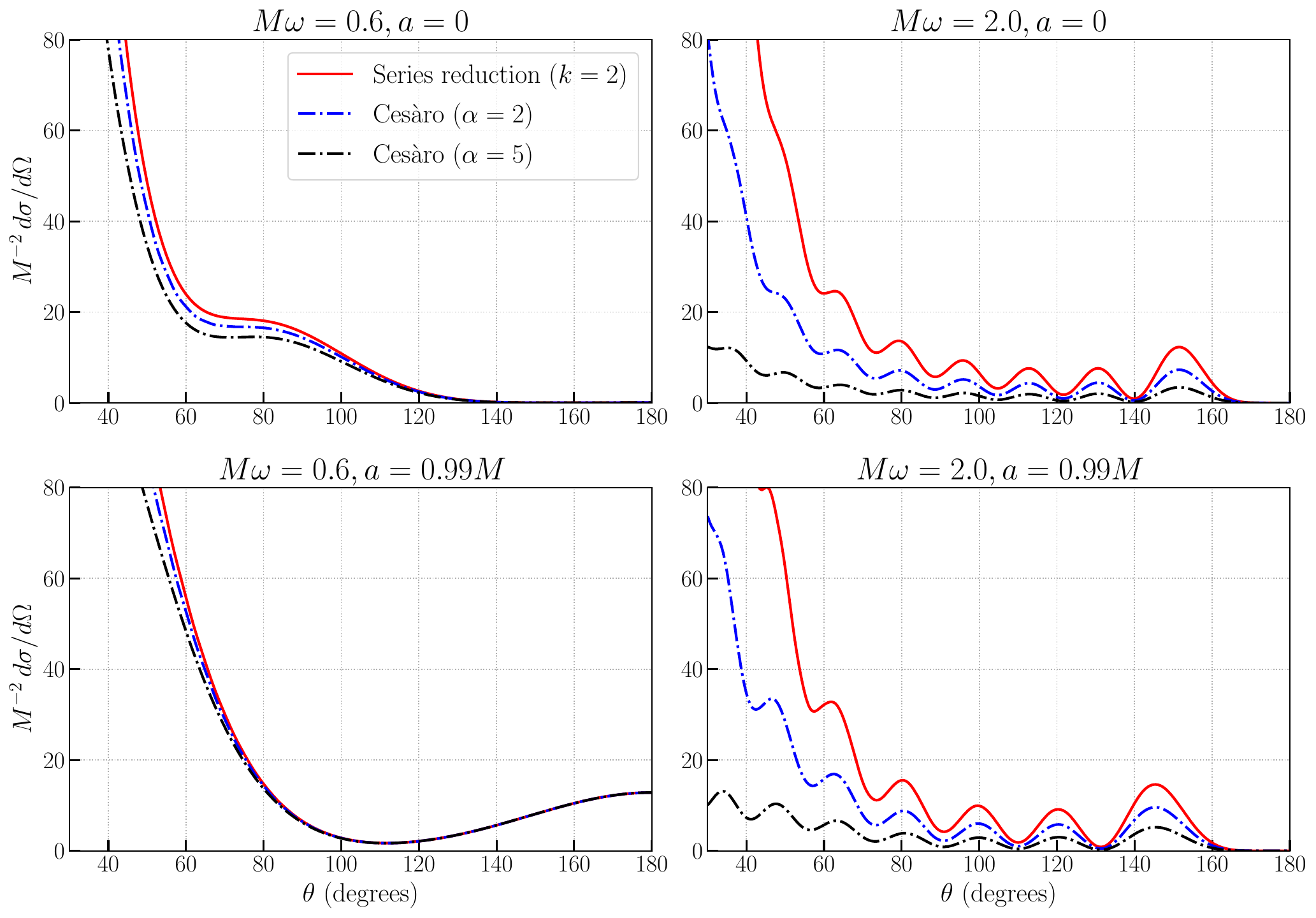}
\caption[Cross sections vs angle]{Plots showing the variations of the on-axis ($\gamma=0$) scattering cross section for different combinations of ($M\omega,\, a$). The plots generated by the series reduction method match very well with that of Ref.~\cite{Dolan:2008kf}. Whereas the plots generated by Cesàro summation (with $\alpha=2,5$) technique systematically tend to disagree especially at low values of $\theta$ and large values of $M\omega$.} 
\centering
\label{sec:crossangle}
\end{figure*}

Some illustrative comparisons with Ref.~\cite{Dolan:2008kf} for the special case of on-axis scattering are shown in Fig.~\ref{sec:crossspin} and Fig.~\ref{sec:crossangle}. In Fig.~\ref{sec:crossspin}, we have plotted the absorption cross section~\cite{Dolan:2008kf}
\begin{equation}
    \sigma_a(\gamma=0) = \frac{4\pi^2}{\omega^2} \sum_{l=2}^{\infty} {}_{-2}S^2_{l2}(0,a\omega) \left[1-|\eta^{\rm P}_{l 2 } e^{2i\delta^{\rm P}_{l 2}}|^2 \right],
    \end{equation}
and differential scattering cross section~\cite{Dolan:2008kf}
\begin{equation}
    \frac{d\sigma(\gamma=0, \theta=\pi)}{d\Omega}=|f(0,\pi,\phi)|^2+|g(0,\pi,\phi)|^2,
\end{equation} 
with frequency ($M \omega$). Here, we have considered no-axis incidence ($\gamma=0$) and scattering angle to be $\theta=\pi$. For both cases, we have used the fact that ${}_{-2}S_{lm}(\gamma=0,a\omega)$ is only non-zero when $m=2$ and as a result, the $m$-sum collapses.

In Fig.~\ref{sec:crossangle}, we have plotted the scattering cross section as a function of scattering angle ($\theta$). The computation of scattering cross section involves summing over $l$-modes appearing via $f$ and $g$ in Eq.~\eqref{eq:psi4sw2}. It is well-known that such a sum does not converge in usual sense and one needs various methods like series reduction (as used in Ref.~\cite{Dolan:2008kf}, for example) as well and Cesàro-$\alpha$ summation~\cite{hardy2024divergent} (as used in Ref.~\cite{Chan:2025wgz}, for example). For more details, we ask our readers to check Appendix-\ref{appC}. It is clear from Fig.~\ref{sec:crossangle} that these two methods do not agree very well (particularly for $\alpha=5$), especially at low scattering angles ($\theta$) and high frequencies ($M \omega$). In this case, the result produced by the series reduction method should be regarded as the more reliable one, since it is more closely tied to the underlying physics by explicitly isolating the partial-wave divergence at $\theta = 0$, and has a well-established history of successful application in computing Coulomb-scattering series~\cite{Yennie:1954zz}.

Here, we must emphasize that, although we have explored various cases spanning the $(a,\omega,\theta)$-parameter space, Fig.~\ref{sec:crossspin} and Fig.~\ref{sec:crossangle} just show a few illustrative examples. In all cases, our plots match very well with that of Ref.~\cite{Dolan:2008kf}. Moreover, the computation of the scattering amplitudes $f$ and $g$ can be computationally intensive. We implemented the MST formalism in C++, based on the publicly available \texttt{Black Hole Perturbation toolkit}~\citep{BHPToolkit}, which is originally implemented in Mathematica. Implementing the MST formalism numerically requires the use of arbitrary-precision arithmetic. For this purpose, we used the \texttt{FLINT (Arb)} library in C~\citep{FLINT}.

\subsection{Scattered wave at observer}
Now, we need to construct the asymptotic metric in the presence of the scattered wave (produced by the lens from the incident positive-helicity wave), which can be approximated as a plane wave in the vicinity of the detector on the earth. To that end, consider a plane wave moving towards the observer located at an angle $(\theta_o,\phi_o)$ with respect to the global frame centred at the lens. Thus, it has a momentum $q^\mu = \{\omega, \vec{q}\}$ with $\vec{q}=\omega(\sin\theta_o \cos\phi_o,\sin\theta_o\sin\phi_o,\cos\theta_o)$. We can decompose the observed wave in a linear combination of both positive and negative helicities ($h=\pm 2$) as
\begin{alignat}{3}
h_{\rm SW}^{\mu\nu} = \Re\left(A\, \xi_{+2}^\mu\, \xi^\nu_{+2}\, e^{i q \cdot x} + B\, \xi_{-2}^\mu\, \xi^\nu_{-2}\, e^{i q \cdot x}\right),
\end{alignat}
where $\xi^\mu_{h=\pm 2}= \{0,\cos\theta_o\cos\phi_o\mp i \sin\phi_o,\pm i \cos\phi_o+\cos\theta_o\sin\phi_o,-\sin\theta_o\}$ are the polarization vectors. The corresponding $\psi_4^{\rm SW}$, in the region around the detector (i.e., $\theta\approx \theta_o$, $\phi\approx \phi_o$), is given by
\begin{alignat}{3} \label{psiobs}
\psi_4^{\rm SW} & \approx -\frac{\omega^2}{2}\left[A e^{-i\omega(t-r_*)} + B^* e^{i\omega (t-r_*)}\right].
\end{alignat}
Comparing with Eq.~\eqref{eq:psi4sw2} and setting $r \to r_*=r_{\rm LO}$, we get $A=\epsilon\, f(\gamma,\theta_o,\phi_o)/r_{\rm LO}$ and $B=\epsilon\, g(\gamma,\theta_o,\phi_o)/r_{\rm LO}$. To decompose the $\psi_4$ scalar into $(+,\times)$-basis, we can use the standard asymptotic relation $2\psi_4^{\rm SW} \approx \partial_t^2(h_+^{\rm SW}-ih_\times^{\rm SW})$. In the monochromatic case, it reduces to $2\psi_4^{\rm SW} \approx -\omega^2\, (h_+^{\rm SW}-i h_\times^{\rm SW})$. Thus, comparing with Eq.~\eqref{psiobs} and taking its real/imaginary parts, we get the following relations:
\begin{equation}
    \begin{split}
        &h_{+}^{\rm SW} =\frac{\epsilon}{r_{\rm LO}}\, \Re\left[(f+g) e^{-i\omega(t-r_*)}\right], \\
        &h_{\times}^{\rm SW} =-\frac{\epsilon}{r_{\rm LO}}\, \Im\left[(f -g)e^{-i\omega(t-r_*)}\right].
    \end{split}
    \label{eq:h2pc}
\end{equation}
We have suppressed the dependence on $\gamma,\theta_o,\phi_o$ in $(f,g)$ for brevity. Now, in the special case of on-axis scattering ($\gamma=0$), we can further write 
\begin{equation} \label{eq:onaxispch}
    \begin{split}
        &h_{+}^{\rm SW} (\gamma=0) = \frac{\epsilon}{r_{\rm LO}}\, \Re\left[F\, e^{2i\phi_o -i\omega(t-r_*)}\right], \\
        &h_{\times}^{\rm SW} (\gamma=0)=-\frac{\epsilon}{r_{\rm LO}}\, \Im\left[\widetilde{F}\, e^{2 i\phi_o -i\omega(t-r_*)}\right].
    \end{split}
\end{equation}
Here, we have defined $F=\tilde{f}+\tilde{g}$ and $\widetilde{F}=\tilde{f}-\tilde{g}$ with $f(\gamma,\theta_o,\phi_o)=\tilde{f}(\gamma,\theta_o)\, e^{2i\phi_o}$, and $g(\gamma,\theta_o,\phi_o)=\tilde{g}(\gamma,\theta_o)\, e^{2i\phi_o}$. 

Note that the above relations are derived considering the initial wave impinging upon the lens to be of positive helicity ($h=+2$). More generally, the initial wave will be a linear combination of both positive and negative helicities. However, without repeating a similar analysis for negative helicity ($h=-2$) incident wave again, the parity and time-reversal symmetries of the system can instead be employed to obtain its scattering amplitudes~\cite{Saketh:2022wap}. In particular, for negative-helicity incident wave $h_{\mr{wave},\, h=-2}^{\mu\nu}=\epsilon~\Re\lb\varepsilon_-^\mu \varepsilon_-^\nu\, e^{ i k\cdot x}\rb$, the outgoing scattered wave will be characterized by $r_{\rm LO}\, A =\epsilon\, g^*(\gamma,\theta_o,\phi_o)|_{a\rightarrow-a}$, and $r_{\rm LO}\, B=\epsilon\, f^*(\gamma, \theta_o,\phi_o)|_{a\rightarrow-a}$. In general, when the incident wave is a linear combination of both helicities as $h_{\mr{wave}}^{\mu\nu}=\epsilon~\Re[(\alpha\, \varepsilon_+^\mu \varepsilon_+^\nu + \beta\, \varepsilon_-^\mu \varepsilon_-^\nu) e^{ i k\cdot x}]$, we have $r_{\rm LO}\, A =\epsilon\, \alpha\, f + \epsilon\, \beta\, g^*(a\rightarrow-a)$, and $r_{\rm LO}\, B=\epsilon\, \alpha\, g + \epsilon\, \beta\, f^*(a\rightarrow-a)$. 

So far, we have considered the case of helicity-mixing alone. Another interesting case is to study the case of polarization-mixing in $(+,\times)$-basis caused by scattering. To analyse these questions, we shall again follow a simpler path by decomposing the initial specific-polarization wave into helicity basis, for which we already know the scattering amplitudes. For illustration, consider the on-axis scattering ($\gamma=0$) of an impinging wave of purely $+$ polarization. It  corresponds to choosing $\alpha=\beta=1/4$ in the helicity-basis, as 
\begin{alignat}{2}
\frac{1}{4}\!\left(\varepsilon_+^{\mu}\varepsilon_+^{\nu}
+\varepsilon_-^{\mu}\varepsilon_-^{\nu}\right)\Big|_{\gamma=0}
&= \frac{1}{2}
\begin{pmatrix}
0 & 0 & 0 & 0\\
0 & 1 & 0 & 0\\
0 & 0 & -1 & 0\\
0 & 0 & 0 & 0
\end{pmatrix}.
\end{alignat}
Then, following previous relations, we have $4\, r_{\rm LO}\, A = \epsilon\, [f + g^*(a\rightarrow-a)]$, and $4\, r_{\rm LO}\, B = \epsilon\, [g +  f^*(a\rightarrow-a)]$. As a result, Eq.~\eqref{eq:onaxispch} gets modified to 
\begin{equation} \label{eq:onaxispcp}
    \begin{split}
        &h_{+}^{\rm SW} (\gamma=0)= \frac{\epsilon}{4\, r_{\rm LO}}\, \Re\Big[\left\{F\, e^{2i\phi_o }+F^*(a \to -a)\, e^{-2i\phi_o} \right\}  \\
        &\kern17em \times e^{-i\omega(t-r_*)}\Big] , \\
        &h_{\times}^{\rm SW} (\gamma=0) = \frac{-\epsilon}{4\, r_{\rm LO}}\, \Im\Big[\left\{\widetilde{F}\, e^{2i\phi_o }-\widetilde{F}^*(a \to -a)\, e^{-2i\phi_o} \right\}  \\
        &\kern17em \times e^{-i\omega(t-r_*)}\Big] .
    \end{split}
\end{equation}
Specializing further to the spinless case, $a=0$, this simplifies to 
\begin{equation} \label{eq:onaxispcpa0}
    \begin{split}
        & h_{+}^{\rm SW} (\gamma=a=0)
        = \frac{\epsilon}{2r_{\rm LO}}
        \bigl[\Re(F)\cos(2\phi_o) - \Im(F)\sin(2\phi_o)\bigr]\\
        &\kern17em \times \cos\!\bigl[\omega(t-r_*)\bigr],
        \\
        & h_{\times}^{\rm SW} (\gamma=a=0)
        = \frac{-\epsilon}{2r_{\rm LO}}
        \bigl[\Im(\widetilde{F})\cos(2\phi_o)+\Re(\widetilde{F})\sin(2\phi_o)\bigr]\\
        & \kern17em \times \cos\!\bigl[\omega(t-r_*)\bigr].
    \end{split}
\end{equation}
This appears inconsistent with the corresponding expressions in Eq.~(35) of Ref.~\cite{Chan:2025wgz}, which seems to be missing the terms containing $\Im(F)$, $\Im(\widetilde{F})$ inside the square braces on the right hand sides of the above expressions. Additionally, Ref.~\cite{Chan:2025wgz} does not seem to differentiate between $F$ and $\widetilde{F}$. Between these two inconsistencies, the second one is nevertheless justifiable for reasonable values of $\theta_o\sim 30^\circ$, for which cases the helicity mixing contributions can be neglected ($|g|\ll |f|$\footnote{As $\theta_o$ increases to $180^\circ$, $|g|$ grows and exceeds $|f|$. The back-scattering cross section at $\theta_o=180^\circ$, presented in the right plot of Fig.~\ref{sec:crossspin}, is almost entirely dictated by $g$.}) and $F\approx \widetilde{F} \approx f$.

\section{Results and Implications} \label{secIII}
To quantify the impact of strong-field lensing on the waveform, we need to compare the scattered wave computed in the previous section with the un-scattered waveform, i.e., the waves reaching directly from the source to the observer. These direct waves will not be subjected to the strong field unless $\theta_o \rightarrow \gamma$, i.e., when the source, lens and observer are collinear. However, our analysis involving asymptotic expansions is not well-defined in this ``shadow'' region, and one needs an alternative implementation. For simplicity, we restrict ourselves to $|\theta_o-\gamma|\geq 30^\circ$ following Ref.~\cite{Chan:2025wgz}, and based on comparisons with Ref.~\cite{Dolan:2008kf}. In this case, the direct wave and the strong-field scattered wave can be treated separately. We have already shown the computation of the scattered wave in the previous section. The correction to the direct wave is expected to be weak (given the large impact parameter), but may be quantified by using the known formula in the point-lens approximation, as previously done for Schwarzschild lenses in Ref.~\cite{Chan:2025wgz}. As we shall show later, the effect of the scattered wave turns out to be important only when the lens and source are sufficiently close to each other, i.e., $r_{\rm SL} \ll r_{\rm SO} \sim r_{\rm LO}$.

\subsection{ Strong-field scattering factor (SFSF)}
For concreteness, let us consider a binary GW source whose radiation field has the approximate form\footnote{In a realistic case, the frequency will evolve as coalescence proceeds. However, the monochromatic results easily generalize to the realistic case in frequency domain. See Appendix.~\ref{appD} for details.}
\begin{alignat}{3}
h^{\mu\nu}
= \Re\left[\frac{\mathcal{A}^+(\hat{n})\xi^\mu_+\xi^\nu_++\mathcal{A}^-(\hat{n})\xi^\mu_-\xi^\nu_-}{r}
  e^{-i\omega(t-r)}\right].
\end{alignat}
Here, $\xi^\mu_{\pm 2}= \{0,\cos\theta\cos\phi\mp i \sin\phi,\pm i \cos\phi+\cos\theta\sin\phi,-\sin\theta\}$, and $\hat{n}$ is the space-like unit vector pointing from the source to the field point. The coefficients $\mc{A}^\pm(\hat{n})$ characterize the polarization content of radiation in the helicity basis in different directions $(\theta,\phi)$ in a source-centred coordinate. Let us choose that the incident wave impinging on the lens to contain only the positive helicity, although similar calculation can be easily done also for negative helicity as well. We further choose the wave-vector to be parallel to the spin axis (i.e., $\gamma=0$), which we identify with the $\hat{z}$-direction as before. In this case, the incident wave is written as 
\begin{alignat}{3}
& h_{\rm wave}^{\mu\nu}
= \Re\ls\frac{\mathcal{A}^+(\hat{n}_{\rm SL})\varepsilon^\mu_+\varepsilon^\nu_+}{r_{\rm SL}}
  e^{-i\omega(t-r_{\rm SL})}\rs,
\end{alignat}
where $\hat{n}_{\rm SL}$ is the unit spatial vector from the source to the lens. Now, identifying $k = \omega\{1,\hat{n}_{\rm SL}\}$, and $x$ as the position 4-vector with respect to the lens, we can write the wave in the standard form
\begin{alignat}{3}
& h_{\rm wave}^{\mu\nu}
= \underbrace{\frac{|\mathcal{A}^+(\hat{n}_{\rm SL})| }{r_{\rm SL}}}_{=\, \epsilon}\Re\ls e^{i(\omega r_{\rm SL}+\psi_{\rm SL})} 
\varepsilon^\mu_+\varepsilon^\nu_+  e^{i k \cdot x}\rs,
\end{alignat}
where $\mc{A}^+(\hat{n}_{\rm SL}) = |\mc{A}^+(\hat{n}_{\rm SL})|\, \exp(i\psi_{\rm SL})$ . We know from previous section that the subsequent scattered wave at the observer's location is given by
\begin{alignat}{3} \label{hobs}
&h_{\rm SW}^{\mu\nu} = \epsilon\, \Re\Bigg[e^{i\psi_{\rm SL}}e^{- i \omega (t-r_{\rm LO}-r_{\rm SL})}\Bigg\{\frac{f(0,\theta_o,\phi_o)}{r_{\rm LO}} \xi_{+2}^\mu\xi^\nu_{+2} \nnm \\
&\kern6em + \frac{g(0,\theta_o,\phi_o)}{r_{\rm LO}}\xi_{-2}^\mu\xi^\nu_{-2}\Bigg\}\Bigg].
\end{alignat}
We can compare this with the direct wave emitted from the source which reaches the observer, given by 
\begin{alignat}{3} \label{hdir}
h^{\mu\nu}_{\rm dir}
= \Re\left[\frac{\mathcal{A}^+(\hat{n}_{\rm SO})\xi^\mu_+\xi^\nu_++\mathcal{A}^- (\hat{n}_{\rm SO})\xi^\mu_-\xi^\nu_-}{r_{\rm SO}}
  e^{-i\omega(t-r_{\rm SO})}\right]
\end{alignat}
where $\hat{n}_{\rm SO}$ is the unit spatial vector from the source to the observer and we have assumed that the polarization vectors appearing in Eq.~\eqref{hobs} and Eq.~\eqref{hdir} are approximately same as $r_{\rm SL} \ll r_{\rm SO} \sim r_{\rm LO}$. Also, as mentioned earlier, $|g|\ll |f|$ for small values of $\theta_0\approx\pi/6$. Thus, focusing only on the contribution of $f$ for a rough (but quite good) estimation, a suitable way to quantify the relative importance of strong-field wave scattering is by the following \textit{strong-field scattering factor (SFSF)}: 
\begin{alignat}{3}
\textrm{SFSF} \approx \frac{|\mc{A}^+(\hat{n}_{\rm SL})|r_{\rm SO}}{|\mc{A}^+(\hat{n}_{\rm SO})| r_{\rm LO} r_{\rm SL}}\, |f(0,\theta_o,\phi_o)|.
\end{alignat}
As a further approximation, we can neglect the variation in the wave amplitude between $\hat{n}_{\rm SL}$ and $\hat{n}_{\rm SO}$, and specialize to the situation where the source and lens are close ($r_{\rm SO}\sim r_{\rm LO} $) to each other. This yields 
\begin{alignat}{3}
\textrm{SFSF} \approx \frac{|f(0,\theta_o,\phi_o)|}{r_{\rm SL}}.
\label{eq:impact}
\end{alignat}
As the strength is governed primarily by $r_{\rm SL}$, a smaller $r_{\rm SL}$ is thus preferred. However, we also require the source-lens separation to be much larger than the the wavelength of the wave, i.e., $ \omega r_{\rm SL} \gg 1$, and much larger than the Schwarzschild radius ($R_L=2M$) of the lens, i.e., $r_{\rm SL}\gg R_L$. These criteria ensure that the incident wave on the lens can be treated as a plane wave.

\begin{figure*}
\includegraphics[width=\textwidth]{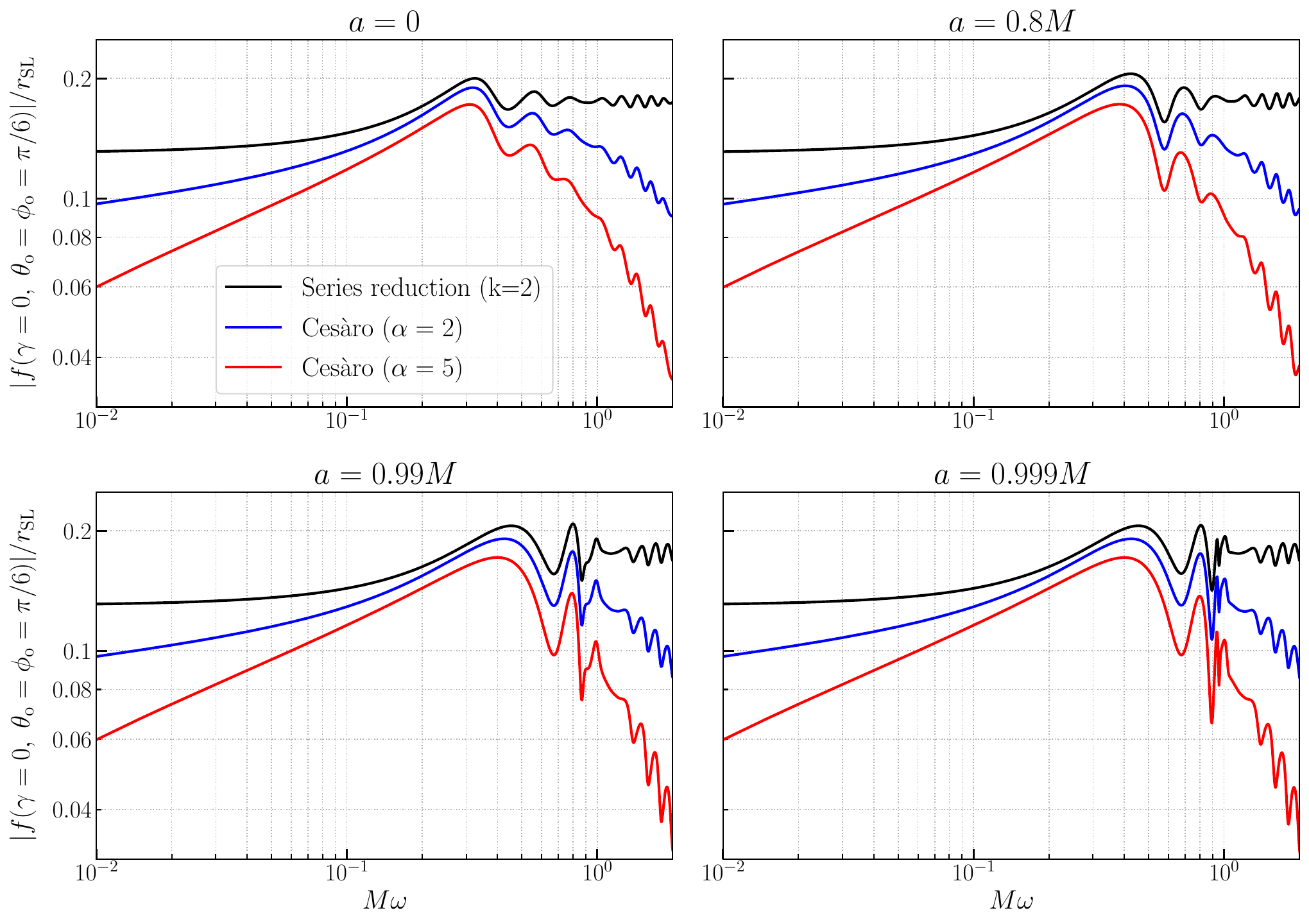}
\caption[Relative impact on waveform (SFSF) due to scattering for $\theta_o=\pi/6$, $\phi_o=\pi/6$, $r_{\rm SL}=100M$, $r_{\rm SO}\approx r_{\rm LO}$, as a function of frequency for various spins.]{Log-Log plots showing the relative impact on the waveform (SFSF) due to on-axis scattering  ($\gamma=0$) for $\theta_o=\phi_o=\pi/6$, $r_{\rm SL}=100M$, $r_{\rm SO}\approx r_{\rm LO}$ and different choices of spin $a$. Spin does not seem to significantly affect the overall scale of the SFSF, but it does contribute appreciably in frequency-modulation. Note that results obtained using Cesàro summation predict a decay at high frequencies, which increases with $\alpha$, in contrast to that obtained via series reduction, which maintains a stable amplitude.} 
\centering
\label{sec:impactangle}
\end{figure*}

We plot Eq.~\eqref{eq:impact} for a particular configuration in Fig.~\ref{sec:impactangle} as a function of frequency, and for various values of spin. We restrict to on-axis scattering($\gamma=0$) and select $\theta_o=\pi/6$, $\phi_o=\pi/6$, $r_{\rm SL}=100M$, and $r_{\rm SO}\approx r_{\rm LO}$. In contrast to the result in Ref.~\cite{Chan:2025wgz}, we do not find SFSF to decay for any value of spin at high frequencies. Our result is consistent with the results in Ref.~\cite{Dolan:2007ut,Dolan:2008kf}, where the differential scattering cross section was also found not to markedly decrease in amplitude for large frequencies. Considering the earlier plots in Fig.~\ref{sec:crossangle} where we showed the comparison of series reduction method with Cesàro summation, it appears that the high-frequency decay seen in Ref.~\cite{Chan:2025wgz} is likely an artefact of the Cesàro resummation procedure, and not a physical effect associated with the absorption of high-frequency GWs. It is also important to emphasize that while a part (with low $l$) of high-frequency radiation can be efficiently absorbed by the BH lens, the degree of absorption also depends on the impact parameter (alternatively, the multipolar index $l$). Thus, in our view, it is not reasonable to conclude that all high-frequency GW radiation will be generically absorbed by the BH lens, leading to a decayed amplitude.

\subsection{Mismatch analysis}
In this section, we assess the \emph{distinguishability} of the lensed waveform (the waveform including strong-field wave-scattering effects) from the unlensed one (without the scattering effects), and the prospects of its \emph{detectability}. 
\subsubsection{Setup}
A natural measure of the \emph{distinguishability} between two waveforms, say $h_1$ and $h_2$, is the \textit{mismatch}, defined as~\cite{Dhurandhar:1992mw}:
\begin{equation}
    \mathscr{M}(h_1, h_2) \coloneqq 1 - \mathcal{O}(h_1, h_2),
\end{equation}
where $\mathcal{O}(h_1, h_2)$ is the \textit{match}, defined as the normalized overlap between the waveforms maximized over any global time ($t_0$) and phase ($\phi_0$) shifts~\cite{Allen:2005fk, Usman:2015kfa}:
\begin{equation}
    \mathcal{O}(h_1, h_2) \coloneqq \max_{t_0, \phi_0}\frac{\braket{h_1|h_2}}{\sqrt{\braket{h_1|h_1}\braket{h_2|h_2}}},
\end{equation}
with the noise-weighted inner product (i.e., overlap) in a frequency window $\nu \in [\nu_{\rm low}, \nu_{\rm high}]$ given by
\begin{equation}
    \braket{h_1|h_2} = 4\, \Re \int_{\nu_{\rm low}}^{\nu_{\rm high}}d\nu\, \frac{\tilde{h}_1(\nu)\tilde{h}_2^*(\nu)}{S_n(\nu)}.
\end{equation}
Here, tildes denote the Fourier-transformed waveforms in the frequency domain, and $S_{n}(\nu)$ is the noise power spectral density (PSD) of the GW detector under consideration~\citep{Finn:1992xs, Cutler:1994ys, Flanagan:1997kp, Flanagan:1997sx}. 
The normalization of the waveforms is done by dividing them by their norm, $||h||=\sqrt{\braket{h|h}}\equiv \rho_{\rm opt}$, where $\rho_{\rm opt}$ defines the optimal SNR. 

Since in a GW search, the goal is essentially to find the template-bank member that has the smallest mismatch with the detector data, we are therefore interested in the mismatch optimized over all of the members of the unlensed waveform family. For a target waveform $h_{\rm target}$ whose match we wish to maximize using the template waveforms $h_{\rm template}$, we define \emph{fitting factor} $(\textrm{FF})$ as the maximized match~\cite{Lindblom:2008cm}:
\begin{equation}
    \textrm{FF}(h_{\rm target}) = \max_{\Theta_{\rm opt}} \mathcal{O}(h_{\rm template}(\Theta_{\rm opt}), h_{\rm target}),
    \label{eq:ff_definition}
\end{equation}
where, motivated by real GW searches, the maximization is done over the $4$D space of binary's intrinsic parameters ($\Theta_{\rm opt}$): chirp mass, mass ratio, and the aligned spin components.
The fully optimized mismatch is, therefore, $1-\textrm{FF}$.

\begin{figure}
\includegraphics[width=\linewidth]{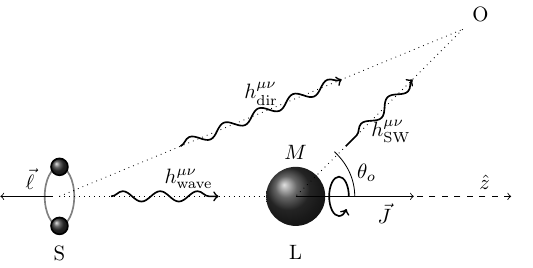}
\caption[Setup]{The source (S) is a quasi-circular binary coalescence of two spinless BHs. The orbital-angular momentum of the source is $\vec{\ell}$, oriented so that it faces away from the lens (note the difference with Fig.~\ref{sec:scatter}). The lens (L) has mass $M$ and spin-angular momentum $|\vec{J}|=M^2\chi$, where $\chi$ is the spin-parameter $\chi\in(-1,1)$. The direction of the lens spin defines the $\hat{z}$-axis and is chosen to be antiparallel (when $\chi>0$) to $\vec{\ell}$, and thus parallel to the source-lens line. The observer (O) is at an angle $\theta_o$ with respect to $\hat{z}$-axis. The GWs emitted by the source reach the observer in two parts, as a direct component $h^{\mu\nu}_{\rm dir}$ and a scattered component $h^{\mu\nu}_{\rm SW}$ which reach roughly at the same time and interfere.} 
\centering
\label{sec:geometry}
\end{figure}

For illustration, we consider a \texttt{GW150914}-like event, consisting of two coalescing Schwarzchild BHs of masses $m_1 = 36.9~\mathrm{M}_\odot$, and $m_2 = 32.8~\mathrm{M}_\odot$\footnote{We use the maximum-likelihood values of the 1D-marginalized posteriors from the `C01:Mixed' channel of `IGWN-GWTC2p1-v2-GW150914\_095045\_PEDataRelease\_mixed\_cosmo.h5'. Posterior samples are available at \href{https://zenodo.org/records/6513631}{Zenodo}.}~\cite{LIGOScientific:2016aoc, LIGOScientific:2018mvr, KAGRA:2021vkt}. The source is at a distance of $r_{\rm SO}=402.6~\rm{Mpc}$. Although including cosmological effects such as redshift are straightforward, we neglect them here for simplicity. The redshift of the source is $0.12$, which is quite small and can be ignored without any major effects for this demonstration. We choose the lens mass as $M=100~\mathrm{M}_\odot$. Moreover, the distance between the source and lens $r_{\rm SL}$ is chosen to be $100M$ or $10^4 ~\mathrm{M}_\odot\sim 4\times 10^{-16} \mathrm{Mpc}$, which is much smaller than the distance between the source and observer. 
We orient our binary source such that its angular momentum points away from the lens (i.e., antiparallel to the source-lens line), this ensures that the lens experiences an approximately plane GW with purely positive helicity ($h=+2$). We further specialize to the case of on-axis ($\gamma=0$) scattering, where the spin of the lens is parallel to the line joining the source and lens.
The source-lens line is set at an angle of $\theta_o$ with respect to the lens-observer line. At the observer, this corresponds to an inclination (angle between the source-observer line and the binary orbital angular momentum) of $\iota = \pi - \theta_o $.
Finally, given the axisymmetry of the setup about the z-axis (the source-lens line which is parallel to the lens spin and the source-orbital angular momentum), we choose the azimuthal angle $\phi_o=0^\circ$ without loss of generality. The setup is visually presented in Fig.~\ref{sec:geometry}.

The observer receives the gravitational radiation as two contributions. The first is the direct component, $h^{\mu\nu}_{\rm dir}$, which is only weakly affected by the lens because of the large impact parameter and approximately follows the straight line between source and observer. Ref.~\cite{Chan:2025wgz} employs the point-lens, scalar-wave approximation to estimate lensing effects on this part. Since the point-lens approximation breaks down for large observation angles (i.e., for large $\theta_o$, where small-angle approximations break down), we do not adopt it and treat the direct wave as essentially unaffected by the lens.
The second is the scattered component, $h^{\mu\nu}_{\rm SW}$, encoding the strong-field influences of the lens. Each frequency mode acquires an additional phase $e^{i\omega (r_{\rm LO}+r_{\rm SL}-r_{\rm SO})}$. In the time domain this corresponds to a delay of approximately $c^{-1}(r_{\rm LO}+r_{\rm SL}-r_{\rm SO})\approx c^{-1}\times 2 r_{\rm SL}\sin^2(\theta_o/2)\sim 0.005~\mathrm{sec}$, much smaller than the $\sim 0.2~\mathrm{sec}$ duration of the \texttt{GW150914} signal observed by LIGO. Thus, the direct and scattered waves arrive nearly simultaneously and interfere, with the modulation depending on parameters such as the lens mass and spin. The observed waveform is, therefore, 
\begin{equation}
    h^{\mu\nu}_{\rm obs}=h^{\mu\nu}_{\rm dir}+h^{\mu\nu}_{\rm SW}.
    \label{eq:lensed_wf}
\end{equation}
For more details see Appendix-\ref{appD}.

We use the software package \texttt{PyCBC} \citep{pycbc_github} to compute the waveforms at the lens and at the observer, and utilize the approximant \texttt{IMRPhenomD}~\citep{PhysRevD.93.044007,Husa:2015iqa} to generate unlensed waveforms. We use the advanced LIGO PSD at its A+ design sensitivity~\citep{LIGOScientific:2014pky}, which is the target sensitivity for the upcoming (O5) LIGO observing runs.
To compute the $\textrm{FF}$ values, we use the Nelder-Mead algorithm~\cite{Nelder:1965zz}, as implemented in the \texttt{optimization} module of the \texttt{Scipy} library \cite{Virtanen:2019joe}, to maximize the \textit{match} between the waveforms.
It is important to note that since we employ a maximization algorithm to compute $\textrm{FF}$, it is not sensitive to the discrete placement of templates. The waveform at the lens is treated as the incident plane wave for the scattering analysis described in earlier sections, and the scattered waveform is computed in the frequency domain as outlined in Appendix-\ref{appD}. We restrict ourselves to cases with $\theta_o > \pi/6$ as it is approximately the smallest angle where the scattering analysis presented in this work may be trusted. The effect is expected to be stronger for smaller scattering angles, as seen from the plots of differential-cross section vs. scattering angle in Fig.~\ref{sec:crossangle}.

\begin{figure*}
\centering
\includegraphics[width=\textwidth]{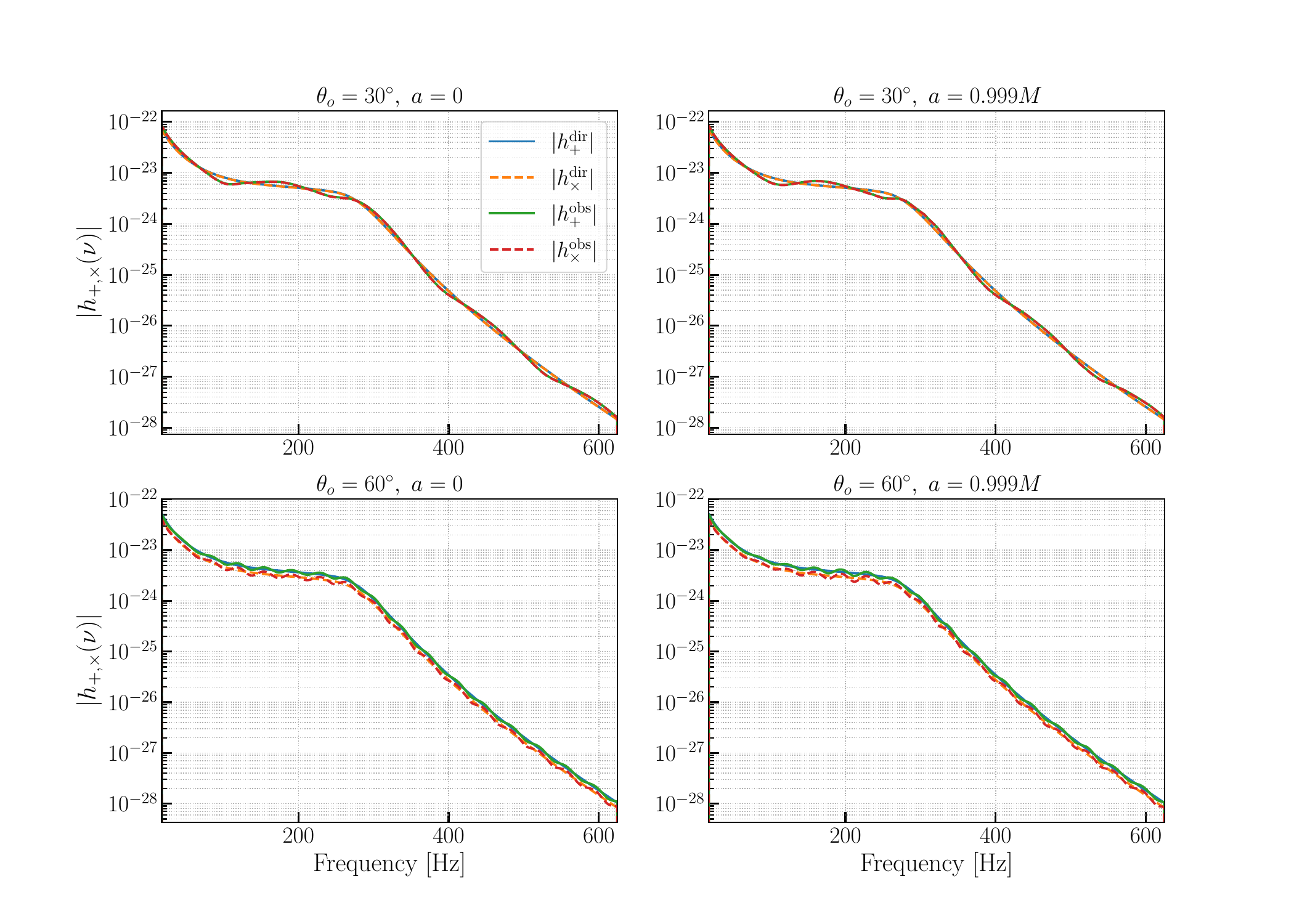}
\vspace{-0.6cm}
\caption[Waveform correction.]{Plots (y-axes in log-scale) illustrating the difference between the magnitude of the total observed waveform ($h_{+,\times}^{\rm obs}$) unscattered, direct component ($h_{+,\times}^{\rm dir}$) for $\theta_o=30^\circ$ (top) and $\theta_o=60^\circ$ (bottom). For $\theta_o=30^\circ$, both polarizations contribute with nearly equal strength, causing the $+,\times$ curves to overlap. At $\theta_o=60^\circ$, the difference in polarization content becomes more pronounced. However, the deviation between the direct and observed waveforms is larger for $\theta_o=30^\circ$, whereas for $\theta_o=60^\circ$ the observed waveform remains close to the direct one, showing only small oscillations about it. The scale of oscillations in the observed waveform is set by the time delay between the direct and scattered components, i.e., a larger delay produces a greater number of oscillations.} 

\label{sec:correction}
\end{figure*}

%%%%%%%%%%%%%%%%%%%%%%% FIG %%%%%%%%%%%%%%%%%%%%%%%%%%%
\begin{figure*}[t]
    \centering
    \subfigure[]{\includegraphics[width=0.48\linewidth]{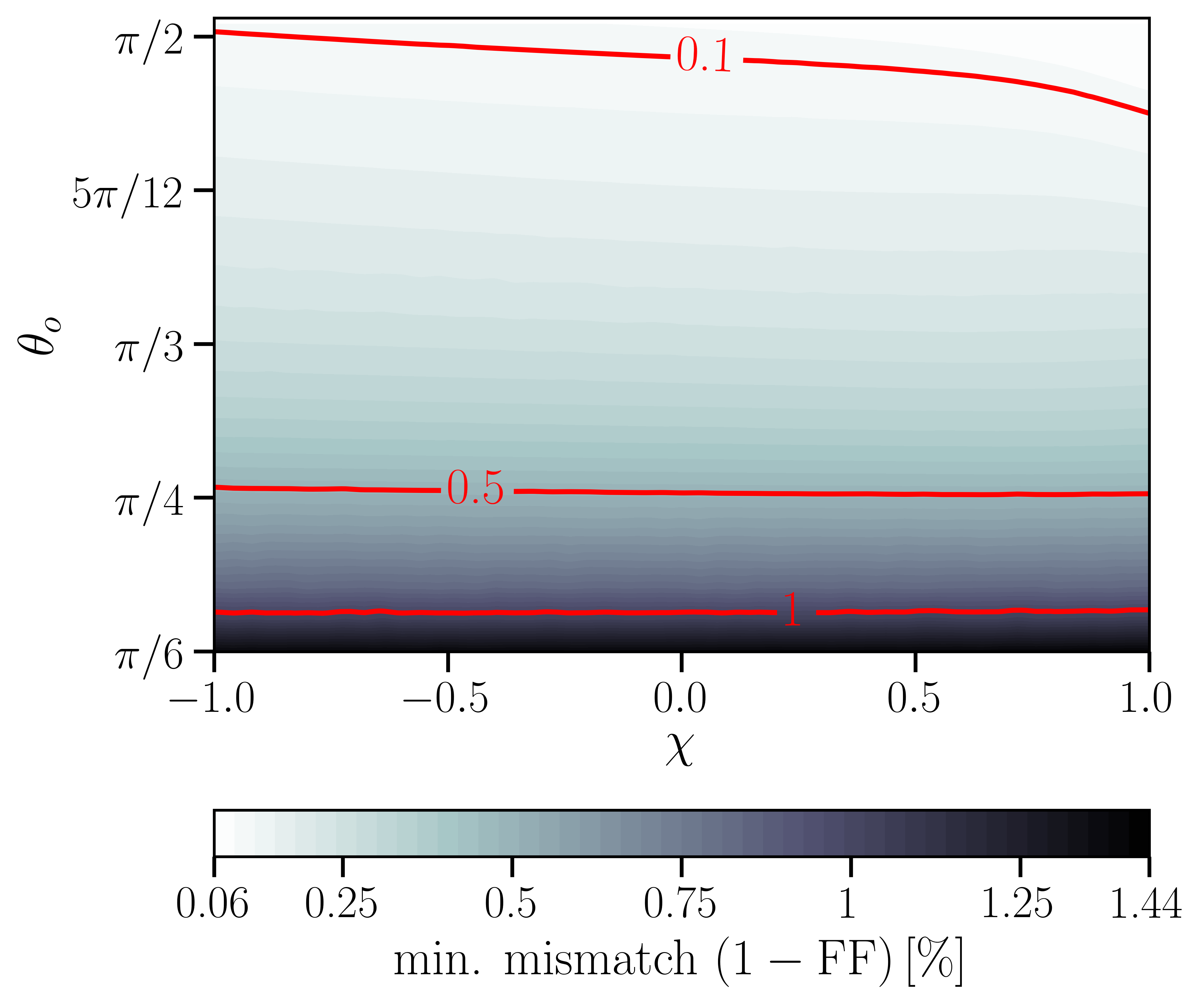}}
    \subfigure[]{\includegraphics[width=0.48\linewidth]{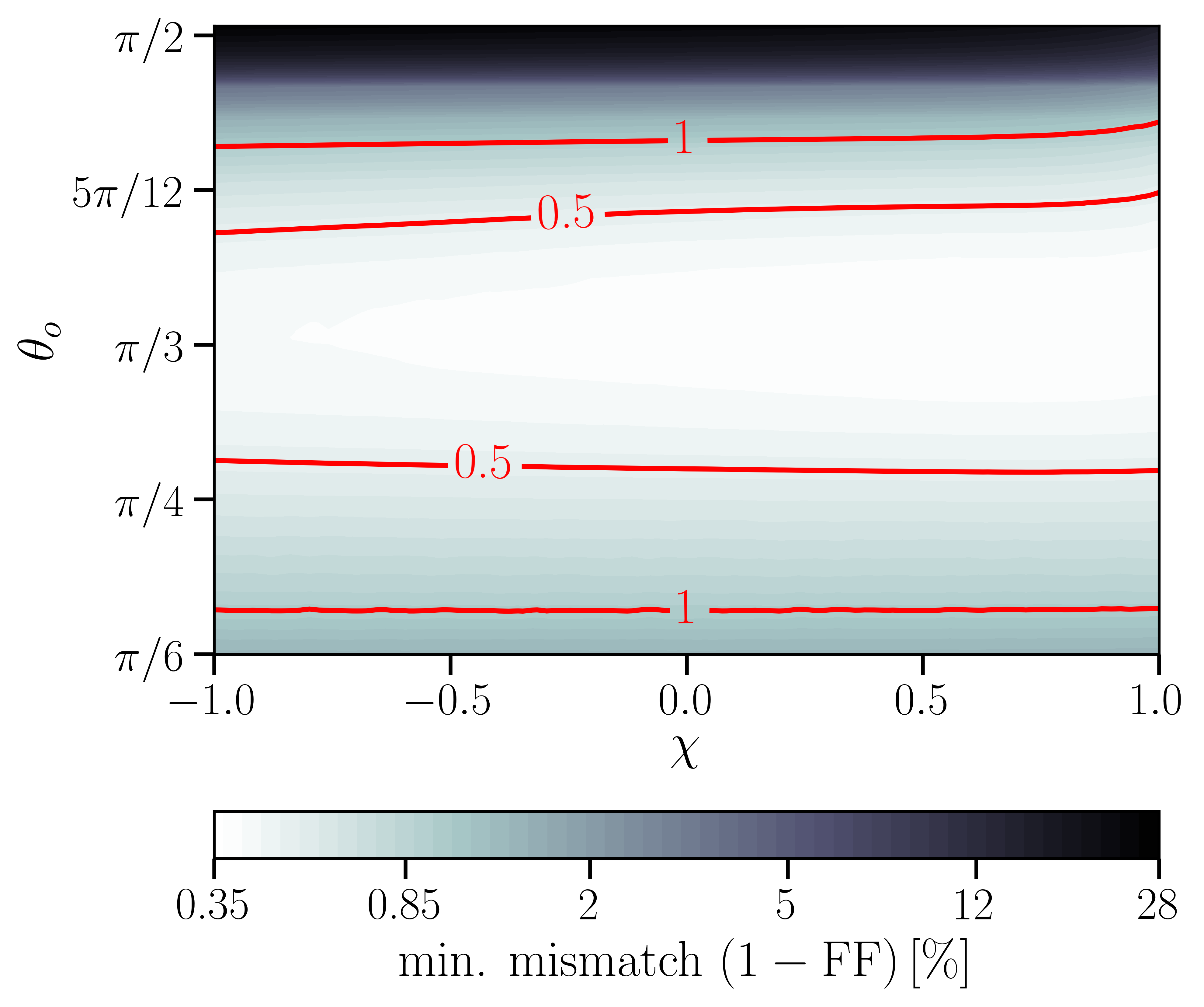}}
    \vspace{-0.6cm}
    \caption{
    Variation in the minimum mismatch ($1-\textrm{FF}$), see Eq.~\eqref{eq:ff_definition}, across the parameter space of $\theta_o$ (scattering angle), and the lens spin $\chi$, for the $+$ and the $\times$ polarizations of the lensed waveform that reaches the observer after scattering shown in the \textit{left} and \textit{right} panels, respectively. Mismatches are computed against unlensed templates and minimised over component masses and aligned spins.}
    \label{fig:min_mismatch}
\end{figure*}
%%%%%%%%%%%%%%%%%%%%%%%%%%%%%%%%%%%%%%%%%%%%%%%%%%%%%%%%

%%%%%%%%%%%%%%%%%%%%%%% FIG %%%%%%%%%%%%%%%%%%%%%%%%%%%
\begin{figure*}[t]
    \centering
    \subfigure[]{\includegraphics[width=0.48\linewidth]{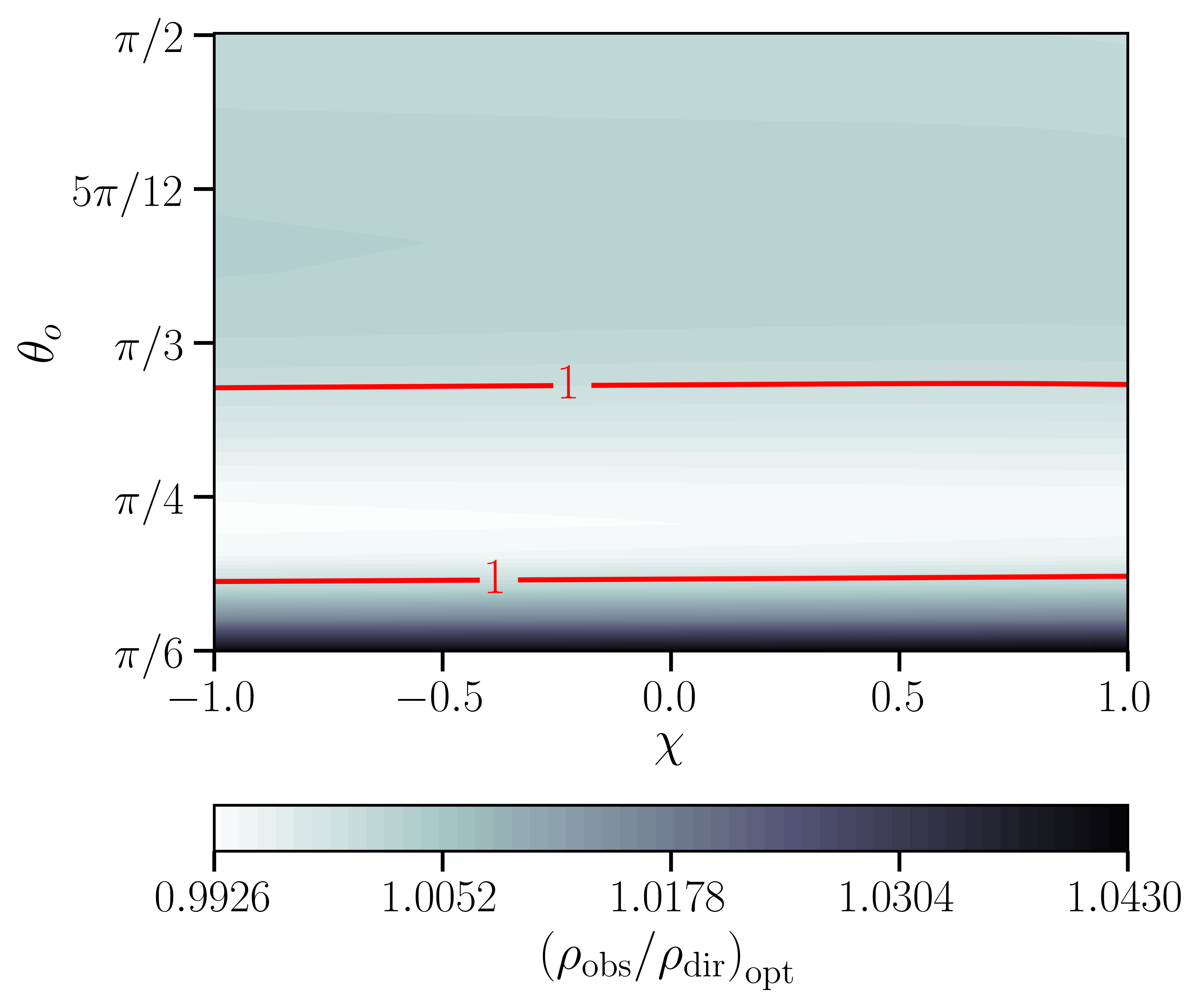}}
    \subfigure[]{\includegraphics[width=0.48\linewidth]{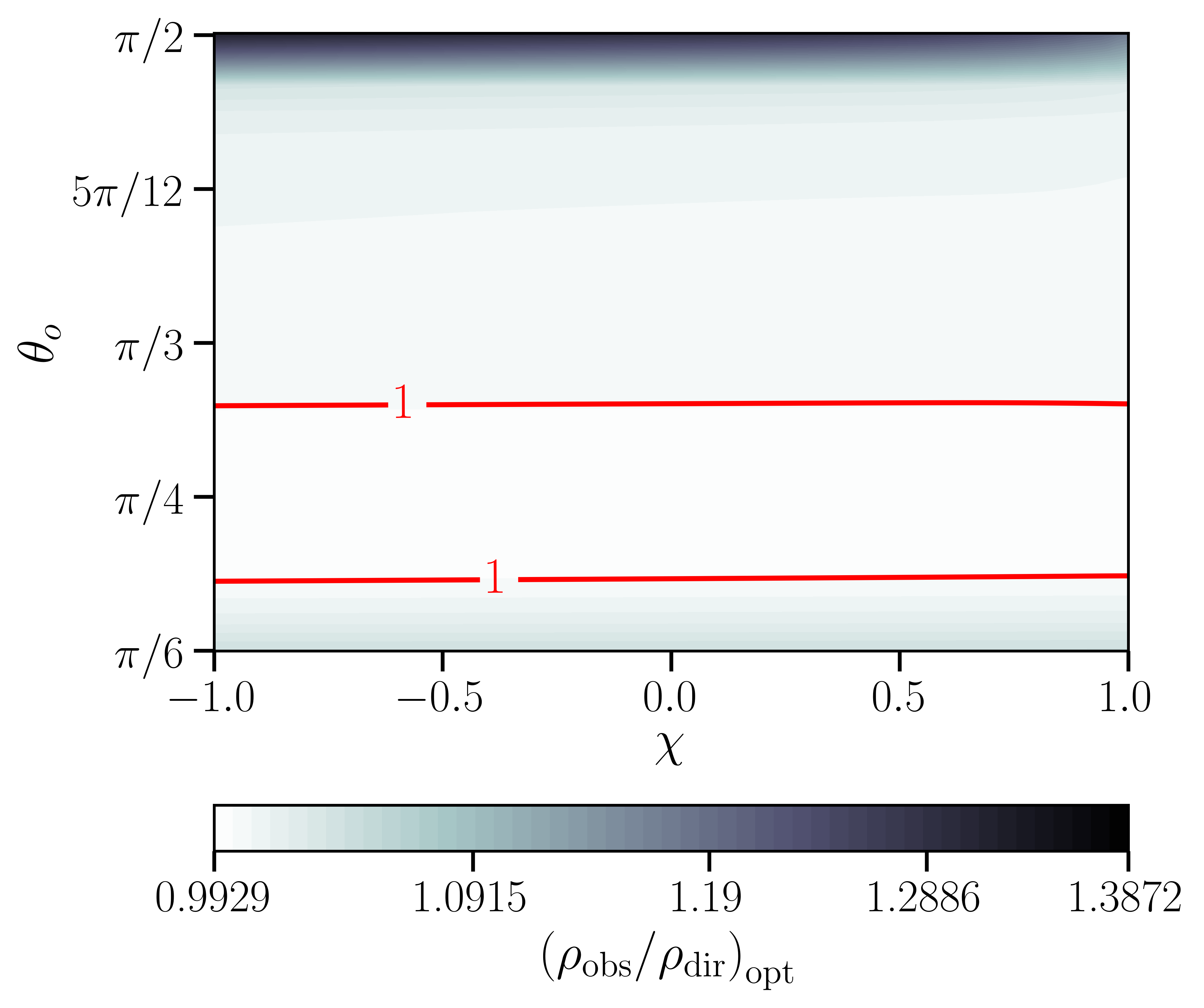}}
    \vspace{-0.6cm}
    \caption{
       Comparison of the optimal SNRs of the lensed and unlensed waveforms for the $+$ and $\times$ polarizations, shown in the left and right panels, respectively.
    }
    \label{fig:opt_snr_ratio}
\end{figure*}
%%%%%%%%%%%%%%%%%%%%%%%%%%%%%%%%%%%%%%%%%%%%%%%%%%%%%%%%

%%%%%%%%%%%%%%%%%%%%%%% FIG %%%%%%%%%%%%%%%%%%%%%%%%%%%
\begin{figure*}[t]
    \centering
    \includegraphics[width=\linewidth]{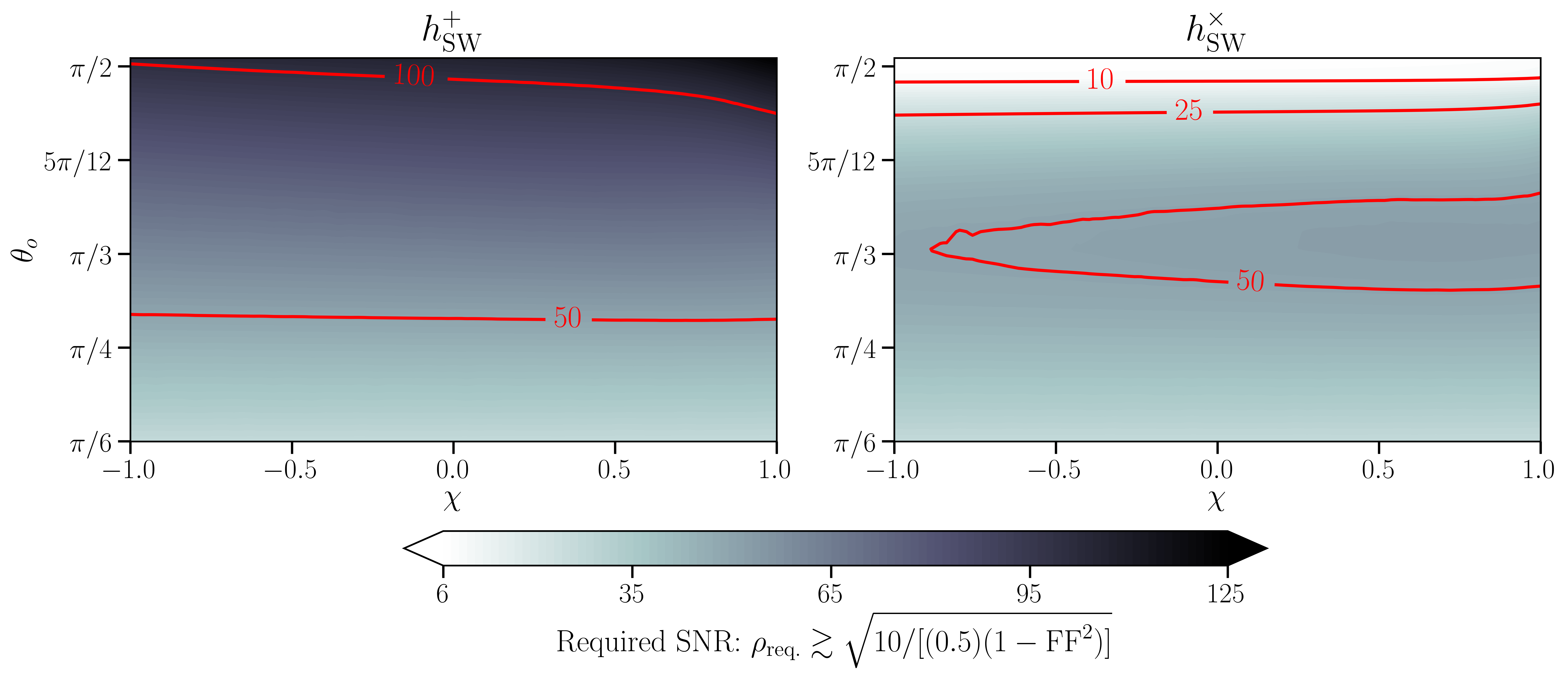}
    \caption{
    Illustrative estimation of the SNR required to identify scattering-induced lensing features in $h^+_{\rm obs}$ and $h^\times_{\rm obs}$, shown in the left and right panels, respectively.
    }
    \label{fig:req_snr}
\end{figure*}
%%%%%%%%%%%%%%%%%%%%%%%%%%%%%%%%%%%%%%%%%%%%%%%%%%%%%%%%

\begin{table}[h!]
\centering
\resizebox{0.85\linewidth}{!}{%
\begin{tabular}{c c c c}
\hline\hline
$\theta_o$ & $\chi$ & $\mathscr{M}(h^+_{\rm dir},\, h^+_{\rm obs})$ & $\mathscr{M}(h^\times_{\rm dir},\, h^
\times_{\rm obs})$ \\
\hline
\multirow{5}{*}{$30^\circ$}
& 0      & 0.0155  & 0.0157\\
& 0.8    & 0.0155 & 0.0157\\
& 0.99   & 0.0155 & 0.0158\\
& 0.999  & 0.0155 & 0.0158\\
& $-0.99$ & 0.0156 & 0.0158\\
\hline
\multirow{5}{*}{$60^\circ$}
& 0      & 0.0026 & 0.0038 \\
& 0.8    & 0.0025 & 0.0036 \\
& 0.99   & 0.0025 & 0.0037\\
& 0.999  & 0.0025 & 0.0037\\
& $-0.99$ & 0.0027 & 0.0040 \\
\hline\hline
\end{tabular}
}
\caption{Mismatch between the direct component and the total waveform as the lens spin-parameter $\chi$ is varied. For a fixed scattering angle ($\theta_o$), spin does not seem to significantly affect the mismatch. This is seen more clearly in the contour plots in Fig.~\ref{fig:min_mismatch}. However, increasing $\theta_o$ leads to a sharp drop in mismatch. The mismatch with cross polarization appears to dominate for these configurations.}
\label{tab:mismatch}
\end{table}

\subsubsection{Results}
Fig.~\ref{sec:correction} shows the frequency-domain waveforms with and without the effect of strong-field scattering. In the top panel, $\theta_o=30^\circ$, and thus the inclination of the source for the observer is $180-30=150^\circ$. At this inclination, the magnitudes of $h_+^{\rm dir}$ and $h_\times^{\rm dir}$ are very close to each other (with equality holding at $180^\circ$). Thus, they appear almost overlapping in the plots. A similar feature is also present for the scattered wave as apparent from Eq.~\eqref{eq:h2pc} since $|g|\ll |f|$ and the waveform at the lens has purely positive helicity as mentioned before. Thus, even in the observed waveform, which includes the scattered component, this feature remains intact and we find that the curves for $h_+^{\rm obs}$ and $h_\times^{\rm obs}$ are overlapping. As a comparison, we also show results for $\theta_o=60^\circ$, where the difference in the polarization content in $+,\times$ basis is more clearly visible. However, the difference between the direct and observed waveforms is much smaller (the oscillations have reduced amplitude) for $\theta_o=60^\circ$. The scale of the oscillatory features in $h_{+,\times}^{\rm obs}$ is determined by the phase generated by the time delay between the two components. As $\theta_o$ increases, the quantity $2r_{\rm SL}\sin^2(\theta_o/2)$ grows, producing a phase that varies more rapidly with frequency. This results in faster oscillations in the observed waveform.

In Fig.~\ref{fig:min_mismatch}, we show the variation in the minimum mismatch ($1-\textrm{FF}$), see Eq.~\eqref{eq:ff_definition}, across the parameter space of $\theta_o$ and $\chi$ for both $h^+_{\rm obs}$ and $h^\times_{\rm obs}$. As expected from Fig.~\ref{sec:impactangle} and demonstrated clearly in Fig.~\ref{sec:correction},  the variation of lens spin parameter induce only very small affects in the mismatch for on-axis scattering. 
However, increasing the angle $\theta_o$ from $\pi/6$ to $\pi/3$ leads to a sharp drop in the mismatch. This is also expected from the plots in Fig.~\ref{sec:crossangle}. This hints that it would be particularly interesting to probe $\theta_o\approx 0$, where the effect could be much stronger, as also suggested in Ref.~\cite{Chan:2025wgz}.
In the case of $h^+_{\rm obs}$, the mismatch values decreases monotonically with increasing $\theta_o$, from $\sim1\%$ at $\pi/6$ to $\sim0.1\%$ at $\pi/2$, for all values of the lens spin as seen from Fig.~\ref{fig:min_mismatch}. In contrast, the mismatch values for $h^+_{\rm obs}$ exhibit qualitatively different behaviour. From $\theta_o=\pi/6$ to $\pi/3$, the (decreasing) trend is similar to that observed for $h^+_{\rm obs}$. However, beyond $\theta_o \approx \pi/3$ the mismatch for $h^\times_{\rm obs}$ starts to increase and within a thin band around $\theta_o\approx \pi/2$, it increases sharply to reach a maximum value of $\sim30\%$ at $\theta_o=\pi/2$. 
This large mismatch arises because near $\theta_o=\pi/2$ the direct cross-polarized component $h^\times_{\rm dir}$ vanishes\footnote{This is not strictly true when higher-mode contributions are significant or when the orbital plane undergoes precession.}, leaving the observed signal dominated primarily by the scattered contribution, i.e., $h^\times_{\rm obs}=h^\times_{\rm dir}+h^\times_{\rm SW}\approx h^\times_{\rm SW}$.
%This behaviour is further corroborated by the results presented in Table~\ref{tab:mismatch}, which lists the mismatch between the plus-polarized direct component and the observed waveform as a function of the lens-spin parameter $\chi=\{-0.999, 0., 0.999\}$ and $\theta=\{\pi/6, \pi/3, \pi/2\}$.

The minimum mismatch also corresponds to the minimum fractional loss in SNR due to missing physics in the template bank~\cite{Ajith:2012mn}, thereby affecting detectability. It is therefore instructive to examine how the optimal SNR itself changes in the presence of lensing. As mentioned before, optimal SNR is quantified by the norm of the total observed waveform including the scattered wave. Thus, the ratio of the optimal SNR between the lensed and unlensed waveforms captures the amplification of the waveform due to strong-field lensing. In Fig.~\ref{fig:opt_snr_ratio}, we compare the optimal SNRs of the lensed and unlensed waveforms.
For the plus-polarization, we observe no significant change in the optimal SNR due to lensing, with the values remaining close to unity. In contrast, for the cross polarization, we find a substantial enhancement in the optimal SNR, up to $\sim 30\%$, near $\theta_o=\pi/2$, for the same reason discussed in the previous paragraph, namely the vanishing of the cross-polarization component in the direct wave for $\theta_o=\pi/2$.

Next, we estimate the SNR required to identify lensing features in the signal. We use the approximation for the Bayes factor in terms of the fitting factor~\cite{Cornish:2011ys, DelPozzo:2014cla},
\begin{equation}
\ln \mathscr{B} \approx \frac{\rho^2}{2}\left(1 - \rm{FF}^2\right),
\end{equation}
and require the lensed-versus-unlensed ln-Bayes factor to be at least $10$ for the identification of lensing features. We adopt a higher threshold than that suggested by, for example, Jeffreys' criterion~\citep{Jeffreys:1939xee}, since this approximation is known to overestimate the Bayes factor (see, e.g., Refs.~\citep{DelPozzo:2014cla, Mishra:2023ddt}).
Fig.~\ref{fig:req_snr} shows the estimated SNR required, $\rho_{\rm req}$, to identify scattering-induced lensing features in $h^+_{\rm obs}$ and $h^\times_{\rm obs}$. We find that $\rho_{\rm req}\gtrsim 25$ near 
$\theta_o=\pi/6$, increasing to $\sim 100$ at $\theta_o=\pi/2$ for $h^+_{\rm obs}$, while it decreases to $\sim 10$ for $h^\times_{\rm obs}$. This behaviour follows directly from the corresponding decrease in the fitting factor values (see Fig.~\ref{fig:min_mismatch}).
We emphasize that the signal observed by a detector is a projection of the plus and cross polarizations. Consequently, near $\theta_o=\pi/2$ although scattering-induced features in $h^\times_{\rm obs}$ become detectable in isolation, the observed signal may still be dominated by $h^+_{\rm obs}$, potentially hindering their detectability. Additionally, note that although the required SNR is lower for $\theta_o=\pi/2$ when the cross-polarization is considered in isolation, detecting a signal that meets this SNR threshold may in fact be more difficult. This is because $h^\times_{\rm obs}$ is dominated by the scattered waveform for $\theta_o=\pi/2$, which has a smaller amplitude than the unlensed waveform due to the increased travel time and the corresponding additional decay [see Eq.~\eqref{eq:impact}], as well as the directional dependence of the scattering amplitude, see Fig.~\ref{sec:crossangle}. Nevertheless, with future detectors it may be possible to observe signals in this regime with SNRs exceeding the required threshold.

Finally, we note that the scattering formalism presented in this work is not valid for small angles ($\theta_o\lesssim \pi/6$), where the modification of the direct component due to lensing cannot be ignored. In fact, the very notion of splitting the observed waveform into direct and scattered components would no longer be feasible. Nevertheless, the convergence of the partial-wave expansion might be improved by adopting an eikonal approach. Such an improvement has indeed been demonstrated recently~\citep{CarrilloGonzalez:2025gqm} by approximating the sum over $l$-modes by an integral over the impact parameter. We leave the numerical study of small $\theta_o$ for future work. 

\section{Conclusion} \label{secIV}
In this work, we presented a detailed analysis of strong-field gravitational lensing of GWs by a Kerr BH in the wave-optics regime, extending previous study for the corresponding non-rotating case. Using MST techniques and high-precision numerical implementations, we quantify the effects via the strong-field scattering factor (SFSF) and the resulting waveform modulations. Our calculations successfully reproduce known Kerr scattering results~\cite{Dolan:2008kf} and clearly demonstrate that the SFSF remains non-decaying even at high frequencies for all spin parameters of the lens, in stark contrast to the conclusion of Ref.~\cite{Chan:2025wgz}. We traced this behaviour to the use Cesàro-$\alpha$ method adopted in Ref.~\cite{Chan:2025wgz}, rather than a physically robust resummation method, like series reduction~\cite{Dolan:2008kf}. The apparent high-frequency suppression reported there might, therefore, be attributed to numerical artefacts rather than genuine absorption of radiation by the lens. Several consistency checks against established results reinforce the reliability of our results. 

By propagating the scattered field to a distant observer and coherently combining it with the direct waveform, we quantify how strong-field wave-optics Kerr lensing modifies incident GW signals. The effects are most pronounced when the source and lens lie sufficiently close to each other, in which case the scattered contribution arrives within milliseconds of the direct signal, which is well inside the duration of typical LIGO event, and produces noticeable amplitude and phase modulations. To quantify this more concretely, we performed a mismatch analysis for a class of source–lens–observer configurations. These results show that strong-field scattering effects could be detectable even with current detectors for a subset of considered configurations, and the improved sensitivity and higher SNR of future observatories will only strengthen this prospect. In particular, as already discussed in the Sec.~\ref{secI}, percentage-level mismatches found for $\theta_o = 30^\circ$ lie near the detectability threshold of the current ground-based detectors and may be observable in sufficiently loud events. Looking ahead, next-generation detectors such as ET, CE, and LISA will achieve SNRs in the hundreds to thousands, reducing the detectable mismatch threshold to $10^{-4}-10^{-6}$. In these regimes, the strong-field Kerr-lensing signatures predicted here become decisively measurable, opening a realistic pathway to probing strong-field wave-optical effects in astrophysical environments.

The present work naturally motivates several future directions. Although our formalism applies to arbitrary scattering angles, we restricted explicit computations to the on-axis configuration for demonstration purposes and numerical simplicity. Extending the analysis to off-axis scattering could unveil richer spin-induced effects relevant for realistic astrophysical scenarios. Another important step is to incorporate the full chirping waveforms with broader frequency content, allowing our framework to address a wider variety of astrophysically viable systems. Such an effort will additionally enable a faithful estimate of lensing-induced biases in parameter estimation and could reveal subtle degeneracies between lensing signatures and intrinsic binary parameters. Ultimately, developing surrogate models tailored to Bayesian inference would allow strong-field lensing effects to be incorporated directly into GW data analysis pipelines.

From an observational standpoint, the rapidly growing catalogue of LVK events, together with with hints that some mergers may occur in AGN disks or hierarchical environments, making our study particularly timely. In summary, our analysis provides a unified and consistent framework for understanding wave-optical lensing in the strong-field region of Kerr BHs, setting the stage for realistic observational applications. As GW astronomy enters an era of higher sensitivity and richer astrophysical contexts, the phenomena investigated here may become essential tools for testing general relativity, probing BH environments, and mapping the strong-field structure of spacetime itself.

\begin{acknowledgments}
We thank Uddeepta Deka for helping us perform the mismatch analysis and providing valuable feedback. We thank Chris Kavanagh and Ajith Parameswaran for useful comments at many phases of the work. We extend our gratitude to Luka Vujeva and Juno Chan for carefully reading an earlier draft and providing valuable feedback. We are also thankful to the Astrophysical Relativity group at ICTS, for many engaging discussions and useful feedback. We acknowledge the use of IUCAA LDG cluster Sarathi for the computational work.
The work utilizes \texttt{Mathematica}~\citep{Mathematica} and the following software packages:
\texttt{Black Hole Perturbation Toolkit}~\citep{BHPToolkit},
\texttt{Cython}~\citep{behnel2011cython},
\texttt{FLINT}~\citep{FLINT}, 
\texttt{Jupyter notebook}~\citep{Kluyver2016jupyter},
\texttt{LALSuite}~\citep{2020ascl.soft12021L},
\texttt{Matplotlib}~\citep{Hunter:2007},
\texttt{NumPy}~\citep{Harris:2020xlr}, 
\texttt{PyCBC}~\citep{pycbc_github}, and
\texttt{SciPy}~\citep{Virtanen:2019joe}.
The research of M.V.S.S is supported by the National Post-Doctoral Fellowship (PDF/2025/004764), ANRF, Government of India. The research of RG and AM is supported by the Department of Atomic Energy, Government of India, under Project No. RTI4001.

\end{acknowledgments}

 \appendix
\section{Spheroidal harmonics decomposition of the incident wave} \label{appA}
In this appendix, our goal is to provide a detailed outline for the computation of $\widetilde{\psi}^{lm}_{P}$ at $r \to \infty$. We follow the approach outlined in Ref.~\cite{Bautista:2022wjf}. For this purpose, we can use the orthogonality relation of the spheroidal harmonics $\int d\Omega\, {}_{-2} S_{l m}(\theta, a\omega) e^{i m \phi}\, {}_{-2} S_{l' m'}(\theta, a\omega) e^{-i m' \phi} = \, \delta_{l l'} \delta_{m m'}$, where $d\Omega$ is the standard area element on a $2$-sphere. Note also that the spheroidal harmonics are all real and hence, we have avoided inserting the complex conjugation sign. Thus, we obtain
\begin{alignat}{3} \label{psiplm}
\widetilde{\psi}^{lm}_{\rm P}(\omega', r)=\int d\Omega~{}_{-2}S_{lm}(\theta,a\omega') e^{
 -i m \phi}\, \widetilde{\psi}_{\rm P},
\end{alignat}
where $\widetilde{\psi}_{\rm P}$ is given by Eq.~\eqref{psip}, which can be rewritten as 
\begin{alignat}{3}
    \widetilde{\psi}_{\rm P} = e^{i(\alpha_{\rm P} \cos\theta+\beta_{\rm P} \sin\theta \cos\phi)} \sum_{m=-2}^{2} A^{\rm P}_m(\theta,\gamma)e^{im\phi}.
    \end{alignat}
Here, we have $\alpha_{\rm P}={\rm P}\, \omega' r \cos\gamma$, $\beta_{\rm P}={\rm P}\, \omega' r \sin\gamma$, and the $10$ coefficients $A^{\rm P}_m$ can be easily identified by projecting onto $e^{-i m \phi}$. Note that the sum runs over $m=\{-2,-1,0,1,2\}$ due to the fact that the maximum powers of $\cos(\phi/2)$ and $\sin(\phi/2)$ in Eq.~\eqref{psipm} is $4$. Then, we can rewrite Eq.~\eqref{psiplm} as
\begin{equation}
    \begin{split}
        &\widetilde{\psi}^{lm}_{\rm P} =  \sum_{m'=-2}^2\int_{0}^{\pi} d\theta\, \sin\theta\, e^{i\alpha_{\rm P} \cos\theta} {}_{-2}S_{lm}(\theta,a\omega) A_{m'}^{\rm P}(\theta,\gamma)\\
        &\kern5em \times \int_{0}^{2\pi} d\phi\, e^{i\beta_{\rm P} \sin\theta\cos\phi}~e^{-i(m-m')\phi}, \\
        &\kern2em = 2\pi \sum_{m'=-2}^{m'=2}  \int e^{i\alpha_{\rm P} \cos\theta} \sin\theta\, d\theta ~ i^{\nu}~J_{\nu}(\beta_{\rm P}\, \sin\theta)\\
        &\kern6em \times A_{m'}^{\rm P}(\theta,\gamma) {}_{-2}S_{lm}(\theta,a\omega'),\\
    \end{split}
\end{equation}
where $\nu=(m-m')$ and we have used the Schl\"afli's integral representation (also known as the Jacobi-Anger expansion) of the Bessel function, namely
\begin{equation}
    i^\nu\, J_\nu(z)=\frac{1}{2\pi} \int_{0}^{2\pi}d\phi\, e^{-i\, \nu\, \phi+i\, z\, \cos\phi}
\end{equation}
with $\nu \in \mathbb{R}$ and $z \in \mathbb{C}$. Now, in the limit $\omega' r \gg 1$, we can approximate the Bessel function as
\begin{equation}
    \begin{split}
        J_\nu(\beta_{\rm P} \sin\theta) \approx \frac{1}{\sqrt{2\pi \beta_{\rm P} \sin\theta}} \left(e^{i \bar{\theta}}+e^{-i \bar{\theta}}\right)
    \end{split}
\end{equation}
where $\bar{\theta}=\beta_{\rm P} \sin\theta-(2\nu+1)\pi/4$, provided $\mathrm{arg}|\beta_{\rm P} \sin\theta| < \pi$. Then, after some straightforward algebra, one obtains 
\begin{equation}
    \begin{split}
        &\widetilde{\psi}^{lm}_{\rm P} \approx \sqrt{\frac{2\pi}{\beta_{\rm P}}} \sum_{m'=-2}^{2} \int_{0}^{\pi} d\theta\, \sqrt{\sin\theta}\, {}_{-2}S_{lm}(\theta,a\omega')\, A_{m'}^{\rm P}(\theta,\gamma)\\
        &\kern1em \times \left[-e^{i {\rm P} \omega' r \cos(\theta-\gamma)} e^{3i\pi/4}+(-1)^\nu e^{i {\rm P} \omega' r \cos(\theta+\gamma) e^{i\pi/4}} \right].
    \end{split}
\end{equation}
Moreover, since in the limit $\omega' r \gg 1$, the above exponential functions are highly oscillatory, we can consider the SPA
\begin{equation}
    \begin{split}
        \int_\mathbb{R} dx\, g(x)\, e^{i k f(x)} \approx &\sum_{x_0|f(x_0)=0} g(x_0)\, e^{i k f(x_0)} \sqrt{\frac{2\pi}{k |f''(x_0)|}} \\
        &\times e^{(i \pi/4)\, \mathrm{sgn}[f''(x_0)]}+\Ord{(k^{-1/2})}.
    \end{split}
\end{equation}
Using this formula, we further get
\begin{equation} \label{eq:fin}
    \begin{split}
        \widetilde{\psi}^{lm}_{\rm P} \approx &\, \frac{2\pi i}{{\rm P} \omega' r} \left[-e^{i {\rm P} \omega' r }\, {}_{-2}S_{lm}(\gamma,a\omega')\, \left(\frac{1+{\rm P}}{2}\right) + \right. \\
        &\left.(-1)^m\, e^{-i {\rm P} \omega' r}\, {}_{-2}S_{lm}(\pi-\gamma,a\omega')\, \left(\frac{1-{\rm P}}{2}\right)\right].
    \end{split}
\end{equation}
Then, specializing for ${\rm P}=\pm 1$, we finally obtain

\begin{equation}
    \begin{split}
        &\widetilde{\psi}_{+}^{lm}(\omega', r) \approx -\frac{2\pi i}{\omega' r}\, e^{i\omega' r}\, {}_{-2}S_{lm}(\gamma,a\omega'),\\
        &\widetilde{\psi}_{-}^{lm}(\omega', r) \approx (-1)^{m+1}\, \frac{2\pi i}{\omega' r}\, e^{i\omega' r}\, {}_{-2}S_{lm}(\pi-\gamma,a\omega').
    \end{split}
\end{equation}
Now, we can use Eq.~\eqref{psip} to write down the parity-specific frequency-domain modes of the incident wave as
\begin{equation}
    \begin{split}
        &\widetilde{\psi}_{+}(\omega', r, \theta, \phi) \approx -\frac{ 2\pi i\, e^{i\omega' r^*}}{\omega' r}\sum_{lm}S_{lm}(\gamma,a\omega') \times\\
        &\kern14em {}_{-2}S_{lm}(\theta,a\omega')\, e^{im\phi} ,\\
        &\widetilde{\psi}_{-}(\omega', r, \theta, \phi) \approx \frac{ 2\pi i\, e^{i\omega' r^*}}{\omega' r}\sum_{lm}(-1)^{m+1}\,S_{lm}(\pi-\gamma,a\omega')  \\
        &\kern13em  \times {}_{-2}S_{lm}(\theta,a\omega')e^{im\phi}.
    \end{split}
\end{equation}
Here, we have replaced $r \to r_*$ in the exponent for the reason discussed earlier in the main text. The above results are consistent with Ref.~\citep{Bautista:2022wjf} for off-axis scattering.

\section{Derivation of the phase formula} \label{appB}
In this appendix, we derive the phase relation presented in Eq.\eqref{sec:eq:ph}. The overall approach can be outlined as follows. The Teukolsky scalar $\psi_0$ characterizes the incoming component ($\psi_0^{\rm in}$) of the radiation at spatial infinity, to leading order in $1/r$, and can be obtained directly from the plane wave given in Eq.~\eqref{h11} using a procedure analogous to that employed for calculating the outgoing component $\psi_4^{\rm out}$ in the main text. Furthermore, as discussed earlier, the subleading term $\psi_4^{\rm in}$ also carries information about the incoming radiation. The connection between this subleading contribution and the plane-wave amplitudes ultimately yields the desired phase formula. We use some of the results given in Ref.~\cite{Bautista:2022wjf}, while correcting for a few typos. 

\textit{(a) Computation of $\psi_0$:} For this purpose, we first compute the asymptotic parity-specific mode expansions of $\psi_0$ for the plane waves given by Eq.~\eqref{h11}:
\begin{alignat}{3} \label{psi0pt}
&(2\,\epsilon\, \omega^2)^{-1}\, \psi_0^{\rm P}(t,r,\theta,\phi) \approx \nnm\\
&-e^{-i\chi^+}\Big[\cos\Big(\frac{\phi}{2}\Big)\sin\Big(\frac{\gamma-\theta}{2}\Big)-i \sin\Big(\frac{\phi}{2}\Big)\sin\Big(\frac{\gamma+\theta}{2}\Big)\Big]^4 \nnm \\
& \mp {\rm P} e^{i\chi^-}\Big[\cos\Big(\frac{\phi}{2}\Big)\cos\Big(\frac{\gamma-\theta}{2}\Big)-i \sin\Big(\frac{\phi}{2}\Big)\cos\Big(\frac{\gamma+\theta}{2}\Big)\Big]^4.
\end{alignat}
Expanding it in frequency basis using similar conventions as for $\psi_4$ in Eq.~\eqref{eq:psi4decomp} we get
\begin{alignat}{3}
\psi_0^{\rm P} =  -\frac{\epsilon\, \omega^2}{4}\int_{-\infty}^\infty d\omega'\, e^{-i\omega' t}\, \widetilde{\psi}_0^{\rm P}(\omega',r,\theta,\phi),
\label{eq:psi0decomp}
\end{alignat}
where to match with the expression in Eq.~\eqref{psi0pt}, we need $\widetilde{\psi}_0^{\rm P}(\omega',r,\theta,\phi) = \widetilde{\psi}'_+(\omega',r,\theta,\phi) \delta(\omega-\omega')+{\rm P}\, \widetilde{\psi}'_-(\omega',r,\theta,\phi) \delta(\omega+\omega')$ with the shorthands
\begin{equation} \label{psi0pm}
    \begin{split}
    &\widetilde{\psi}'_{+}(\omega', r, \theta, \phi) \approx 8e^{i \omega' r (\cos\gamma\cos\theta+ \sin\gamma\sin\theta\cos\phi)} \times \\
    &\kern2em\Big[\cos\Big(\frac{\phi}{2}\Big)\sin\Big(\frac{\gamma-\theta}{2}\Big)-i \sin\Big(\frac{\phi}{2}\Big)\sin\Big(\frac{\gamma+\theta}{2}\Big)\Big]^4,\\[0.7em]
    &\widetilde{\psi}'_{-}(\omega', r, \theta, \phi) \approx 8e^{-i \omega' r (\cos\gamma\cos\theta+ \sin\gamma\sin\theta\cos\phi)} \times \\
    &\kern2em \Big[\cos\Big(\frac{\phi}{2}\Big)\cos\Big(\frac{\gamma-\theta}{2}\Big)-i \sin\Big(\frac{\phi}{2}\Big)\cos\Big(\frac{\gamma+\theta}{2}\Big)\Big]^4,
    \end{split}
\end{equation}
We can now further decompose $\widetilde{\psi}'_{\rm P}$ in spheroidal harmonics, but this time with spin weight $s=+2$, as
\begin{alignat}{3}
&\tilde{\psi}'_{\rm P} = \sum_{lm} \widetilde{\psi}'^{lm}_{\rm P}\, {}_{2}S_{l m}(\theta,a\omega')\, e^{im\phi} \nonumber \\
&\implies \widetilde{\psi}'^{lm}_{\rm P} = \int d\Omega\, \widetilde{\psi}'_{\rm P}\,{}_{2}S_{l m}(\theta,a
\omega')e^{-im\phi}.
\end{alignat}
Now, following the similar computations as demonstrated in Appendix-\ref{appA}, and obtain
\begin{equation}
    \begin{split}
        &\widetilde{\psi}'^{lm}_{+}(\omega', r) \approx (-1)^m\,\frac{8\pi i}{\omega' r}\, e^{-i\omega' r}\, {}_{2}S_{lm}(\pi-\gamma,a\omega'),\\
        &\widetilde{\psi}'^{lm}_{-}(\omega', r) \approx \frac{8\pi i}{\omega' r}\, e^{-i\omega' r}\, {}_{2}S_{lm}(\gamma,a\omega').
    \end{split}
\end{equation}
Now, using Eq.~\eqref{eq:psi0decomp}, we can rewrite the ingoing part of $\psi_0$ by adding the specific-parity modes in the following suggestive form
\begin{equation} \label{psi0in}
    \begin{split}
        &\psi_0^{\rm in} \approx \frac{-\epsilon\, \omega^2}{4}\int_{-\infty}^{\infty}   d\omega'\sum_{{\rm P} lm} K'^{\rm in}_{l m \rm P}(\omega')\, \frac{e^{-i\omega'(t+r^*)}}{\omega' r} \\
        &\kern12em \times {}_{2}S_{lm}(\theta,a\omega')\, e^{im\phi},
    \end{split}
\end{equation}
where $K'^{\rm in}_{lm \rm P} =  \widehat{\psi}'^{lm}_+(\omega')\, \delta(\omega'-\omega) + {\rm P}\, \widehat{\psi}'^{lm}_{-}(\omega')\, \delta(\omega'+\omega)$ with $\widehat{\psi}_{\rm \rm P}'^{lm}(\omega') = \widetilde{\psi}_{\rm P}'^{lm}\, \omega' r\, e^{i \omega' r_*}$.

\textit{(b) Obtaining $\psi_4^{\rm in}$ from $\psi_0^{\rm in}$:} Now that we have found the leading piece of $\psi_0^{\rm in}$, we can apply the following Teukolsky-Starobinsky identity in the Geroch-Held-Penrose (GHP) notation~\cite{Bautista:2022wjf} as
\begin{alignat}{3} \label{TS}
 \thorn^4\zeta^4 \psi_4 =\eth'^4 \zeta^4\psi_0 -3M\, \mc{L}_t \bar{\psi}_0,
\end{alignat}
where $\zeta=r-i a \cos\theta$, $\thorn = \ell^\mu \partial_\mu$ in Kinnersley tetrad, $\eth' = \bar{m}^\mu\partial_\mu+p\beta'-q\bar{\beta}$ with $\beta=\cot\theta/(2 \sqrt{2}\, \bar{\zeta})$, $\beta'=[r\cot\theta-ia\sin\theta(\csc^2\theta+1)]/(2 \sqrt{2}\, \zeta)$, and $\{p,q\}$ are the GHP weights of the function being acted upon by the operator (here $\eth$). Specifically, the GHP weights of $\psi_0$ and $\zeta$ are $\{4,0\}$ and $\{0,0\}$, respectively. When two functions are multiplied, their GHP weights are added. Finally, $\mc{L}_t$ is the Lie-derivative with respect to the time-like killing vector. Since $\psi_0$ is a scalar function, and we have aligned our time-like direction with $(\partial_t)^\alpha$, $\mc{L}_t$ is simply the time-derivative.

Let us now evaluate both sides of the above identity for the following asymptotic $\omega'lm$-mode expansion of $\psi_4$: 
\begin{equation} \label{psi4refinc}
    \begin{split}
        &\zeta^4\, \psi_4 \approx \epsilon\, \int d\omega' \sum_{lm} e^{-i\omega' t+i m \phi}\, {}_{-2}S_{lm}(\theta,a\omega') \times\\
        &\kern6em  \Big[r^3\, B^{(\rm refl)}_{l m \omega'}\, e^{i\omega' r} + B^{(\rm inc)}_{l m \omega'}\, \frac{e^{-i \omega' r}}{r} \Big].
    \end{split}
\end{equation}
Then, the left hand side of Eq.~\eqref{TS}, for the choice of tetrads given by Eq.~\eqref{tetrad}, asymptotically reduces (in leading contribution) to \begin{alignat}{3}
(\partial_t+\partial_r)^4 \zeta^4 \psi_4 \approx &16\epsilon \sum_{l m}\int d\omega'\, \omega'^4\, r^{-1} B^{(\rm inc)}_{l m\omega'}\, e^{-i\omega'(t+r)} \nonumber\\ &\kern4em \times {}_{-2}S_{l m}(\theta,a\omega')e^{i m\phi}.
\end{alignat}
Note that the leading piece (outgoing) of $\psi_4$ is annihilated by $\thorn^4$. Now, on the right hand side of Eq.~\eqref{TS}, the first term can be evaluated by using the relation~\cite{Bautista:2022wjf}
\begin{alignat}{3}
\eth'^4\zeta^4\, {}_2 S_{l m}(\theta,a\omega)\, e^{im\phi} = \frac{1}{4}\Re(C) ~{}_{-2}S_{l m}(\theta,a\omega)\, e^{im\phi},
\end{alignat}
where the Teukolsky-Starobinsky constant $C$ was introduced earlier in Eq.~\eqref{sec:eq:ph}. Note that the operator $\eth'$ only involves angular derivatives as the tetrad vector $m$ has no time-like or radial components far away from the BH. Hence, one can straightforwardly apply the above result on Eq.~\eqref{psi0in}. 

Finally, we have $-3M\, \mc{L}_t\bar{\psi_{0}}=-3M\, (d\bar{\psi}_0/dt)$, where $\bar{\psi}_0$ can be obtained by taking the complex conjugation of Eq.~\eqref{psi0in}. This can be further simplified to
\begin{alignat}{3}
&-\frac{3i M \omega^2}{4} \epsilon \sum_{{\rm P} l m}\int d\omega'\, (-1)^l\, K'^{\rm in}_{lm{\rm P}}(\omega')\, \frac{e^{-i\omega'(t+r)}}{r} \nonumber\\ &\kern14em {}_{-2}S_{l m}(\theta,a\omega')e^{im\phi} \nonumber
\end{alignat}
by first switching the dummy variables $\omega'\rightarrow -\omega'$ and $m\rightarrow -m$, and then applying the identity ${}_sS_{l m}(\theta,a\omega) = (-1)^{m+s}{}_{-s}S_{l -m}(\theta,-a\omega)$. Moreover, we have used $\bar{K}'^{\rm in}_{l-mP}(-\omega') = (-1)^{l+m+1} {\rm P} K'^{\rm in}_{lm{\rm P}}(\omega')$, which follows easily from the definition of $K'^{\rm in}_{lm{\rm P}}$ given below Eq.~\eqref{psi0in}. 

Now, collecting all parts of the Teukolsky-Starobinsky identity together, we get the relation between $B^{(\rm inc)}_{l m \omega'}$ and $K'^{\rm in}_{l m {\rm P}}(\omega')$ as
\begin{alignat}{3} \label{BincKin}
B^{(\rm inc)}_{l m \omega'} = -\frac{\omega^2}{4} \sum_{\rm P} \Big[\frac{\Re(C)+12 i M \omega'\, {\rm P} (-1)^{l}}{16\omega'^5}\Big]K^{\rm in}_{l m {\rm P}}(\omega'),
\end{alignat}
where we have used $K^{\rm in}_{l m {\rm P}}(\omega')=K'^{\rm in}_{l m {\rm P}}(\omega')/4$ for future reference. Note also that the above expression has a relative $(-1)^l$ compared to the expression in Ref.~\cite{Dolan:2008kf}, which is due to the different convention of parity operators being used.\\ 

\textit{(c) Obtaining the phase formulae:} We can now rewrite the asymptotic form of the incident $\psi_4$ given by Eq.~\eqref{psi4refinc} with parity modes separated and including the subleading (ingoing) term as well:
\begin{equation} \label{eq:st2:psi4:cor}
    \begin{split}
        &\frac{\zeta^4}{r^4}\psi_4^{\rm PW} \approx 
        \frac{-\epsilon\, \omega^2}{4} \int_{-\infty}^{\infty}   d\omega'\, {}_{-2}S_{lm}(\theta,a\omega')\, e^{im\phi} \times \\
        &\sum_{{\rm P} l m} \Bigg[K^{\rm out}_{l m {\rm P}}(\omega')\, \frac{ e^{-i\omega'(t-r^*)}}{\omega' r}+\frac{\Re(C)+12i M \omega'\, {\rm P} (-1)^l}{16(\omega')^4 r^4}\\
        &\kern12em \times K^{\rm in}_{l m {\rm P}}(\omega')\,  \frac{e^{-i\omega'(t+r^*)}}{\omega' r}\Bigg],
    \end{split}
\end{equation}
where we have used Eq.~\eqref{BincKin} and identified $B^{\rm refl}$ in terms of $K^{\rm out}$.

Finally, to define the phase, we need to understand the relationship connecting the in-out coefficients, as given below Eq.~\eqref{psi4out} and Eq.~\eqref{psi0in}, for the case of incident and scattered plane waves. It can be easily verified that for the incident plane wave, we have the relation $K^{\rm out}_{l m {\rm P}}(\omega') = (-1)^{l+1}\, K^{\rm in}_{l m {\rm P}}(\omega')$. However, as discussed earlier, scattering produces a relative phase shift in the outgoing part of the wave. This suggests the above relation should be modified as $K^{\rm out}_{l m {\rm P}}(\omega') = (-1)^{l+1}\, \eta^{\rm P}_{l m}\, \exp(2i\delta^{\rm P}_{l m})\, K^{\rm in}_{l m {\rm P}}(\omega')$ after scattering. Thus, we have 
\begin{alignat}{3}
\eta^{\rm P}_{l m} e^{2i\delta^{\rm P}_{l m}}(\omega') = (-1)^{l+1} \frac{\Re(C)+12iM\omega' ~{\rm P} (-1)^{l}}{16\omega'^{4}} \frac{B^{(\text{refl})}_{l m\omega'}}{B^{(\text{inc})}_{l m \omega'}}
\end{alignat}
as used in the main text. Note that the ratio of $B's$ in the right hand side of the above expression is independent of parity, and can be computed using the MST approach.

\section{Two resummation techniques} \label{appC}
In this appendix, we briefly describe two resummation techniques, namely the Cesàro-$\alpha$ summation and series reduction procedure, used to resum the multipolar ($l$) spheroidal-harmonic expansion of a series, especially while reconstructing a plane wave from partial waves. When such a reconstruction is done from a truncated set of multipolar modes, i.e., $l \in [0,l_{\rm max}]$, the straightforward sum typically fails to converge and usually show unphysical oscillations about the true value. This is a well-know difficulty in partial-wave analysis, where the following techniques can provide a controlled way to extract the physically meaningful sum.

\textit{(a) Cesàro-$\alpha$ summation:} Let us first discuss the methodology behind the Cesàro-$\alpha$ sum~\cite{hardy2024divergent}. Given a sequence of terms $\{a_l\}$ to be summed, the series $\sum_{l=0}^{\infty} a_l$ is said to be Cesàro-$\alpha$ summable to a value of $S$ if the following series of weighted-terms (equivalently, of averaged partial sums) converges:
\begin{equation}
    C^{\alpha}[a_l] := \sum_{l=0}^{l_{\rm max}} \frac{\binom{l_{\rm max}}{l}}{\binom{l_{\rm max}+\alpha}{l}}\, a_l\, \xrightarrow[ l_{\rm max} \to \infty]{} S,
\end{equation}
for some $\alpha>-1$. For $\alpha=0$, this reduces to the ordinary sum. Whereas larger values of $\alpha$ usually apply stronger smoothening by suppressing oscillations in the ordinary sum. In practice, however, one keeps on increase the value of $\alpha$ unless a numerically stable sum [i.e., when the relative error between Cesàro-$\alpha$ and Cesàro-$(\alpha+1)$ sum drops below a prefixed threshold] of the underlying truncated (at $l_{\rm max}$) series is attained.

Although larger $\alpha$ values usually regularize the sum better, care must be taken when applying it on such truncated series. Choosing too large a value of $\alpha$ can over-suppress the high-$l$ contributions that could be physically important. In such cases, the resummed result may deviate from the correct value due to artificially induced smoothening/distortion effects. Hence, it is always best practice to start from lower values, say $\alpha =2$, and then gradually increasing it to higher values, say $\alpha=5$, while keeping track of the relative error in the subsequent increase in $\alpha$.

One should, however, keep in mind that Cesàro summation is essentially a numerical regularization scheme. Thus, in a given physical problem (e.g., BH scattering), some other methods like series reduction, which we shall discuss next, can work better if they factor out the known physical divergence(s) of the underlying sum for achieving a better convergence property.

\begin{figure}[h!]
\includegraphics[width=\linewidth]{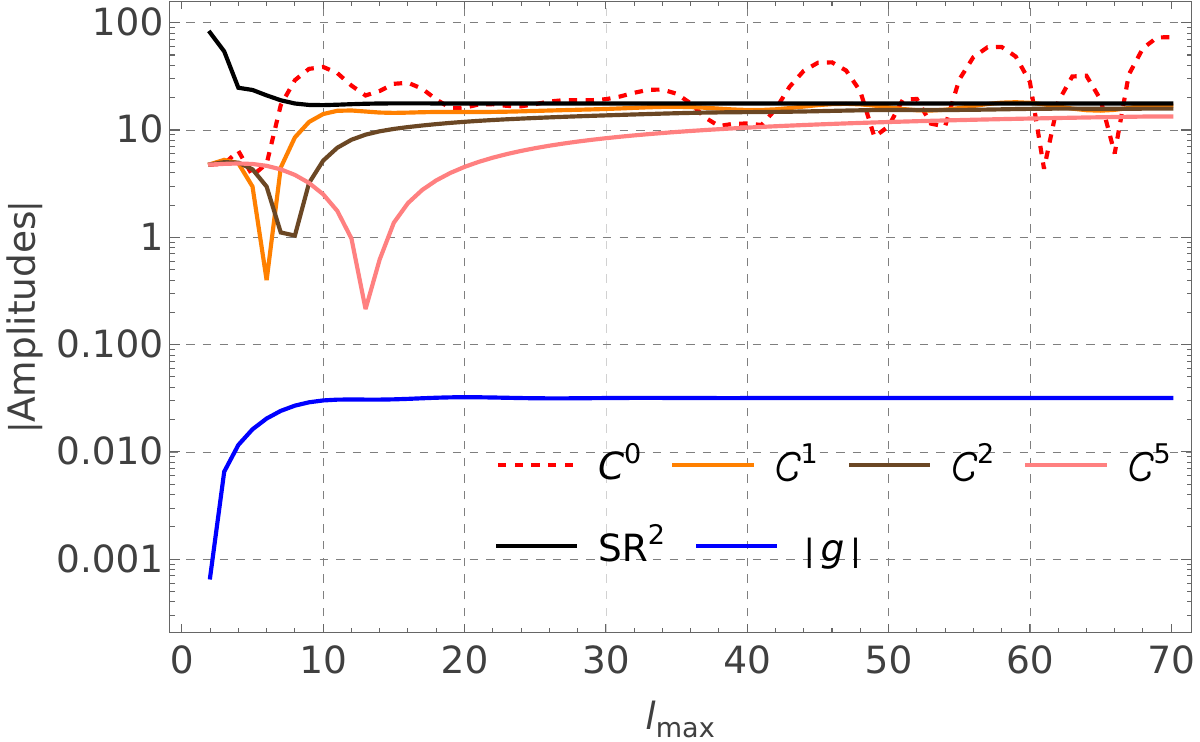}
\caption[Plots quantifying the effect of Cesàro and series reduction]{Plots showing the absolute amplitudes $|f(\gamma=0,\theta_o=\phi_o=\pi/6)|$ and $|g(\gamma=0,\theta_o=\phi_o=\pi/6)|$ for the spinless case ($a=0$) at $M\omega=0.5$, resummed using series reduction of order 2, and Cesàro summation for $\alpha=\{0,1,2,5\}$. The x-axis shows $l_{\rm max}$ which sets the limit for summation over $l$ in Eqs.~\eqref{eq:psi4sw2}. Note that $g$, the helicity-reversing amplitude, smoothly converges as $l_{\rm max}$ is increased. However, the raw result for $f$, i.e.,($C^0$) (shown by the dotted-red line) is highly oscillatory and does not converge as $l_{\rm max}$ is increased. However, upon doing series reduction of order $2$ ($SR^2$) or Cesàro summation ($C^0$,~$C^1$,~$C^2$,~$C^5$), it shows much better convergece properties for large $l_{\rm max}$.} 
\centering
\label{sec:cesarosr}
\end{figure}

\textit{(b) Series reduction:} 
In series reduction, one analytically captures its divergent behaviour at some parameter location(s) by properly factoring them out, leaving behind a well defined convergent sum. In the scattering problem, it turns out that the helicity-preserving scattering amplitude $f(\gamma,\theta_o,\phi_o)$ requires series reduction, whereas the helicity-reversing scattering amplitude $g(\gamma,\theta_o,\phi_o)$, does not. This follows from the fact that $f$ is divergent when $\theta_o=\gamma$, whereas $g$ is regular there. This can be seen even from the leading-order analytical expressions for the scattering amplitudes~\cite{Saketh:2022wap}. 

Thus, to perform series reduction, we first write $f$ as an expansion in spin-weighted spherical harmonics. As shown in Eq.~\eqref{eq:psi4sw2}, $f$ is naturally expressed as an expansion in spin-weighted spheroidal harmonics. However, spin-weighted spheroidal harmonics further admit an expansion in spin weighted spherical harmonics (as any other regular function defined on a 2-sphere). This expansion can be found using the matrix method implemented in the \texttt{SpinWeightedSpheroidalHarmonics} package of the \texttt{Black Hole Perturbation Toolkit}~\citep{BHPToolkit}.

Thus, we have an expansion of the form 
\begin{alignat}{3}
f(\gamma,\theta_o,\phi_o) = \sum_{l m}\mc{F}_{l m}\, {}_{-2}Y_{l m}(\theta_o,\phi_o).
\end{alignat}
As a demonstration, we show series reduction for the on-axis case, where $\gamma=0$, albeit the principles remain similar. In this case, the above function is divergent for $\theta_o=0$. Note that one can always perform a rotation so the divergence is centred at 0. Thus, we rewrite the function as 
\begin{alignat}{3}
f(0,\theta_o,\phi_o) = \frac{\hat{f}(\theta_o,\phi_o)}{(1-\cos\theta_o)^k},\quad \hat{f}=\sum_{l m}\hat{\mc{F}}_{l m} ~{}_{-2}Y_{l m}(\theta_o,\phi_o).
\end{alignat}
For a suitable value of $k$ (referred to as the order of series reduction), the function $\hat{f}(\theta_o,\phi_o)$ is expected to have a well-defined convergent spherical expansion. For the specific problem of interest~\cite{Dolan:2008kf}, order $k=2$ appears to be a good choice.

To proceed further, we need to compute the coefficients $\hat{\mc{F}}_{l m}$, in terms of the knowns, i.e., the coefficients $\mc{F}_{l m}$. It is sufficient to solve the problem for $k=1$, as one can then just iterate as required to obtain the relation for any $k\in \mathbb{Z}^+$. We thus write
\begin{alignat}{3}
\hat{\mc{F}}_{l m}&=\int d\theta_o\, \sin\theta_o\, \int d\phi_o\, \hat{f}~{}_{-2}Y_{l m}^* \nnm\\
&=\sum_{l' m'}\mc{F}_{l' m'}\int d\theta_o \sin\theta_o\, \int d\phi_o\, {}_{-2}Y_{l' m'} \times \\
&\kern10em {}_{-2}Y_{l m}^* (1-\cos\theta_o). \nnm 
\end{alignat}
This can be evaluated analytically using the Wigner-3j symbols. It is clear from angular-momentum considerations that there can be a non-zero value only for $l=\{l',l'+1,l'-1\}$ and $m'=m$. Specifically, we have
\begin{alignat}{3}
&\hat{\mc{F}}_{l m} = \mc{F}_{l m} \left(1-\frac{2m}{l(l+1)}\right)-\frac{A_l}{l+1}\mc{F}_{l+1,m}-\frac{B_l}{l}\mc{F}_{l-1,m}, \nnm \\
&A_l = \sqrt{\frac{(l-1)(l+3)(l-m+1)(l+m+1)}{4l(l+2)+3}}, \\
&B_{l} =\sqrt{\frac{(l^2-4)(l^2 - m^2)}{4l^2-1}}.
\end{alignat}
The above relation lets us generate the list of the reduced series $\hat{F}_{l m}$. Doing it twice yields the result for $k=2$. The reduced series is found to converge much faster. In this work, we compute the coefficients up to $l=70$.

As a demonstration, we show for particular case ($a=0$, $M\omega=0.5$, $\gamma=0$, $\theta_o=\phi_o=\pi/6$) in Fig.~\ref{sec:cesarosr} how the amplitudes ($f$ and $g$) obtained directly via computation compares with that obtained either via series reduction (at k=2), and Cesàro summation (for $\alpha=1,2,3,5$.). Note that all methods succeed in smoothening the data and giving a convergent result for $f$ and $g$ as $l_{\rm max}$ is increased. However, increasing $\alpha$ does not cause the series-reduction result and Cesàro summation results to agree with each other. For example, in Fig.~\ref{sec:cesarosr}, Cesàro summation with $\alpha=5$ (solid pink) disagrees more with the result obtained via series-reduction technique than the corresponding $\alpha=2$ (solid orange) case.

\section{Computing correction to waveform in frequency domain}
\label{appD}
In the main text, we have already shown (via SFSF) how a monochromatic-plane wave is affected by a Kerr BH lens. Here, we generalize this to the case where the plane wave is instead a superposition of frequencies, as expected from realistic GW signals that evolve in time. It is simplest to work in the frequency domain. We again restrict to the case where the lens experiences purely positive helicity GW radiation, for simplicity. It is straightforward to generalize this, following the discussion below Eq.~\eqref{eq:onaxispch}, in cases where the incident wave is a linear combination of both helicities.

The frequency-domain waveform at the lens location in the $(+,\times)$-basis can be easily obtained using the software package \texttt{PyCBC} and is related to the time-domain waveform via a Fourier transform. If the time-domain waveform is schematically given by
\begin{equation}
        h^{\mu\nu}_{\rm wave}(t,\vec{r})=h_+(t-r,\vec{r})\, \eta_+^{\mu\nu} + h_\times (t-r,r)\, \eta_\times^{\mu\nu},
\end{equation}
where $\eta_{+,\times}^{\mu\nu}$ are the basis polarization tensors. We define the frequency domain waveforms as
\begin{equation}
    h_{+,\times}(\nu,\vec{r})=\int_{-\infty}^\infty h_{+,\times}(u,\vec{r})\, e^{2i \pi \nu u}du, 
\end{equation}
where $u=t-r$ is the retarded time. It is convenient to write them as $h_+(\nu)=A_+ e^{i\psi_+}$ and $h_\times(\nu)=A_\times e^{i\psi_\times}$. We can invert these relations as
\begin{equation}
    \begin{split}
        &h_{\rm wave}^{\mu\nu} (u,\vec{r}) = \int_{-\infty}^\infty \frac{d\nu}{2\pi}e^{-2\pi i \nu u}\, \left[h_+(\nu,\vec{r})  \eta_+^{\mu\nu} +(+\leftrightarrow \times)\right] \\ 
        &= 2 \Re\left\{\int_{0}^\infty \frac{d\nu}{2\pi}e^{-2\pi i \nu u}\, \left[h_+(\nu,\vec{r})  \eta_+^{\mu\nu} +(+\leftrightarrow \times)\right]\right\}
    \end{split}
\end{equation}
Now, in the special case where $A_+ = A_\times$, and $\psi_\times=\pi/2+\psi_+$. We have
\begin{equation}
   h_{\rm wave}^{\mu\nu}(u,\vec{r}) = 2\Re \left\{\int_0^\infty \frac{d\nu}{2\pi}e^{-2\pi i \nu u}\, A_+ e^{i\psi_+}\, \left[\eta_+^{\mu\nu} + i \eta_\times^{\mu\nu}\right]\right\}.
\end{equation}
Now, using the relation $\varepsilon_{+}^{\mu}\varepsilon_{+}^\nu = \eta_{+}^{\mu\nu} + i \eta_\times^{\mu\nu} $, where $\varepsilon_+^\mu\varepsilon_{+}^\nu$ is the polarization tensor for a positive helicity wave, defined in the main text. We see, thus, that a purely positive helicity GW perturbation, conveniently rewritten as 
\begin{equation}
\begin{split}
      h_{\rm wave}^{\mu\nu}(u,\vec{r}) &= 2\Re\left\{\int_0^\infty \frac{d\nu}{2\pi}e^{-2\pi i \nu u } A_+ e^{i\psi_+} \varepsilon^{\mu}_+\varepsilon^{\nu}_{+}\right\} \\ & = 2 \Re\left\{\int_0^\infty \frac{d\nu}{2\pi}e^{-2\pi i \nu u } h_+(\nu,\vec{r}) \varepsilon^{\mu}_+\varepsilon^{\nu}_{+}\right\}
\end{split}
\end{equation}

In the main text, we showed in Eq.~\eqref{eq:h2pc} that for a monochromatic plane-wave with positive helicity, i.e, when $h_+(\nu>0,\vec{r})=\pi \delta(\nu-\nu_0)=i h_\times(\nu>0)$, the $+$ and $\times$ polarizations in the scattered wave are simply given by 
\begin{equation}
    \begin{split}
        & h_+^{\rm SW}(u,\vec{r}_{\rm LO}) = \Re\left\{\frac{f+g}{r_{\rm LO}}e^{-i2\pi \nu_0 u}\right\},\\
        & h_\times^{\rm SW}(u,\vec{r}_{\rm LO}) = -\Im\left\{\frac{f-g}{r_{\rm LO}}e^{-i2\pi \nu_0 u}\right\}.
    \end{split}
\end{equation}
In the frequency domain, these can be rewritten as
\begin{equation}
\begin{split}
        h^{\rm SW}_+(\nu>0) &= \frac{\pi}{ r_{\rm LO}}(f+g)\delta(\nu-\nu_0) \\
    h^{\rm SW}_\times(\nu>0) &= \frac{i \pi}{ r_{\rm LO}}(f-g)\delta(\nu-\nu_0).
\end{split}
\end{equation}

To generalize them for arbitrary $h_+(\nu>0)$, we simply rewrite them as 
\begin{equation}
\begin{split}
        h^{\rm SW}_+(\nu>0) &= \frac{1}{ r_{\rm LO}}(f+g) h_+(\nu>0)\\
    h^{\rm SW}_\times(\nu>0) &= \frac{1}{ r_{\rm LO}}(f-g)h_\times(\nu>0).
\end{split}
\end{equation}
Finally, before adding the scattered wave to the direct wave from the source to observer, we need to account for the time-delay between them. The direct wave travels roughly along the straight line from source to observer, whereas the scattered wave first reaches the lens, and subsequently makes its way to the observer. Thus, the time-delay is given by $\Delta t = r_{\rm LO}+r_{\rm SL}- r_{\rm SO}\approx r_{\rm SL}[1-\cos\theta_o] = 2 r_{\rm SL} \sin^2(\theta_o/2)$, where we have used $r_{\rm SL}\ll r_{\rm SO}\sim r_{\rm LO}$. In the frequency domain, this time-delay can be taken into account simply through a relative phase before adding the two contributions as
\begin{equation}
\begin{split}
        h^{\rm obs}_+(\nu>0) &= h^{\rm dir}_+(\nu>0) + e^{2 i\pi \nu \Delta t}h^{\rm SW}_+(\nu>0),\\
    h^{\rm obs}_\times(\nu>0) &= h^{\rm dir}_\times(\nu>0)  + e^{2 i\pi \nu \Delta t}h^{\rm SW}_\times(\nu>0).
\end{split}
\end{equation}
The direct waveform, $h_{+,\times}^{\rm dir}$ can also be computed with ease using \texttt{PyCBC}. This concludes the derivation.
    \bibliography{ref}

@article{LIGOScientific:2016aoc,
    author = "Abbott, B. P. and others",
    collaboration = "LIGO Scientific, Virgo",
    title = "{Observation of Gravitational Waves from a Binary Black Hole Merger}",
    eprint = "1602.03837",
    archivePrefix = "arXiv",
    primaryClass = "gr-qc",
    reportNumber = "LIGO-P150914",
    doi = "10.1103/PhysRevLett.116.061102",
    journal = "Phys. Rev. Lett.",
    volume = "116",
    number = "6",
    pages = "061102",
    year = "2016"
}

@article{LIGOScientific:2014pky,
    author = "Aasi, J. and others",
    collaboration = "LIGO Scientific",
    title = "{Advanced LIGO}",
    eprint = "1411.4547",
    archivePrefix = "arXiv",
    primaryClass = "gr-qc",
    doi = "10.1088/0264-9381/32/7/074001",
    journal = "Class. Quant. Grav.",
    volume = "32",
    pages = "074001",
    year = "2015"
}

@article{VIRGO:2014yos,
    author = "Acernese, F. and others",
    collaboration = "VIRGO",
    title = "{Advanced Virgo: a second-generation interferometric gravitational wave detector}",
    eprint = "1408.3978",
    archivePrefix = "arXiv",
    primaryClass = "gr-qc",
    doi = "10.1088/0264-9381/32/2/024001",
    journal = "Class. Quant. Grav.",
    volume = "32",
    number = "2",
    pages = "024001",
    year = "2015"
}

@article{LIGOScientific:2016lio,
    author = "Abbott, B. P. and others",
    collaboration = "LIGO Scientific, Virgo",
    title = "{Tests of general relativity with GW150914}",
    eprint = "1602.03841",
    archivePrefix = "arXiv",
    primaryClass = "gr-qc",
    reportNumber = "LIGO-P1500213",
    doi = "10.1103/PhysRevLett.116.221101",
    journal = "Phys. Rev. Lett.",
    volume = "116",
    number = "22",
    pages = "221101",
    year = "2016",
    note = "[Erratum: Phys.Rev.Lett. 121, 129902 (2018)]"
}

@article{Liao:2017ioi,
    author = "Liao, Kai and Fan, Xi-Long and Ding, Xu-Heng and Biesiada, Marek and Zhu, Zong-Hong",
    title = "{Precision cosmology from future lensed gravitational wave and electromagnetic signals}",
    eprint = "1703.04151",
    archivePrefix = "arXiv",
    primaryClass = "astro-ph.CO",
    doi = "10.1038/s41467-017-01152-9",
    journal = "Nature Commun.",
    volume = "8",
    number = "1",
    pages = "1148",
    year = "2017",
    note = "[Erratum: Nature Commun. 8, 2136 (2017)]"
}

@article{LIGOScientific:2017zic,
    author = "Abbott, B. P. and others",
    collaboration = "LIGO Scientific, Virgo, Fermi-GBM, INTEGRAL",
    title = "{Gravitational Waves and Gamma-rays from a Binary Neutron Star Merger: GW170817 and GRB 170817A}",
    eprint = "1710.05834",
    archivePrefix = "arXiv",
    primaryClass = "astro-ph.HE",
    reportNumber = "LIGO-P1700308",
    doi = "10.3847/2041-8213/aa920c",
    journal = "Astrophys. J. Lett.",
    volume = "848",
    number = "2",
    pages = "L13",
    year = "2017"
}

@article{LIGOScientific:2018mvr,
    author = "Abbott, B. P. and others",
    collaboration = "LIGO Scientific, Virgo",
    title = "{GWTC-1: A Gravitational-Wave Transient Catalog of Compact Binary Mergers Observed by LIGO and Virgo during the First and Second Observing Runs}",
    eprint = "1811.12907",
    archivePrefix = "arXiv",
    primaryClass = "astro-ph.HE",
    reportNumber = "LIGO-P1800307",
    doi = "10.1103/PhysRevX.9.031040",
    journal = "Phys. Rev. X",
    volume = "9",
    number = "3",
    pages = "031040",
    year = "2019"
}

@article{LIGOScientific:2019fpa,
    author = "Abbott, B. P. and others",
    collaboration = "LIGO Scientific, Virgo",
    title = "{Tests of General Relativity with the Binary Black Hole Signals from the LIGO-Virgo Catalog GWTC-1}",
    eprint = "1903.04467",
    archivePrefix = "arXiv",
    primaryClass = "gr-qc",
    reportNumber = "LIGO-P1800316",
    doi = "10.1103/PhysRevD.100.104036",
    journal = "Phys. Rev. D",
    volume = "100",
    number = "10",
    pages = "104036",
    year = "2019"
}

@article{LIGOScientific:2020ibl,
    author = "Abbott, R. and others",
    collaboration = "LIGO Scientific, Virgo",
    title = "{GWTC-2: Compact Binary Coalescences Observed by LIGO and Virgo During the First Half of the Third Observing Run}",
    eprint = "2010.14527",
    archivePrefix = "arXiv",
    primaryClass = "gr-qc",
    reportNumber = "P2000061",
    doi = "10.1103/PhysRevX.11.021053",
    journal = "Phys. Rev. X",
    volume = "11",
    pages = "021053",
    year = "2021"
}

@article{LIGOScientific:2020kqk,
    author = "Abbott, R. and others",
    collaboration = "LIGO Scientific, Virgo",
    title = "{Population Properties of Compact Objects from the Second LIGO-Virgo Gravitational-Wave Transient Catalog}",
    eprint = "2010.14533",
    archivePrefix = "arXiv",
    primaryClass = "astro-ph.HE",
    reportNumber = "LIGO-P2000077",
    doi = "10.3847/2041-8213/abe949",
    journal = "Astrophys. J. Lett.",
    volume = "913",
    number = "1",
    pages = "L7",
    year = "2021"
}

@article{LIGOScientific:2021usb,
    author = "Abbott, R. and others",
    collaboration = "LIGO Scientific, VIRGO",
    title = "{GWTC-2.1: Deep extended catalog of compact binary coalescences observed by LIGO and Virgo during the first half of the third observing run}",
    eprint = "2108.01045",
    archivePrefix = "arXiv",
    primaryClass = "gr-qc",
    reportNumber = "LIGO-P2100063",
    doi = "10.1103/PhysRevD.109.022001",
    journal = "Phys. Rev. D",
    volume = "109",
    number = "2",
    pages = "022001",
    year = "2024"
}

@article{LIGOScientific:2021sio,
    author = "Abbott, R. and others",
    collaboration = "LIGO Scientific, VIRGO, KAGRA",
    title = "{Tests of General Relativity with GWTC-3}",
    journal = "arXiv:2112.06861",
    eprint = "2112.06861",
    archivePrefix = "arXiv",
    primaryClass = "gr-qc",
    reportNumber = "LIGO-P2100275",
    month = "12",
    year = "2021"
}

@article{KAGRA:2021vkt,
    author = "Abbott, R. and others",
    collaboration = "KAGRA, VIRGO, LIGO Scientific",
    title = "{GWTC-3: Compact Binary Coalescences Observed by LIGO and Virgo during the Second Part of the Third Observing Run}",
    eprint = "2111.03606",
    archivePrefix = "arXiv",
    primaryClass = "gr-qc",
    reportNumber = "LIGO-P2000318",
    doi = "10.1103/PhysRevX.13.041039",
    journal = "Phys. Rev. X",
    volume = "13",
    number = "4",
    pages = "041039",
    year = "2023"
}

@article{LIGOScientific:2025slb,
    author = "Abac, A. G. and others",
    collaboration = "LIGO Scientific, VIRGO, KAGRA",
    title = "{GWTC-4.0: Updating the Gravitational-Wave Transient Catalog with Observations from the First Part of the Fourth LIGO-Virgo-KAGRA Observing Run}",
    journal = "arXiv:2508.18082",
    eprint = "2508.18082",
    archivePrefix = "arXiv",
    primaryClass = "gr-qc",
    reportNumber = "LIGO-P2400386",
    month = "8",
    year = "2025"
}

@article{LIGOScientific:2025obp,
    collaboration = "LIGO Scientific, VIRGO, KAGRA",
    title = "{Black Hole Spectroscopy and Tests of General Relativity with GW250114}",
    journal = "arXiv:2509.08099",
    eprint = "2509.08099",
    archivePrefix = "arXiv",
    primaryClass = "gr-qc",
    reportNumber = "LIGO P2500461",
    month = "9",
    year = "2025"
}

@article{Siegel:2025xgb,
    author = "Siegel, Harrison and Khusid, Nicole M. and Isi, Maximiliano and Farr, Will M.",
    title = "{GW231123 ringdown: interpretation as multimodal Kerr signal}",
    journal = "arXiv:2511.02691",
    eprint = "2511.02691",
    archivePrefix = "arXiv",
    primaryClass = "gr-qc",
    reportNumber = "LIGO document P2500629",
    month = "11",
    year = "2025"
}

@article{KAGRA:2021duu,
    author = "Abbott, R. and others",
    collaboration = "KAGRA, VIRGO, LIGO Scientific",
    title = "{Population of Merging Compact Binaries Inferred Using Gravitational Waves through GWTC-3}",
    eprint = "2111.03634",
    archivePrefix = "arXiv",
    primaryClass = "astro-ph.HE",
    reportNumber = "LIGO-P2100239 ; Data release: https://zenodo.org/record/5655785, LIGO-P2100239",
    doi = "10.1103/PhysRevX.13.011048",
    journal = "Phys. Rev. X",
    volume = "13",
    number = "1",
    pages = "011048",
    year = "2023"
}

@article{LIGOScientific:2025pvj,
    author = "Abac, A. G. and others",
    collaboration = "LIGO Scientific, VIRGO, KAGRA",
    title = "{GWTC-4.0: Population Properties of Merging Compact Binaries}",
    eprint = "2508.18083",
    journal = "arXiv:2508.18083",
    archivePrefix = "arXiv",
    primaryClass = "astro-ph.HE",
    reportNumber = "LIGO-P2400004",
    month = "8",
    year = "2025"
}

@article{Mehta:2021fgz,
    author = "Mehta, Ajit Kumar and Buonanno, Alessandra and Gair, Jonathan and Miller, M. Coleman and Farag, Ebraheem and deBoer, R. J. and Wiescher, M. and Timmes, F. X.",
    title = "{Observing Intermediate-mass Black Holes and the Upper Stellar-mass gap with LIGO and Virgo}",
    eprint = "2105.06366",
    archivePrefix = "arXiv",
    primaryClass = "gr-qc",
    doi = "10.3847/1538-4357/ac3130",
    journal = "Astrophys. J.",
    volume = "924",
    number = "1",
    pages = "39",
    year = "2022"
}

@article{Edelman_2021,
doi = {10.3847/2041-8213/abfdb3},
url = {https://doi.org/10.3847/2041-8213/abfdb3},
year = {2021},
month = {may},
publisher = {The American Astronomical Society},
volume = {913},
number = {2},
pages = {L23},
author = {Edelman, Bruce and Doctor, Zoheyr and Farr, Ben},
title = {Poking Holes: Looking for Gaps in LIGO/Virgo’s Black Hole Population},
journal = {The Astrophysical Journal Letters},
}

@article{Zevin:2020gma,
    author = "Zevin, Michael and Spera, Mario and Berry, Christopher P. L. and Kalogera, Vicky",
    title = "{Exploring the Lower Mass Gap and Unequal Mass Regime in Compact Binary Evolution}",
    eprint = "2006.14573",
    archivePrefix = "arXiv",
    primaryClass = "astro-ph.HE",
    doi = "10.3847/2041-8213/aba74e",
    journal = "Astrophys. J. Lett.",
    volume = "899",
    pages = "L1",
    year = "2020"
}

@article{Pierra:2024fbl,
    author = "Pierra, Gr{\'e}goire and Mastrogiovanni, Simone and Perri{\`e}s, St{\'e}phane",
    title = "{The spin magnitude of stellar-mass black holes evolves with the mass}",
    eprint = "2406.01679",
    archivePrefix = "arXiv",
    primaryClass = "gr-qc",
    doi = "10.1051/0004-6361/202452545",
    journal = "Astron. Astrophys.",
    volume = "692",
    pages = "A80",
    year = "2024"
}

@article{Chattopadhyay:2022cnp,
    author = "Chattopadhyay, Debatri and Stevenson, Simon and Broekgaarden, Floor and Antonini, Fabio and Belczynski, Krzysztof",
    title = "{Modelling the formation of the first two neutron star{\textendash}black hole mergers, GW200105 and GW200115: metallicity, chirp masses, and merger remnant spins}",
    eprint = "2203.05850",
    archivePrefix = "arXiv",
    primaryClass = "astro-ph.HE",
    reportNumber = "Volume 513, Issue 4, Pages 5780--5789",
    doi = "10.1093/mnras/stac1283",
    journal = "Mon. Not. Roy. Astron. Soc.",
    volume = "513",
    number = "4",
    pages = "5780--5789",
    year = "2022"
}

@article{Qin:2024ojw,
    author = "Qin, Ying and others",
    title = "{Origin of the black hole spin in lower-mass-gap black hole-neutron star binaries}",
    eprint = "2409.14476",
    archivePrefix = "arXiv",
    primaryClass = "astro-ph.HE",
    doi = "10.1051/0004-6361/202452335",
    journal = "Astron. Astrophys.",
    volume = "691",
    pages = "L19",
    year = "2024"
}

@article{Li:2022gly,
    author = "Li, Guo-Peng",
    title = "{Constraining hierarchical mergers of binary black holes detectable with LIGO-Virgo}",
    eprint = "2208.11894",
    archivePrefix = "arXiv",
    primaryClass = "astro-ph.HE",
    doi = "10.1051/0004-6361/202244257",
    journal = "Astron. Astrophys.",
    volume = "666",
    pages = "A194",
    year = "2022"
}

@article{Tagawa:2021ofj,
    author = "Tagawa, Hiromichi and Haiman, Zolt{\'a}n and Bartos, Imre and Kocsis, Bence and Omukai, Kazuyuki",
    title = "{Signatures of hierarchical mergers in black hole spin and mass distribution}",
    eprint = "2104.09510",
    archivePrefix = "arXiv",
    primaryClass = "astro-ph.HE",
    doi = "10.1093/mnras/stab2315",
    journal = "Mon. Not. Roy. Astron. Soc.",
    volume = "507",
    number = "3",
    pages = "3362--3380",
    year = "2021"
}

@article{Somiya:2011np,
    author = "Somiya, Kentaro",
    editor = "Hannam, Mark and Sutton, Patrick and Hild, Stefan and van den Broeck, Chris",
    collaboration = "KAGRA",
    title = "{Detector configuration of KAGRA: The Japanese cryogenic gravitational-wave detector}",
    eprint = "1111.7185",
    archivePrefix = "arXiv",
    primaryClass = "gr-qc",
    doi = "10.1088/0264-9381/29/12/124007",
    journal = "Class. Quant. Grav.",
    volume = "29",
    pages = "124007",
    year = "2012"
}

@article{Aso:2013eba,
    author = "Aso, Yoichi and Michimura, Yuta and Somiya, Kentaro and Ando, Masaki and Miyakawa, Osamu and Sekiguchi, Takanori and Tatsumi, Daisuke and Yamamoto, Hiroaki",
    collaboration = "KAGRA",
    title = "{Interferometer design of the KAGRA gravitational wave detector}",
    eprint = "1306.6747",
    archivePrefix = "arXiv",
    primaryClass = "gr-qc",
    doi = "10.1103/PhysRevD.88.043007",
    journal = "Phys. Rev. D",
    volume = "88",
    number = "4",
    pages = "043007",
    year = "2013"
}

@article{KAGRA:2018plz,
    author = "Akutsu, T. and others",
    collaboration = "KAGRA",
    title = "{KAGRA: 2.5 Generation Interferometric Gravitational Wave Detector}",
    eprint = "1811.08079",
    archivePrefix = "arXiv",
    primaryClass = "gr-qc",
    reportNumber = "JGW-P1809243",
    doi = "10.1038/s41550-018-0658-y",
    journal = "Nature Astron.",
    volume = "3",
    number = "1",
    pages = "35--40",
    year = "2019"
}

@article{KAGRA:2020tym,
    author = "Akutsu, T. and others",
    collaboration = "KAGRA",
    title = "{Overview of KAGRA: Detector design and construction history}",
    eprint = "2005.05574",
    archivePrefix = "arXiv",
    primaryClass = "physics.ins-det",
    doi = "10.1093/ptep/ptaa125",
    journal = "PTEP",
    volume = "2021",
    number = "5",
    pages = "05A101",
    year = "2021"
}

@misc{LigoIndia,
    author = "Iyer, Bala and Souradeep, Tarun and Unnikrishnan, CS and Dhurandhar, Sanjeev and Raja, Sandhi and Sengupta, Anand",
    title = "{LIGO-India Tech. rep. }",
    year = 2011}

@article{Unnikrishnan:2013qwa,
    author = "Unnikrishnan, C. S.",
    title = "{IndIGO and LIGO-India: Scope and plans for gravitational wave research and precision metrology in India}",
    eprint = "1510.06059",
    archivePrefix = "arXiv",
    primaryClass = "physics.ins-det",
    doi = "10.1142/S0218271813410101",
    journal = "Int. J. Mod. Phys. D",
    volume = "22",
    pages = "1341010",
    year = "2013"
}

@article{Ajith:2024inj,
    author = "Ajith, P. and others",
    title = "{Gravitational physics in the context of Indian astronomy: A vision document}",
    eprint = "2501.04333",
    archivePrefix = "arXiv",
    primaryClass = "astro-ph.IM",
    doi = "10.1007/s12036-024-10031-x",
    journal = "J. Astrophys. Astron.",
    volume = "46",
    number = "1",
    pages = "6",
    year = "2025"
}

@article{Punturo:2010zz,
    author = "Punturo, M. and others",
    editor = "Ricci, Fulvio",
    title = "{The Einstein Telescope: A third-generation gravitational wave observatory}",
    doi = "10.1088/0264-9381/27/19/194002",
    journal = "Class. Quant. Grav.",
    volume = "27",
    pages = "194002",
    year = "2010"
}

@article{Hild:2010id,
    author = "Hild, S. and others",
    title = "{Sensitivity Studies for Third-Generation Gravitational Wave Observatories}",
    eprint = "1012.0908",
    archivePrefix = "arXiv",
    primaryClass = "gr-qc",
    doi = "10.1088/0264-9381/28/9/094013",
    journal = "Class. Quant. Grav.",
    volume = "28",
    pages = "094013",
    year = "2011"
}

@article{Reitze:2019iox,
    author = "Reitze, David and others",
    title = "{Cosmic Explorer: The U.S. Contribution to Gravitational-Wave Astronomy beyond LIGO}",
    eprint = "1907.04833",
    archivePrefix = "arXiv",
    primaryClass = "astro-ph.IM",
    reportNumber = "LIGO-P1900316",
    journal = "Bull. Am. Astron. Soc.",
    volume = "51",
    number = "7",
    pages = "035",
    year = "2019"
}

@article{LIGOScientific:2016wof,
    author = "Abbott, Benjamin P and others",
    collaboration = "LIGO Scientific",
    title = "{Exploring the Sensitivity of Next Generation Gravitational Wave Detectors}",
    eprint = "1607.08697",
    archivePrefix = "arXiv",
    primaryClass = "astro-ph.IM",
    reportNumber = "LIGO-P1600143",
    doi = "10.1088/1361-6382/aa51f4",
    journal = "Class. Quant. Grav.",
    volume = "34",
    number = "4",
    pages = "044001",
    year = "2017"
}

@article{Regimbau:2016ike,
    author = "Regimbau, T. and Evans, M. and Christensen, N. and Katsavounidis, E. and Sathyaprakash, B. and Vitale, S.",
    title = "{Digging deeper: Observing primordial gravitational waves below the binary black hole produced stochastic background}",
    eprint = "1611.08943",
    archivePrefix = "arXiv",
    primaryClass = "astro-ph.CO",
    doi = "10.1103/PhysRevLett.118.151105",
    journal = "Phys. Rev. Lett.",
    volume = "118",
    number = "15",
    pages = "151105",
    year = "2017"
}

@article{LISA:2017,
    author = {Amaro-Seoane, Pau and others},
    title = "{Laser Interferometer Space Antenna}",
    journal = "arXiv:1702.00786",
    eprint = "1702.00786",
    archivePrefix = "arXiv",
    primaryClass = "astro-ph.IM",
    year = "2017"
}

@ARTICLE{1936Sci....84..506E,
       author = "Einstein, Albert",
        title = "{Lens-Like Action of a Star by the Deviation of Light in the Gravitational Field}",
      journal = {Science},
         year = 1936,
        month = dec,
       volume = {84},
       number = {2188},
        pages = {506-507},
          doi = {10.1126/science.84.2188.506},
       adsurl = {https://ui.adsabs.harvard.edu/abs/1936Sci....84..506E}
}

@article{PhysRev.51.290,
  title = {Nebulae as Gravitational Lenses},
  author = {Zwicky, F.},
  journal = {Phys. Rev.},
  volume = {51},
  issue = {4},
  pages = {290--290},
  numpages = {0},
  year = {1937},
  month = {Feb},
  publisher = {American Physical Society},
  doi = {10.1103/PhysRev.51.290},
  url = {https://link.aps.org/doi/10.1103/PhysRev.51.290}
}

@BOOK{1992grle.book.....S,
       author = "Schneider, Peter and Ehlers, J{\"u}rgen and Falco, Emilio E.",
        title = "{Gravitational Lenses}",
         year = 1992,
          doi = {10.1007/978-3-662-03758-4},
       adsurl = {https://ui.adsabs.harvard.edu/abs/1992grle.book.....S}
}

@ARTICLE{1971NCimB...6..225L,
       author = "Lawrence, J. K.",
        title = "{Focusing of gravitational radiation by interior gravitational fields.}",
      journal = {Nuovo Cimento B Serie},
         year = 1971,
        month = jan,
       volume = {6B},
        pages = {225-235},
          doi = {10.1007/BF02735388},
       adsurl = {https://ui.adsabs.harvard.edu/abs/1971NCimB...6..225L}
}

@article{Ohanian:1974ys,
    author = "Ohanian, H. C.",
    title = "{On the focusing of gravitational radiation}",
    doi = "10.1007/BF01810927",
    journal = "Int. J. Theor. Phys.",
    volume = "9",
    pages = "425--437",
    year = "1974"
}

@article{Bernardeau:1999mh,
    author = "Bernardeau, F.",
    title = "{Gravitational lenses}",
    journal = "arXiv:astro-ph/9901117",
    eprint = "astro-ph/9901117",
    archivePrefix = "arXiv",
    month = "1",
    year = "1999"
}

@article{Dai:2020tpj,
    author = "Dai, Liang and Zackay, Barak and Venumadhav, Tejaswi and Roulet, Javier and Zaldarriaga, Matias",
    title = "{Search for Lensed Gravitational Waves Including Morse Phase Information: An Intriguing Candidate in O2}",
    journal = "arXiv:2007.12709",
    eprint = "2007.12709",
    archivePrefix = "arXiv",
    primaryClass = "astro-ph.HE",
    month = "7",
    year = "2020"
}

@article{Ezquiaga:2020gdt,
    author = "Ezquiaga, Jose Mar\'\i{}a and Holz, Daniel E. and Hu, Wayne and Lagos, Macarena and Wald, Robert M.",
    title = "{Phase effects from strong gravitational lensing of gravitational waves}",
    eprint = "2008.12814",
    archivePrefix = "arXiv",
    primaryClass = "gr-qc",
    doi = "10.1103/PhysRevD.103.064047",
    journal = "Phys. Rev. D",
    volume = "103",
    number = "6",
    pages = "064047",
    year = "2021"
}

@article{Vijaykumar:2022dlp,
    author = "Vijaykumar, Aditya and Mehta, Ajit Kumar and Ganguly, Apratim",
    title = "{Detection and parameter estimation challenges of type-II lensed binary black hole signals}",
    eprint = "2202.06334",
    archivePrefix = "arXiv",
    primaryClass = "gr-qc",
    doi = "10.1103/PhysRevD.108.043036",
    journal = "Phys. Rev. D",
    volume = "108",
    number = "4",
    pages = "043036",
    year = "2023"
}

@article{Takahashi:2016jom,
    author = "Takahashi, Ryuichi",
    title = "{Arrival time differences between gravitational waves and electromagnetic signals due to gravitational lensing}",
    eprint = "1606.00458",
    archivePrefix = "arXiv",
    primaryClass = "astro-ph.CO",
    doi = "10.3847/1538-4357/835/1/103",
    journal = "Astrophys. J.",
    volume = "835",
    number = "1",
    pages = "103",
    year = "2017"
}

@article{Dai:2017huk,
    author = "Dai, Liang and Venumadhav, Tejaswi",
    title = "{On the waveforms of gravitationally lensed gravitational waves}",
    journal = "arXiv:1702,04724",
    eprint = "1702.04724",
    archivePrefix = "arXiv",
    primaryClass = "gr-qc",
    month = "2",
    year = "2017"
}

@article{Haris:2018vmn,
    author = "Haris, K. and Mehta, Ajit Kumar and Kumar, Sumit and Venumadhav, Tejaswi and Ajith, Parameswaran",
    title = "{Identifying strongly lensed gravitational wave signals from binary black hole mergers}",
    journal = "arXiv:1807.07062",
    eprint = "1807.07062",
    archivePrefix = "arXiv",
    primaryClass = "gr-qc",
    reportNumber = "LIGO- P1800155",
    month = "7",
    year = "2018"
}

@article{Li:2018prc,
    author = "Li, Shun-Sheng and Mao, Shude and Zhao, Yuetong and Lu, Youjun",
    title = "{Gravitational lensing of gravitational waves: A statistical perspective}",
    eprint = "1802.05089",
    archivePrefix = "arXiv",
    primaryClass = "astro-ph.CO",
    doi = "10.1093/mnras/sty411",
    journal = "Mon. Not. Roy. Astron. Soc.",
    volume = "476",
    number = "2",
    pages = "2220--2229",
    year = "2018"
}

@article{Smith:2017jdz,
    author = "Smith, G. P. and others",
    editor = "Gonz\'alez, Gabriela and Hynes, Robert",
    title = "{Strong-lensing of Gravitational Waves by Galaxy Clusters}",
    eprint = "1803.07851",
    archivePrefix = "arXiv",
    primaryClass = "astro-ph.CO",
    doi = "10.1017/S1743921318003757",
    journal = "IAU Symp.",
    volume = "338",
    pages = "98--102",
    year = "2017"
}

@article{Broadhurst:2018saj,
    author = "Broadhurst, Tom and Diego, Jose M. and Smoot, George",
    title = "{Reinterpreting Low Frequency LIGO/Virgo Events as Magnified Stellar-Mass Black Holes at Cosmological Distances}",
    journal = "arXiv:1802.05273",
    eprint = "1802.05273",
    archivePrefix = "arXiv",
    primaryClass = "astro-ph.CO",
    month = "2",
    year = "2018"
}

@article{Broadhurst:2019ijv,
    author = "Broadhurst, Tom and Diego, Jose M. and Smoot, George F.",
    title = "{Twin LIGO/Virgo Detections of a Viable Gravitationally-Lensed Black Hole Merger}",
    journal = "arXiv:1901.03190",
    eprint = "1901.03190",
    archivePrefix = "arXiv",
    primaryClass = "astro-ph.CO",
    month = "1",
    year = "2019"
}

@article{Broadhurst:2020cvm,
    author = "Broadhurst, Tom and Diego, Jose M. and Smoot, George F.",
    title = "{Interpreting LIGO/Virgo ''Mass-Gap'' events as lensed Neutron Star-Black Hole binaries}",
    journal = "arXiv:2006.13219",
    eprint = "2006.13219",
    archivePrefix = "arXiv",
    primaryClass = "astro-ph.CO",
    month = "6",
    year = "2020"
}

@article{Caliskan:2022wbh,
    author = "\c{C}al\i{}\c{s}kan, Mesut and Ezquiaga, Jose Mar\'\i{}a and Hannuksela, Otto A. and Holz, Daniel E.",
    title = "{Lensing or luck? False alarm probabilities for gravitational lensing of gravitational waves}",
    eprint = "2201.04619",
    archivePrefix = "arXiv",
    primaryClass = "astro-ph.CO",
    doi = "10.1103/PhysRevD.107.063023",
    journal = "Phys. Rev. D",
    volume = "107",
    number = "6",
    pages = "063023",
    year = "2023"
}

@article{Barsode:2024zwv,
    author = "Barsode, Ankur and Goyal, Srashti and Ajith, Parameswaran",
    title = "{Fast and Efficient Bayesian Method to Search for Strongly Lensed Gravitational Waves}",
    eprint = "2412.01278",
    archivePrefix = "arXiv",
    primaryClass = "gr-qc",
    doi = "10.3847/1538-4357/adae10",
    journal = "Astrophys. J.",
    volume = "980",
    number = "2",
    pages = "258",
    year = "2025"
}

@ARTICLE{1986ApJ...307...30D,
       author = "Deguchi, S. and Watson, W.~D.",
       title = "{Diffraction in Gravitational Lensing for Compact Objects of Low Mass}",
      journal = {\apj},
         year = 1986,
        month = aug,
       volume = {307},
        pages = {30},
          doi = {10.1086/164389},
       adsurl = {https://ui.adsabs.harvard.edu/abs/1986ApJ...307...30D}
}

@article{Nakamura:1997sw,
    author = "Nakamura, Takahiro T.",
    title = "{Gravitational lensing of gravitational waves from inspiraling binaries by a point mass lens}",
    reportNumber = "UTAP-272-97, YITP-97-61",
    doi = "10.1103/PhysRevLett.80.1138",
    journal = "Phys. Rev. Lett.",
    volume = "80",
    pages = "1138--1141",
    year = "1998"
}

@article{Nakamura:1999uwi,
    author = "Nakamura, Takahiro T. and Deguchi, Shuji",
    title = "{Wave Optics in Gravitational Lensing}",
    doi = "10.1143/ptps.133.137",
    journal = "Prog. Theor. Phys. Suppl.",
    volume = "133",
    pages = "137--153",
    year = "1999"
}

@ARTICLE{2003ApJ...595.1039T,
       author = "Takahashi, Ryuichi and Nakamura, Takahiro T.",
        title = "{Wave Effects in the Gravitational Lensing of Gravitational Waves from Chirping Binaries}",
      journal = {\apj},
         year = 2003,
        month = oct,
       volume = {595},
       number = {2},
        pages = {1039-1051},
          doi = {10.1086/377430},
archivePrefix = {arXiv},
       eprint = {astro-ph/0305055},
 primaryClass = {astro-ph},
       adsurl = {https://ui.adsabs.harvard.edu/abs/2003ApJ...595.1039T}
}

@article{Jung:2017flg,
    author = "Jung, Sunghoon and Shin, Chang Sub",
    title = "{Gravitational-Wave Fringes at LIGO: Detecting Compact Dark Matter by Gravitational Lensing}",
    eprint = "1712.01396",
    archivePrefix = "arXiv",
    primaryClass = "astro-ph.CO",
    doi = "10.1103/PhysRevLett.122.041103",
    journal = "Phys. Rev. Lett.",
    volume = "122",
    number = "4",
    pages = "041103",
    year = "2019"
}

@article{Cremonese:2021ahz,
    author = "Cremonese, Paolo and Mota, David Fonseca and Salzano, Vincenzo",
    title = "{Characteristic Features of Gravitational Wave Lensing as Probe of Lens Mass Model}",
    eprint = "2111.01163",
    archivePrefix = "arXiv",
    primaryClass = "astro-ph.CO",
    doi = "10.1002/andp.202300040",
    journal = "Annalen Phys.",
    volume = "535",
    number = "6",
    pages = "2300040",
    year = "2023"
}

@article{Cremonese:2021puh,
    author = "Cremonese, Paolo and Ezquiaga, Jose Mar\'\i{}a and Salzano, Vincenzo",
    title = "{Breaking the mass-sheet degeneracy with gravitational wave interference in lensed events}",
    eprint = "2104.07055",
    archivePrefix = "arXiv",
    primaryClass = "astro-ph.CO",
    doi = "10.1103/PhysRevD.104.023503",
    journal = "Phys. Rev. D",
    volume = "104",
    number = "2",
    pages = "023503",
    year = "2021"
}

@article{Mishra:2021xzz,
    author = "Mishra, Anuj and Meena, Ashish Kumar and More, Anupreeta and Bose, Sukanta and Bagla, Jasjeet Singh",
    title = "{Gravitational lensing of gravitational waves: effect of microlens population in lensing galaxies}",
    eprint = "2102.03946",
    archivePrefix = "arXiv",
    primaryClass = "astro-ph.CO",
    doi = "10.1093/mnras/stab2875",
    journal = "Mon. Not. Roy. Astron. Soc.",
    volume = "508",
    number = "4",
    pages = "4869--4886",
    year = "2021"
}

@article{Shan:2020esq,
    author = "Shan, Xikai and Wei, Chengliang and Hu, Bin",
    title = "{Lensing magnification: gravitational waves from coalescing stellar-mass binary black holes}",
    eprint = "2012.08381",
    archivePrefix = "arXiv",
    primaryClass = "astro-ph.CO",
    doi = "10.1093/mnras/stab2567",
    journal = "Mon. Not. Roy. Astron. Soc.",
    volume = "508",
    number = "1",
    pages = "1253--1261",
    year = "2021"
}

@article{Shan:2023ngi,
    author = "Shan, Xikai and Chen, Xuechun and Hu, Bin and Cai, Rong-Gen",
    title = "{Microlensing sheds light on the detection of strong lensing gravitational waves}",
    journal = "arXiv:2301.06117",
    eprint = "2301.06117",
    archivePrefix = "arXiv",
    primaryClass = "astro-ph.IM",
    month = "1",
    year = "2023"
}

@article{Meena:2022unp,
    author = "Meena, Ashish Kumar and Mishra, Anuj and More, Anupreeta and Bose, Sukanta and Bagla, Jasjeet Singh",
    title = "{Gravitational lensing of gravitational waves: Probability of microlensing in galaxy-scale lens population}",
    eprint = "2205.05409",
    archivePrefix = "arXiv",
    primaryClass = "astro-ph.GA",
    doi = "10.1093/mnras/stac2721",
    journal = "Mon. Not. Roy. Astron. Soc.",
    volume = "517",
    number = "1",
    pages = "872--884",
    year = "2022"
}

@article{Bondarescu:2022srx,
    author = "Bondarescu, Ruxandra and Ubach, Helena and Bulashenko, Oleg and Lundgren, Andrew P.",
    title = "{Compact binaries through a lens: Silent versus detectable microlensing for the LIGO-Virgo-KAGRA gravitational wave observatories}",
    eprint = "2211.13604",
    archivePrefix = "arXiv",
    primaryClass = "gr-qc",
    doi = "10.1103/PhysRevD.108.084033",
    journal = "Phys. Rev. D",
    volume = "108",
    number = "8",
    pages = "084033",
    year = "2023"
}

@article{Mishra:2023ddt,
    author = "Mishra, Anuj and Meena, Ashish Kumar and More, Anupreeta and Bose, Sukanta",
    title = "{Exploring the impact of microlensing on gravitational wave signals: Biases, population characteristics, and prospects for detection}",
    eprint = "2306.11479",
    archivePrefix = "arXiv",
    primaryClass = "astro-ph.CO",
    doi = "10.1093/mnras/stae836",
    journal = "Mon. Not. Roy. Astron. Soc.",
    volume = "531",
    number = "1",
    pages = "764--787",
    year = "2024"
}

@article{Mishra:2023vzo,
    author = "Mishra, Anuj and Krishnendu, N. V. and Ganguly, Apratim",
    title = "{Unveiling microlensing biases in testing general relativity with gravitational waves}",
    eprint = "2311.08446",
    archivePrefix = "arXiv",
    primaryClass = "gr-qc",
    doi = "10.1103/PhysRevD.110.084009",
    journal = "Phys. Rev. D",
    volume = "110",
    number = "8",
    pages = "084009",
    year = "2024"
}

@article{Rao:2025poe,
    author = "Rao, Nishkal and Mishra, Anuj and Ganguly, Apratim and More, Anupreeta",
    title = "{Comprehensive analysis of time-domain overlapping gravitational wave transients: A Lensing Study}",
    journal = "arXiv:2510.17787",
    eprint = "2510.17787",
    archivePrefix = "arXiv",
    primaryClass = "gr-qc",
    reportNumber = "LIGO-P2500640",
    month = "10",
    year = "2025"
}

@article{Caliskan:2022hbu,
    author = "{\c{C}}al{\i}{\c{s}}kan, Mesut and Ji, Lingyuan and Cotesta, Roberto and Berti, Emanuele and Kamionkowski, Marc and Marsat, Sylvain",
    title = "{Observability of lensing of gravitational waves from massive black hole binaries with LISA}",
    eprint = "2206.02803",
    archivePrefix = "arXiv",
    primaryClass = "astro-ph.CO",
    doi = "10.1103/PhysRevD.107.043029",
    journal = "Phys. Rev. D",
    volume = "107",
    number = "4",
    pages = "043029",
    year = "2023"
}

@article{Pijnenburg:2024btj,
    author = "Pijnenburg, Martin and Cusin, Giulia and Pitrou, Cyril and Uzan, Jean-Philippe",
    title = "{Wave optics lensing of gravitational waves: Theory and phenomenology of triple systems in the LISA band}",
    eprint = "2404.07186",
    archivePrefix = "arXiv",
    primaryClass = "gr-qc",
    doi = "10.1103/PhysRevD.110.044054",
    journal = "Phys. Rev. D",
    volume = "110",
    number = "4",
    pages = "044054",
    year = "2024"
}

@article{Virbhadra:1999nm,
    author = "Virbhadra, K. S. and Ellis, George F. R.",
    title = "{Schwarzschild black hole lensing}",
    eprint = "astro-ph/9904193",
    archivePrefix = "arXiv",
    doi = "10.1103/PhysRevD.62.084003",
    journal = "Phys. Rev. D",
    volume = "62",
    pages = "084003",
    year = "2000"
}

@article{Bozza:2002zj,
    author = "Bozza, V.",
    title = "{Gravitational lensing in the strong field limit}",
    eprint = "gr-qc/0208075",
    archivePrefix = "arXiv",
    doi = "10.1103/PhysRevD.66.103001",
    journal = "Phys. Rev. D",
    volume = "66",
    pages = "103001",
    year = "2002"
}

@article{Kumar_2022,
doi = {10.3847/1538-4357/ac912c},
url = {https://doi.org/10.3847/1538-4357/ac912c},
year = {2022},
month = {oct},
publisher = {The American Astronomical Society},
volume = {938},
number = {2},
pages = {104},
author = {Kumar, Jitendra and Ul Islam, Shafqat and Ghosh, Sushant G.},
title = {Testing Strong Gravitational Lensing Effects of Supermassive Compact Objects with Regular Spacetimes},
journal = {The Astrophysical Journal},
}

@article{EventHorizonTelescope:2019dse,
    author = "Akiyama, Kazunori and others",
    collaboration = "Event Horizon Telescope",
    title = "{First M87 Event Horizon Telescope Results. I. The Shadow of the Supermassive Black Hole}",
    eprint = "1906.11238",
    archivePrefix = "arXiv",
    primaryClass = "astro-ph.GA",
    doi = "10.3847/2041-8213/ab0ec7",
    journal = "Astrophys. J. Lett.",
    volume = "875",
    pages = "L1",
    year = "2019"
}

@article{EventHorizonTelescope:2022wkp,
    author = "Akiyama, Kazunori and others",
    collaboration = "Event Horizon Telescope",
    title = "{First Sagittarius A* Event Horizon Telescope Results. I. The Shadow of the Supermassive Black Hole in the Center of the Milky Way}",
    eprint = "2311.08680",
    archivePrefix = "arXiv",
    primaryClass = "astro-ph.HE",
    doi = "10.3847/2041-8213/ac6674",
    journal = "Astrophys. J. Lett.",
    volume = "930",
    number = "2",
    pages = "L12",
    year = "2022"
}

@article{Baraldo:1999ny,
    author = "Baraldo, Christian and Hosoya, Akio and Nakamura, Takahiro T.",
    title = "{Gravitationally induced interference of gravitational waves by a rotating massive object}",
    doi = "10.1103/PhysRevD.59.083001",
    journal = "Phys. Rev. D",
    volume = "59",
    pages = "083001",
    year = "1999"
}

@article{Biesiada:2021pzo,
    author = "Biesiada, Marek and Harikumar, Sreekanth",
    title = "{Gravitational Lensing of Continuous Gravitational Waves}",
    eprint = "2111.05963",
    archivePrefix = "arXiv",
    primaryClass = "gr-qc",
    doi = "10.3390/universe7120502",
    journal = "Universe",
    volume = "7",
    number = "12",
    pages = "502",
    year = "2021"
}

@article{Dalang:2021qhu,
    author = "Dalang, Charles and Cusin, Giulia and Lagos, Macarena",
    title = "{Polarization distortions of lensed gravitational waves}",
    eprint = "2104.10119",
    archivePrefix = "arXiv",
    primaryClass = "gr-qc",
    doi = "10.1103/PhysRevD.105.024005",
    journal = "Phys. Rev. D",
    volume = "105",
    number = "2",
    pages = "024005",
    year = "2022"
}

@article{Tambalo:2022plm,
    author = "Tambalo, Giovanni and Zumalac{\'a}rregui, Miguel and Dai, Liang and Cheung, Mark Ho-Yeuk",
    title = "{Lensing of gravitational waves: Efficient wave-optics methods and validation with symmetric lenses}",
    eprint = "2210.05658",
    archivePrefix = "arXiv",
    primaryClass = "gr-qc",
    doi = "10.1103/PhysRevD.108.043527",
    journal = "Phys. Rev. D",
    volume = "108",
    number = "4",
    pages = "043527",
    year = "2023"
}

@article{Harikumar:2023gzh,
    author = {Harikumar, Sreekanth and J{\"a}rv, Laur and Saal, Margus and Wojnar, Aneta and Biesiada, Marek},
    title = "{Propagation and lensing of gravitational waves in Palatini f(R{\textasciicircum}) gravity}",
    eprint = "2312.09908",
    archivePrefix = "arXiv",
    primaryClass = "gr-qc",
    doi = "10.1103/PhysRevD.109.124014",
    journal = "Phys. Rev. D",
    volume = "109",
    number = "12",
    pages = "124014",
    year = "2024"
}

@article{Leung:2023lmq,
    author = "Leung, Calvin and Jow, Dylan and Saha, Prasenjit and Dai, Liang and Oguri, Masamune and Koopmans, L{\'e}on V. E.",
    title = "{Wave Optics, Interference, and Decoherence in Strong Gravitational Lensing}",
    eprint = "2304.01202",
    archivePrefix = "arXiv",
    primaryClass = "astro-ph.HE",
    doi = "10.1007/s11214-025-01157-7",
    journal = "Space Sci. Rev.",
    volume = "221",
    number = "2",
    pages = "29",
    year = "2025"
}

@article{Deka:2025vzx,
    author = "Deka, Uddeepta and Prabhu, Gopalkrishna and Shaikh, Md Arif and Kapadia, Shasvath J. and Varma, Vijay and Field, Scott E.",
    title = "{Surrogate modeling of gravitational waves microlensed by spherically symmetric potentials}",
    eprint = "2501.02974",
    archivePrefix = "arXiv",
    primaryClass = "gr-qc",
    doi = "10.1103/PhysRevD.111.104042",
    journal = "Phys. Rev. D",
    volume = "111",
    number = "10",
    pages = "104042",
    year = "2025"
}

@article{Villarrubia-Rojo:2024xcj,
    author = "Villarrubia-Rojo, Hector and Savastano, Stefano and Zumalac{\'a}rregui, Miguel and Choi, Lyla and Goyal, Srashti and Dai, Liang and Tambalo, Giovanni",
    title = "{Gravitational lensing of waves: Novel methods for wave-optics phenomena}",
    eprint = "2409.04606",
    archivePrefix = "arXiv",
    primaryClass = "gr-qc",
    doi = "10.1103/PhysRevD.111.103539",
    journal = "Phys. Rev. D",
    volume = "111",
    number = "10",
    pages = "103539",
    year = "2025"
}

@article{Shan:2022xfx,
    author = "Shan, Xikai and Li, Guoliang and Chen, Xuechun and Zheng, Wenwen and Zhao, Wen",
    title = "{Wave effect of gravitational waves intersected with a microlens field: A new algorithm and supplementary study}",
    eprint = "2208.13566",
    archivePrefix = "arXiv",
    primaryClass = "astro-ph.CO",
    doi = "10.1007/s11433-022-1985-3",
    journal = "Sci. China Phys. Mech. Astron.",
    volume = "66",
    number = "3",
    pages = "239511",
    year = "2023"
}

@article{Liu:2023ikc,
    author = "Liu, Anna and Wong, Isaac C. F. and Leong, Samson H. W. and More, Anupreeta and Hannuksela, Otto A. and Li, Tjonnie G. F.",
    title = "{Exploring the hidden Universe: a novel phenomenological approach for recovering arbitrary gravitational-wave millilensing configurations}",
    eprint = "2302.09870",
    archivePrefix = "arXiv",
    primaryClass = "gr-qc",
    doi = "10.1093/mnras/stad1302",
    journal = "Mon. Not. Roy. Astron. Soc.",
    volume = "525",
    number = "3",
    pages = "4149--4160",
    year = "2023"
}

@article{Bardeen1972RotatingBH,
  title={Rotating Black Holes: Locally Nonrotating Frames, Energy Extraction, and Scalar Synchrotron Radiation},
  author={James M. Bardeen and William H. Press and Saul A. Teukolsky},
  journal={The Astrophysical Journal},
  year={1972},
  volume={178},
  pages={347-370},
  url={https://api.semanticscholar.org/CorpusID:122183015}
}

@book{Misner:1973prb,
    author = "Misner, Charles W. and Thorne, K. S. and Wheeler, J. A.",
    title = "{Gravitation}",
    isbn = "978-0-7167-0344-0, 978-0-691-17779-3",
    publisher = "W. H. Freeman",
    address = "San Francisco",
    year = "1973"
}

@article{Kerr:1963ud,
    author = "Kerr, Roy P.",
    title = "{Gravitational field of a spinning mass as an example of algebraically special metrics}",
    doi = "10.1103/PhysRevLett.11.237",
    journal = "Phys. Rev. Lett.",
    volume = "11",
    pages = "237--238",
    year = "1963"
}

@article{Farooqui:2013rga,
    author = "Farooqui, Anusar and Kamran, Niky and Panangaden, Prakash",
    title = "{An Exact Expression for Photon Polarization in Kerr Geometry}",
    eprint = "1306.6292",
    archivePrefix = "arXiv",
    primaryClass = "math-ph",
    doi = "10.4310/ATMP.2014.v18.n3.a3",
    journal = "Adv. Theor. Math. Phys.",
    volume = "18",
    number = "3",
    pages = "659--686",
    year = "2014"
}

@article{Gelles:2021kti,
    author = "Gelles, Zachary and Himwich, Elizabeth and Palumbo, Daniel C. M. and Johnson, Michael D.",
    title = "{Polarized image of equatorial emission in the Kerr geometry}",
    eprint = "2105.09440",
    archivePrefix = "arXiv",
    primaryClass = "gr-qc",
    doi = "10.1103/PhysRevD.104.044060",
    journal = "Phys. Rev. D",
    volume = "104",
    number = "4",
    pages = "044060",
    year = "2021"
}

@article{Frolov:2025bva,
    author = "Frolov, Valeri P. and Koek, Alex",
    title = "{Spinoptics in the Kerr spacetime: Polarized wave scattering}",
    eprint = "2503.19208",
    archivePrefix = "arXiv",
    primaryClass = "gr-qc",
    doi = "10.1103/trh5-4sgq",
    journal = "Phys. Rev. D",
    volume = "111",
    number = "10",
    pages = "104081",
    year = "2025"
}

@article{Bozza:2005tg,
    author = "Bozza, V. and De Luca, F. and Scarpetta, G. and Sereno, M.",
    title = "{Analytic Kerr black hole lensing for equatorial observers in the strong deflection limit}",
    eprint = "gr-qc/0507137",
    archivePrefix = "arXiv",
    doi = "10.1103/PhysRevD.72.083003",
    journal = "Phys. Rev. D",
    volume = "72",
    pages = "083003",
    year = "2005"
}

@article{Sereno:2007gd,
    author = "Sereno, M. and De Luca, F.",
    title = "{Primary caustics and critical points behind a Kerr black hole}",
    eprint = "0710.5923",
    archivePrefix = "arXiv",
    primaryClass = "astro-ph",
    doi = "10.1103/PhysRevD.78.023008",
    journal = "Phys. Rev. D",
    volume = "78",
    pages = "023008",
    year = "2008"
}

@article{Sereno:2006ss,
    author = "Sereno, Mauro and De Luca, Fabiana",
    title = "{Analytical Kerr black hole lensing in the weak deflection limit}",
    eprint = "astro-ph/0609435",
    archivePrefix = "arXiv",
    doi = "10.1103/PhysRevD.74.123009",
    journal = "Phys. Rev. D",
    volume = "74",
    pages = "123009",
    year = "2006"
}

@article{Teukolsky:1972my,
    author = "Teukolsky, S. A.",
    title = "{Rotating black holes - separable wave equations for gravitational and electromagnetic perturbations}",
    reportNumber = "OAP-291",
    doi = "10.1103/PhysRevLett.29.1114",
    journal = "Phys. Rev. Lett.",
    volume = "29",
    pages = "1114--1118",
    year = "1972"
}

@article{Teukolsky:1974yv,
    author = "Teukolsky, S. A. and Press, W. H.",
    title = "{Perturbations of a rotating black hole. III - Interaction of the hole with gravitational and electromagnetic radiation}",
    doi = "10.1086/153180",
    journal = "Astrophys. J.",
    volume = "193",
    pages = "443--461",
    year = "1974"
}

@book{Futterman:1988ni,
    author = "Futterman, J. A. H. and Handler, F. A. and Matzner, R. A.",
    title = "{SCATTERING FROM BLACK HOLES}",
    doi = "10.1017/CBO9780511735615",
    isbn = "978-1-139-24539-5, 978-0-521-11210-9",
    publisher = "Cambridge University Press",
    series = "Cambridge Monographs on Mathematical Physics",
    month = "5",
    year = "2012"
}

@article{Glampedakis:2001cx,
    author = "Glampedakis, Kostas and Andersson, Nils",
    title = "{Scattering of scalar waves by rotating black holes}",
    eprint = "gr-qc/0102100",
    archivePrefix = "arXiv",
    doi = "10.1088/0264-9381/18/10/309",
    journal = "Class. Quant. Grav.",
    volume = "18",
    pages = "1939--1966",
    year = "2001"
}

@article{Dolan:2007ut,
    author = "Dolan, Sam R.",
    title = "{Scattering of long-wavelength gravitational waves}",
    eprint = "0710.4252",
    archivePrefix = "arXiv",
    primaryClass = "gr-qc",
    doi = "10.1103/PhysRevD.77.044004",
    journal = "Phys. Rev. D",
    volume = "77",
    pages = "044004",
    year = "2008"
}

@article{Kubota:2024zkv,
    author = "Kubota, Kei-ichiro and Arai, Shun and Motohashi, Hayato and Mukohyama, Shinji",
    title = "{Spin wave optics for gravitational waves lensed by a Kerr black hole}",
    eprint = "2408.03289",
    archivePrefix = "arXiv",
    primaryClass = "gr-qc",
    reportNumber = "YITP-24-89, IPMU24-0031",
    doi = "10.1103/PhysRevD.110.124011",
    journal = "Phys. Rev. D",
    volume = "110",
    number = "12",
    pages = "124011",
    year = "2024"
}

@article{Leite:2017hkm,
    author = "Leite, Luiz C. S. and Benone, Carolina L. and Crispino, Lu{\'\i}s C. B.",
    title = "{Scalar absorption by charged rotating black holes}",
    eprint = "1708.03370",
    archivePrefix = "arXiv",
    primaryClass = "gr-qc",
    doi = "10.1103/PhysRevD.96.044043",
    journal = "Phys. Rev. D",
    volume = "96",
    number = "4",
    pages = "044043",
    year = "2017"
}

@ARTICLE{2012MNRAS.425..460M,
       author = {{McKernan}, B. and {Ford}, K.~E.~S. and {Lyra}, W. and {Perets}, H.~B.},
        title = "{Intermediate mass black holes in AGN discs - I. Production and growth}",
      journal = {MNRAS},
         year = 2012,
        month = sep,
       volume = {425},
       number = {1},
        pages = {460-469},
          doi = {10.1111/j.1365-2966.2012.21486.x},
archivePrefix = {arXiv},
       eprint = {1206.2309},
 primaryClass = {astro-ph.GA},
       adsurl = {https://ui.adsabs.harvard.edu/abs/2012MNRAS.425..460M}
}

@article{Tagawa:2019osr,
    author = "Tagawa, Hiromichi and Haiman, Zoltan and Kocsis, Bence",
    title = "{Formation and Evolution of Compact Object Binaries in AGN Disks}",
    eprint = "1912.08218",
    archivePrefix = "arXiv",
    primaryClass = "astro-ph.GA",
    doi = "10.3847/1538-4357/ab9b8c",
    journal = "Astrophys. J.",
    volume = "898",
    number = "1",
    pages = "25",
    year = "2020"
}

@article{Mckernan:2017ssq,
    author = "Mckernan, B. and others",
    title = "{Constraining Stellar-mass Black Hole Mergers in AGN Disks Detectable with LIGO}",
    eprint = "1702.07818",
    archivePrefix = "arXiv",
    primaryClass = "astro-ph.HE",
    doi = "10.3847/1538-4357/aadae5",
    journal = "Astrophys. J.",
    volume = "866",
    number = "1",
    pages = "66",
    year = "2018"
}

@article{Grobner:2020drr,
    author = {Gr{\"o}bner, Matthias and Ishibashi, Wako and Tiwari, Shubhanshu and Haney, Maria and Jetzer, Philippe},
    title = "{Binary black hole mergers in AGN accretion discs: gravitational wave rate density estimates}",
    eprint = "2005.03571",
    archivePrefix = "arXiv",
    primaryClass = "astro-ph.GA",
    doi = "10.1051/0004-6361/202037681",
    journal = "Astron. Astrophys.",
    volume = "638",
    pages = "A119",
    year = "2020"
}

@article{Rodriguez:2018rmd,
    author = "Rodriguez, Carl L. and Loeb, Abraham",
    title = "{Redshift Evolution of the Black Hole Merger Rate from Globular Clusters}",
    eprint = "1809.01152",
    archivePrefix = "arXiv",
    primaryClass = "astro-ph.HE",
    doi = "10.3847/2041-8213/aae377",
    journal = "Astrophys. J. Lett.",
    volume = "866",
    number = "1",
    pages = "L5",
    year = "2018"
}

@article{Antonini:2020xnd,
    author = "Antonini, Fabio and Gieles, Mark",
    title = "{Merger rate of black hole binaries from globular clusters: theoretical error bars and comparison to gravitational wave data from GWTC-2}",
    eprint = "2009.01861",
    archivePrefix = "arXiv",
    primaryClass = "astro-ph.HE",
    doi = "10.1103/PhysRevD.102.123016",
    journal = "Phys. Rev. D",
    volume = "102",
    pages = "123016",
    year = "2020"
}

@article{Bautista:2022wjf,
    author = "Bautista, Yilber Fabian and Guevara, Alfredo and Kavanagh, Chris and Vines, Justin",
    title = "{Scattering in black hole backgrounds and higher-spin amplitudes. Part II}",
    eprint = "2212.07965",
    archivePrefix = "arXiv",
    primaryClass = "hep-th",
    doi = "10.1007/JHEP05(2023)211",
    journal = "JHEP",
    volume = "05",
    pages = "211",
    year = "2023"
}

@article{Dolan:2008kf,
    author = "Dolan, Sam R.",
    title = "{Scattering and Absorption of Gravitational Plane Waves by Rotating Black Holes}",
    eprint = "0801.3805",
    archivePrefix = "arXiv",
    primaryClass = "gr-qc",
    doi = "10.1088/0264-9381/25/23/235002",
    journal = "Class. Quant. Grav.",
    volume = "25",
    pages = "235002",
    year = "2008"
}

@article{Chan:2025wgz,
    author = "Chan, Juno C. L. and Dyson, Conor and Garcia, Matilde and Redondo-Yuste, Jaime and Vujeva, Luka",
    title = "{Lensing and wave optics in the strong field of a black hole}",
    eprint = "2502.14073",
    archivePrefix = "arXiv",
    primaryClass = "gr-qc",
    doi = "10.1103/6h6r-46cd",
    journal = "Phys. Rev. D",
    volume = "112",
    number = "6",
    pages = "064009",
    year = "2025"
}

@article{Saketh:2022wap,
    author = "Saketh, M. V. S. and Vines, Justin",
    title = "{Scattering of gravitational waves off spinning compact objects with an effective worldline theory}",
    eprint = "2208.03170",
    archivePrefix = "arXiv",
    primaryClass = "gr-qc",
    doi = "10.1103/PhysRevD.106.124026",
    journal = "Phys. Rev. D",
    volume = "106",
    number = "12",
    pages = "124026",
    year = "2022"
}

@article{Antonini:2013tea,
    author = "Antonini, Fabio and Murray, Norman and Mikkola, Seppo",
    title = "{Black hole triple dynamics: breakdown of the orbit average approximation and implications for gravitational wave detections}",
    eprint = "1308.3674",
    archivePrefix = "arXiv",
    primaryClass = "astro-ph.HE",
    doi = "10.1088/0004-637X/781/1/45",
    journal = "Astrophys. J.",
    volume = "781",
    pages = "45",
    year = "2014"
}

@article{Silsbee:2016djf,
    author = "Silsbee, Kedron and Tremaine, Scott",
    title = "{Lidov-Kozai Cycles with Gravitational Radiation: Merging Black Holes in Isolated Triple Systems}",
    eprint = "1608.07642",
    archivePrefix = "arXiv",
    primaryClass = "astro-ph.HE",
    doi = "10.3847/1538-4357/aa5729",
    journal = "Astrophys. J.",
    volume = "836",
    number = "1",
    pages = "39",
    year = "2017"
}

@article{Fragione:2019zhm,
    author = "Fragione, Giacomo and Loeb, Abraham",
    title = "{Black hole{\textendash}neutron star mergers from triples}",
    eprint = "1903.10511",
    archivePrefix = "arXiv",
    primaryClass = "astro-ph.GA",
    doi = "10.1093/mnras/stz1131",
    journal = "Mon. Not. Roy. Astron. Soc.",
    volume = "486",
    number = "3",
    pages = "4443--4450",
    year = "2019"
}

@article{Liu:2020gif,
    author = "Liu, Bin and Lai, Dong",
    title = "{Hierarchical Black-Hole Mergers in Multiple Systems: Constrain the Formation of GW190412, GW190814 and GW190521-like events}",
    eprint = "2009.10068",
    archivePrefix = "arXiv",
    primaryClass = "astro-ph.HE",
    doi = "10.1093/mnras/stab178",
    journal = "Mon. Not. Roy. Astron. Soc.",
    volume = "502",
    number = "2",
    pages = "2049--2064",
    year = "2021"
}

@article{Graham:2020gwr,
    author = "Graham, M. J. and others",
    title = "{Candidate Electromagnetic Counterpart to the Binary Black Hole Merger Gravitational Wave Event S190521g}",
    eprint = "2006.14122",
    archivePrefix = "arXiv",
    primaryClass = "astro-ph.HE",
    doi = "10.1103/PhysRevLett.124.251102",
    journal = "Phys. Rev. Lett.",
    volume = "124",
    number = "25",
    pages = "251102",
    year = "2020"
}

@article{Morton:2023wxg,
    author = "Morton, Sophia L. and Rinaldi, Stefano and Torres-Orjuela, Alejandro and Derdzinski, Andrea and Vaccaro, Maria Paola and Del Pozzo, Walter",
    title = "{GW190521: A binary black hole merger inside an active galactic nucleus?}",
    eprint = "2310.16025",
    archivePrefix = "arXiv",
    primaryClass = "gr-qc",
    doi = "10.1103/PhysRevD.108.123039",
    journal = "Phys. Rev. D",
    volume = "108",
    number = "12",
    pages = "123039",
    year = "2023"
}

@article{Leong:2024nnx,
    author = "Leong, Samson H. W. and Janquart, Justin and Sharma, Aditya Kumar and Martens, Paul and Ajith, Parameswaran and Hannuksela, Otto A.",
    title = "{Constraining Binary Mergers in Active Galactic Nuclei Disks Using the Nonobservation of Lensed Gravitational Waves}",
    eprint = "2408.13144",
    archivePrefix = "arXiv",
    primaryClass = "astro-ph.HE",
    reportNumber = "LIGO DCC - P2400346",
    doi = "10.3847/2041-8213/ad9ead",
    journal = "Astrophys. J. Lett.",
    volume = "979",
    number = "2",
    pages = "L27",
    year = "2025"
}

@article{Yennie:1954zz,
    author = "Yennie, D. R. and Ravenhall, D. G. and Wilson, R. N.",
    title = "{Phase-Shift Calculation of High-Energy Electron Scattering}",
    doi = "10.1103/PhysRev.95.500",
    journal = "Phys. Rev.",
    volume = "95",
    pages = "500--512",
    year = "1954"
}

@misc{Mathematica, 
author = {}, 
title = {Wolfram Research{,} Inc. Mathematica, {V}ersion 13.2}, 
note = {Champaign, IL, 2022}}

@article{Virtanen:2019joe,
    author = "Virtanen, Pauli and others",
    title = "{SciPy 1.0--Fundamental Algorithms for Scientific Computing in Python}",
    eprint = "1907.10121",
    archivePrefix = "arXiv",
    primaryClass = "cs.MS",
    doi = "10.1038/s41592-019-0686-2",
    journal = "Nature Meth.",
    volume = "17",
    pages = "261",
    year = "2020"
}

@misc{pycbc_github,
  author = {Alex Nitz and others},
  title = {PyCBC Software},
  year = {2024},
  doi  = {10.5281/zenodo.596388},
  howpublished = {\url{https://github.com/gwastro/pycbc}},
}

@Article{Hunter:2007,
  Author    = {Hunter, J. D.},
  Title     = {Matplotlib: A 2D graphics environment},
  Journal   = {Computing in Science \& Engineering},
  Volume    = {9},
  Number    = {3},
  Pages     = {90--95},
  publisher = {IEEE COMPUTER SOC},
  doi       = {10.1109/MCSE.2007.55},
  year      = 2007
}

@article{behnel2011cython,
title={Cython: The best of both worlds},
author={Behnel, Stefan and Bradshaw, Robert and Citro, Craig and Dalcin, Lisandro and Seljebotn, Dag Sverre and Smith, Kurt},
journal={Computing in Science \& Engineering},
volume={13},
number={2},
pages={31--39},
year={2011},
publisher={IEEE}
}

@article{Harris:2020xlr,
    author = "Harris, Charles R. and others",
    title = "{Array programming with NumPy}",
    eprint = "2006.10256",
    archivePrefix = "arXiv",
    primaryClass = "cs.MS",
    doi = "10.1038/s41586-020-2649-2",
    journal = "Nature",
    volume = "585",
    number = "7825",
    pages = "357--362",
    year = "2020"
}

@MISC{2020ascl.soft12021L,
       author = {{LIGO Scientific Collaboration}},
        title = "{LALSuite: LIGO Scientific Collaboration Algorithm Library Suite}",
 howpublished = {Astrophysics Source Code Library, record ascl:2012.021},
         year = 2020,
        month = dec,
          eid = {ascl:2012.021},
        pages = {ascl:2012.021},
archivePrefix = {ascl},
       eprint = {2012.021},
       adsurl = {https://ui.adsabs.harvard.edu/abs/2020ascl.soft12021L}
}

@conference{Kluyver2016jupyter,
    Title = {Jupyter Notebooks -- a publishing format for reproducible computational workflows},
    Author = {Thomas Kluyver and others},
    Booktitle = {Positioning and Power in Academic Publishing: Players, Agents and Agendas},
    Editor = {F. Loizides and B. Schmidt},
    Organization = {IOS Press},
    Pages = {87 - 90},
    Year = {2016}
}

@article{MST,
    author = {Mano, Shuhei and Suzuki, Hisao and Takasugi, Eiichi},
    title = {Analytic Solutions of the Regge-Wheeler Equation and the Post-Minkowskian Expansion},
    journal = {Progress of Theoretical Physics},
    volume = {96},
    number = {3},
    pages = {549-565},
    year = {1996},
    month = {09},
    issn = {0033-068X},
    doi = {10.1143/PTP.96.549},
    url = {https://doi.org/10.1143/PTP.96.549},
}

@article{MT,
    author = {Shuhei, Mano and Eiichi, Takasugi},
    title = {Analytic Solutions of the Teukolsky Equation and Their Properties},
    journal = {Progress of Theoretical Physics},
    volume = {97},
    number = {2},
    pages = {213-232},
    year = {1997},
    month = {02},
    issn = {0033-068X},
    doi = {10.1143/PTP.97.213},
    url = {https://doi.org/10.1143/PTP.97.213},
}

@article{FLINT,
  author = {Fredrik Johansson},
  journal = {{IEEE} Transactions on Computers},
  title = {{A}rb: Efficient Arbitrary-Precision Midpoint-Radius Interval Arithmetic},
  year = {2017},
  volume = {66},
  number = {8},
  pages = {1281-1292},
  doi = {10.1109/TC.2017.2690633}
}

@book{hardy2024divergent,
  title={Divergent Series},
  author={Hardy, G.H.},
  isbn={9781470477851},
  lccn={91075377},
  series={AMS Chelsea Publishing},
  url={https://books.google.co.in/books?id=XZ0TEQAAQBAJ},
  year={2024},
  publisher={American Mathematical Society}
}

@article{Ubach:2025dob,
    author = "Ubach, Helena and Gieles, Mark and Miralda-Escud{\'e}, Jordi",
    title = "{Constraining the environment of compact binary mergers with self-lensing signatures}",
    eprint = "2505.04794",
    archivePrefix = "arXiv",
    primaryClass = "astro-ph.HE",
    doi = "10.1103/ql7q-t6wc",
    journal = "Phys. Rev. D",
    volume = "112",
    number = "8",
    pages = "083026",
    year = "2025"
}

@article{Finn:1992xs,
    author = "Finn, Lee Samuel and Chernoff, David F.",
    title = "{Observing binary inspiral in gravitational radiation: One interferometer}",
    eprint = "gr-qc/9301003",
    archivePrefix = "arXiv",
    reportNumber = "PRINT-93-0138 (NORTHWESTERN)",
    doi = "10.1103/PhysRevD.47.2198",
    journal = "Phys. Rev. D",
    volume = "47",
    pages = "2198--2219",
    year = "1993"
}

@article{Cutler:1994ys,
    author = "Cutler, Curt and Flanagan, Eanna E.",
    title = "{Gravitational waves from merging compact binaries: How accurately can one extract the binary's parameters from the inspiral wave form?}",
    eprint = "gr-qc/9402014",
    archivePrefix = "arXiv",
    reportNumber = "GRP-369",
    doi = "10.1103/PhysRevD.49.2658",
    journal = "Phys. Rev. D",
    volume = "49",
    pages = "2658--2697",
    year = "1994"
}

@article{Flanagan:1997kp,
    author = "Flanagan, Eanna E. and Hughes, Scott A.",
    title = "{Measuring gravitational waves from binary black hole coalescences: 2. The Waves' information and its extraction, with and without templates}",
    eprint = "gr-qc/9710129",
    archivePrefix = "arXiv",
    doi = "10.1103/PhysRevD.57.4566",
    journal = "Phys. Rev. D",
    volume = "57",
    pages = "4566--4587",
    year = "1998"
}

@article{Flanagan:1997sx,
    author = "Flanagan, Eanna E. and Hughes, Scott A.",
    title = "{Measuring gravitational waves from binary black hole coalescences: 1. Signal-to-noise for inspiral, merger, and ringdown}",
    eprint = "gr-qc/9701039",
    archivePrefix = "arXiv",
    reportNumber = "GRP-456",
    doi = "10.1103/PhysRevD.57.4535",
    journal = "Phys. Rev. D",
    volume = "57",
    pages = "4535--4565",
    year = "1998"
}

@misc{BHPToolkit,
  title = {{Black Hole Perturbation Toolkit}},
  howpublished = {(\href{http://bhptoolkit.org/}{bhptoolkit.org})},
}

@article{ReggePhysRev.108.1063,
  title = {Stability of a Schwarzschild Singularity},
  author = {Regge, Tullio and Wheeler, John A.},
  journal = {Phys. Rev.},
  volume = {108},
  issue = {4},
  pages = {1063--1069},
  numpages = {0},
  year = {1957},
  month = {Nov},
  publisher = {American Physical Society},
  doi = {10.1103/PhysRev.108.1063},
  url = {https://link.aps.org/doi/10.1103/PhysRev.108.1063}
}

@article{ZerilliPhysRevLett.24.737,
  title = {Effective Potential for Even-Parity Regge-Wheeler Gravitational Perturbation Equations},
  author = {Zerilli, Frank J.},
  journal = {Phys. Rev. Lett.},
  volume = {24},
  issue = {13},
  pages = {737--738},
  numpages = {0},
  year = {1970},
  month = {Mar},
  publisher = {American Physical Society},
  doi = {10.1103/PhysRevLett.24.737},
  url = {https://link.aps.org/doi/10.1103/PhysRevLett.24.737}
}

@article{Teuk1PhysRevLett.29.1114,
  title = {Rotating Black Holes: Separable Wave Equations for Gravitational and Electromagnetic Perturbations},
  author = {Teukolsky, Saul A.},
  journal = {Phys. Rev. Lett.},
  volume = {29},
  issue = {16},
  pages = {1114--1118},
  numpages = {0},
  year = {1972},
  month = {Oct},
  publisher = {American Physical Society},
  doi = {10.1103/PhysRevLett.29.1114},
  url = {https://link.aps.org/doi/10.1103/PhysRevLett.29.1114}
}

@article{Deka:2024ecp,
    author = "Deka, Uddeepta and Chakraborty, Sumanta and Kapadia, Shasvath J. and Shaikh, Md Arif and Ajith, Parameswaran",
    title = "{Probing the charge of compact objects with gravitational microlensing of gravitational waves}",
    eprint = "2401.06553",
    archivePrefix = "arXiv",
    primaryClass = "gr-qc",
    doi = "10.1103/PhysRevD.111.064028",
    journal = "Phys. Rev. D",
    volume = "111",
    number = "6",
    pages = "064028",
    year = "2025"
}

@article{PhysRevD.93.044007,
  title = {Frequency-domain gravitational waves from nonprecessing black-hole binaries. II. A phenomenological model for the advanced detector era},
  author = {Khan, Sebastian and Husa, Sascha and Hannam, Mark and Ohme, Frank and P\"urrer, Michael and Forteza, Xisco Jim\'enez and Boh\'e, Alejandro},
  journal = {Phys. Rev. D},
  volume = {93},
  issue = {4},
  pages = {044007},
  numpages = {27},
  year = {2016},
  month = {Feb},
  publisher = {American Physical Society},
  doi = {10.1103/PhysRevD.93.044007},
  url = {https://link.aps.org/doi/10.1103/PhysRevD.93.044007}
}

@article{Husa:2015iqa,
    author = {Husa, Sascha and Khan, Sebastian and Hannam, Mark and P{\"u}rrer, Michael and Ohme, Frank and Jim{\'e}nez Forteza, Xisco and Boh{\'e}, Alejandro},
    title = "{Frequency-domain gravitational waves from nonprecessing black-hole binaries. I. New numerical waveforms and anatomy of the signal}",
    eprint = "1508.07250",
    archivePrefix = "arXiv",
    primaryClass = "gr-qc",
    doi = "10.1103/PhysRevD.93.044006",
    journal = "Phys. Rev. D",
    volume = "93",
    number = "4",
    pages = "044006",
    year = "2016"
}

@article{Santos:2025ass,
    author = "Santos, Jo{\~a}o S. and Cardoso, Vitor and Nat{\'a}rio, Jos{\'e} and van de Meent, Maarten",
    title = "{Gravitational Waves from Binary Extreme Mass Ratio Inspirals: Doppler Shift and Beaming, Resonant Excitation, Helicity Oscillations, and Self-Lensing}",
    eprint = "2506.14868",
    archivePrefix = "arXiv",
    primaryClass = "gr-qc",
    doi = "10.1103/qq3m-6phh",
    journal = "Phys. Rev. Lett.",
    volume = "135",
    number = "21",
    pages = "211402",
    year = "2025"
}

@article{CarrilloGonzalez:2025gqm,
    author = "Carrillo Gonzalez, Mariana and De Luca, Valerio and Garoffolo, Alice and Parra-Martinez, Julio and Trodden, Mark",
    title = "{A scattering perspective on gravitational lensing}",
    journal = "arXiv:2511.15797",
    eprint = "2511.15797",
    archivePrefix = "arXiv",
    primaryClass = "hep-th",
    month = "11",
    year = "2025"
}

@article{Liu:2023emk,
    author = "Liu, Anna and Kim, Kyungmin",
    title = "{Can we discern millilensed gravitational-wave signals from signals produced by precessing binary black holes with ground-based detectors?}",
    eprint = "2301.07253",
    archivePrefix = "arXiv",
    primaryClass = "gr-qc",
    reportNumber = "LIGO-P2200398",
    doi = "10.1103/PhysRevD.110.123008",
    journal = "Phys. Rev. D",
    volume = "110",
    number = "12",
    pages = "123008",
    year = "2024"
}

@article{Liu:2024xxn,
    author = "Liu, Anna and Chandramouli, Rohit S. and Hannuksela, Otto A. and Yunes, Nicol{\'a}s and Li, Tjonnie G. F.",
    title = "{Millilensing induced systematic biases in parametrized tests of general relativity}",
    eprint = "2410.21738",
    archivePrefix = "arXiv",
    primaryClass = "gr-qc",
    doi = "10.1103/bsml-2rdh",
    journal = "Phys. Rev. D",
    volume = "112",
    number = "6",
    pages = "064043",
    year = "2025"
}

@article{Ezquiaga:2025gkd,
    author = "Ezquiaga, Jose Mar{\'\i}a and Lo, Rico K. L. and Vujeva, Luka",
    title = "{Diffraction around caustics in gravitational wave lensing}",
    eprint = "2503.22648",
    archivePrefix = "arXiv",
    primaryClass = "gr-qc",
    doi = "10.1103/tkk2-x9st",
    journal = "Phys. Rev. D",
    volume = "112",
    number = "4",
    pages = "043544",
    year = "2025"
}

@article{Lo:2024wqm,
    author = "Lo, Rico K. L. and Vujeva, Luka and Ezquiaga, Jose Mar{\'\i}a and Chan, Juno C. L.",
    title = "{Observational Signatures of Highly Magnified Gravitational Waves from Compact Binary Coalescence}",
    eprint = "2407.17547",
    archivePrefix = "arXiv",
    primaryClass = "gr-qc",
    doi = "10.1103/PhysRevLett.134.151401",
    journal = "Phys. Rev. Lett.",
    volume = "134",
    number = "15",
    pages = "151401",
    year = "2025"
}

@article{Vujeva:2025nwg,
    author = "Vujeva, Luka and Ezquiaga, Jose Mar{\'\i}a and Gilman, Daniel and Goyal, Srashti and Zumalac{\'a}rregui, Miguel",
    title = "{Dark Matter Subhalos and Higher Order Catastrophes in Gravitational Wave Lensing}",
    eprint = "2510.14953",
    journal = "arXiv",
    primaryClass = "astro-ph.CO",
    month = "10",
    year = "2025"
}

@article{Bulashenko:2025vdx,
    author = "Bulashenko, Oleg and Villanueva, Nino and Nerin, Roberto Bada and Font, Jos{\'e} A.",
    title = "{On the observation of cosmic strings via gravitational-wave lensing}",
    eprint = "2510.20442",
    journal = "arXiv",
    primaryClass = "gr-qc",
    reportNumber = "LIGO-P2500643, VIR-0974A-25",
    month = "10",
    year = "2025"
}

@article{Vujeva:2025kko,
    author = "Vujeva, Luka and Ezquiaga, Jose Mar{\'\i}a and Lo, Rico K. L. and Chan, Juno C. L.",
    title = "{Effects of galaxy cluster structure on lensed gravitational waves}",
    eprint = "2501.02096",
    archivePrefix = "arXiv",
    primaryClass = "astro-ph.CO",
    doi = "10.1103/1hmj-pxjr",
    journal = "Phys. Rev. D",
    volume = "112",
    number = "6",
    pages = "063044",
    year = "2025"
}

@article{Lindblom:2008cm,
    author = "Lindblom, Lee and Owen, Benjamin J. and Brown, Duncan A.",
    title = "{Model Waveform Accuracy Standards for Gravitational Wave Data Analysis}",
    eprint = "0809.3844",
    archivePrefix = "arXiv",
    primaryClass = "gr-qc",
    doi = "10.1103/PhysRevD.78.124020",
    journal = "Phys. Rev. D",
    volume = "78",
    pages = "124020",
    year = "2008"
}

@article{Cornish:2011ys,
    author = "Cornish, Neil and Sampson, Laura and Yunes, Nicolas and Pretorius, Frans",
    title = "{Gravitational Wave Tests of General Relativity with the Parameterized Post-Einsteinian Framework}",
    eprint = "1105.2088",
    archivePrefix = "arXiv",
    primaryClass = "gr-qc",
    doi = "10.1103/PhysRevD.84.062003",
    journal = "Phys. Rev. D",
    volume = "84",
    pages = "062003",
    year = "2011"
}

@article{Thompson:2025hhc,
    author = "Thompson, Jonathan E. and Hoy, Charlie and Fauchon-Jones, Edward and Hannam, Mark",
    title = "{Use and interpretation of signal-model indistinguishability measures for gravitational-wave astronomy}",
    eprint = "2506.10530",
    archivePrefix = "arXiv",
    primaryClass = "gr-qc",
    reportNumber = "LIGO-P2500361",
    doi = "10.1103/ddz7-x9zz",
    journal = "Phys. Rev. D",
    volume = "112",
    number = "6",
    pages = "064011",
    year = "2025"
}

@article{Damour:2010zb,
    author = "Damour, Thibault and Nagar, Alessandro and Trias, Miguel",
    title = "{Accuracy and effectualness of closed-form, frequency-domain waveforms for non-spinning black hole binaries}",
    eprint = "1009.5998",
    archivePrefix = "arXiv",
    primaryClass = "gr-qc",
    reportNumber = "LIGO-P1000099-V3",
    doi = "10.1103/PhysRevD.83.024006",
    journal = "Phys. Rev. D",
    volume = "83",
    pages = "024006",
    year = "2011"
}

@article{Nelder:1965zz,
    author = "Nelder, J. A. and Mead, R.",
    title = "{A Simplex Method for Function Minimization}",
    doi = "10.1093/comjnl/7.4.308",
    journal = "Comput. J.",
    volume = "7",
    pages = "308--313",
    year = "1965"
}

@article{Dhurandhar:1992mw,
    author = "Dhurandhar, S. V. and Sathyaprakash, B. S.",
    title = "{Choice of filters for the detection of gravitational waves from coalescing binaries. 2. Detection in colored noise}",
    reportNumber = "IC-92-345",
    doi = "10.1103/PhysRevD.49.1707",
    journal = "Phys. Rev. D",
    volume = "49",
    pages = "1707--1722",
    year = "1994"
}

@article{Allen:2005fk,
    author = "Allen, Bruce and Anderson, Warren G. and Brady, Patrick R. and Brown, Duncan A. and Creighton, Jolien D. E.",
    title = "{FINDCHIRP: An Algorithm for detection of gravitational waves from inspiraling compact binaries}",
    eprint = "gr-qc/0509116",
    archivePrefix = "arXiv",
    doi = "10.1103/PhysRevD.85.122006",
    journal = "Phys. Rev. D",
    volume = "85",
    pages = "122006",
    year = "2012"
}

@article{Usman:2015kfa,
    author = "Usman, Samantha A. and others",
    title = "{The PyCBC search for gravitational waves from compact binary coalescence}",
    eprint = "1508.02357",
    archivePrefix = "arXiv",
    primaryClass = "gr-qc",
    reportNumber = "LIGO-P1500086",
    doi = "10.1088/0264-9381/33/21/215004",
    journal = "Class. Quant. Grav.",
    volume = "33",
    number = "21",
    pages = "215004",
    year = "2016"
}

@article{DelPozzo:2014cla,
    author = "Del Pozzo, Walter and Grover, Katherine and Mandel, Ilya and Vecchio, Alberto",
    title = "{Testing general relativity with compact coalescing binaries: comparing exact and predictive methods to compute the Bayes factor}",
    eprint = "1408.2356",
    archivePrefix = "arXiv",
    primaryClass = "gr-qc",
    doi = "10.1088/0264-9381/31/20/205006",
    journal = "Class. Quant. Grav.",
    volume = "31",
    number = "20",
    pages = "205006",
    year = "2014"
}

@book{Jeffreys:1939xee,
    author = "Jeffreys, Harold",
    title = "{The Theory of Probability}",
    isbn = "978-0-19-850368-2, 978-0-19-853193-7",
    series = "Oxford Classic Texts in the Physical Sciences",
    year = "1939"
}

@article{Ajith:2012mn,
    author = "Ajith, P. and Fotopoulos, N. and Privitera, S. and Neunzert, A. and Weinstein, A. J.",
    title = "{Effectual template bank for the detection of gravitational waves from inspiralling compact binaries with generic spins}",
    eprint = "1210.6666",
    archivePrefix = "arXiv",
    primaryClass = "gr-qc",
    reportNumber = "LIGO-P1200106-V2, LIGO-P1200106-V3",
    doi = "10.1103/PhysRevD.89.084041",
    journal = "Phys. Rev. D",
    volume = "89",
    number = "8",
    pages = "084041",
    year = "2014"
}

\end{document}